\documentclass[a4paper,10pt]{article}
\pdfoutput=1 

\usepackage{jheppub} 

\usepackage[T1]{fontenc} 

\usepackage{graphicx,float}
\usepackage{verbatim,epsfig}
\usepackage{epstopdf}
\usepackage[utf8x]{inputenc}
\usepackage[colorlinks=true,linkcolor=blue,citecolor=blue]{hyperref}
  \usepackage{bm}
   \usepackage{amsmath}
    \usepackage{amssymb}
     \usepackage{pifont}
\usepackage{rotating}

\newcommand{\nn}{\nonumber}

\newcommand{\ot}{\leftarrow}

\renewcommand{\(}{\left(}
\renewcommand{\)}{\right)}
\renewcommand{\[}{\left[}
\renewcommand{\]}{\right]}

\newcommand{\GeV}{\text{GeV}}

\renewcommand{\vec}[1]{\bm{#1}}

\newcommand{\specialcell}[2][c]{\begin{tabular}[#1]{@{}l@{}}#2\end{tabular}}
\newcommand{\specialcellcenter}[2][c]{\begin{tabular}[#1]{@{}c@{}}#2\end{tabular}}

\title{Analysis of vector boson production within TMD factorization}

\author[a]{Ignazio Scimemi}
\author[b]{and Alexey Vladimirov}

\affiliation[a]{Departamento de F\' isica Te\'orica II, Universidad Complutense de Madrid,\\
Ciudad Universitaria, 28040 Madrid, Spain}
\affiliation[b]{Institut f\"ur Theoretische Physik, Universit\"at Regensburg,\\
D-93040 Regensburg, Germany}

\emailAdd{ignazios@fis.ucm.es} \emailAdd{aleksey.vladimirov@gmail.com}

\abstract{We present a  comprehensive analysis and extraction of the unpolarized transverse momentum dependent (TMD) parton distribution functions, which are fundamental constituents of the TMD factorization theorem.  We provide a general review of the theory of TMD distributions, and present a new scheme of scale fixation. This scheme, called the $\zeta$-prescription, allows to minimize the impact of perturbative logarithms in a large range of scales and does not generate undesired power corrections. Within $\zeta$-prescription we consistently include the perturbatively calculable parts up to next-to-next-to-leading order (NNLO), and perform the global fit of the Drell-Yan and Z-boson production, which include the data of E288, Tevatron and LHC experiments. The non-perturbative part of the TMDs are explored checking  a variety of models. We support the obtained results by a study of theoretical uncertainties, perturbative convergence, and a dedicated study of the range of applicability of the TMD factorization theorem. The considered non-perturbative models present significant differences in the fitting behavior, which allow us to clearly disfavor most of them. The numerical evaluations are provided by the \texttt{arTeMiDe} code, which is introduced in this work and that can be used  for current/future TMD phenomenology.}

\begin{document}
\maketitle
\flushbottom

\section{Introduction}

The transverse momentum dependent (TMD) distributions are universal functions that describe the interactions of partons in a hadron. The TMD distributions naturally appear within the TMD factorization theorem for the differential cross section of double-inclusive hard processes. A lot of efforts have been made to arrive to a comprehensive picture of TMD factorization (for the latest works see \cite{Collins:2011zzd,GarciaEchevarria:2011rb,Echevarria:2012js,Echevarria:2014rua,Gaunt:2014ska,Becher:2010tm,Chiu:2012ir,Mantry:2010bi}). In this work we perform a detailed comparison of the experimental measurements with the theory expectations based on our studies of higher-order perturbative expansions and power corrections for unpolarized TMDPDFs made in refs.~\cite{Echevarria:2015usa,Echevarria:2015byo,Echevarria:2016scs,Scimemi:2016ffw}.

Among many different spin (in)dependent TMD distributions, the unpolarized TMD parton distribution functions (TMDPDFs) play a central role. From the practical point of view, their precise knowledge is required to extract further TMD distributions and perform other precision measurements. The ideal process to study the unpolarized TMDPDFs is the unpolarized vector boson production. The data on the $q_T$-dependent cross-section for the Drell-Yan process are collected by many experiments, including the precise measurements done by Tevatron and LHC. The theoretical descriptions of Drell-Yan data were made by many groups within one or another form of TMD factorization (see e.g. \cite{Landry:1999an,Qiu:2000hf,Landry:2002ix,Watt:2003vf,Catani:2007vq,Bozzi:2010xn,Mantry:2010bi,Su:2014wpa,DAlesio:2014mrz,Catani:2015vma,Bacchetta:2017gcc}).

This work presents a number of differences with the respect to the previous literature.  The collection of the improvements  forms a completely new point of view in the TMD phenomenology. The main difference of the present work with respect to  the  more standard ones (here we  consider as the most spread out,  and {\it de facto} standard, analyses those based on the codes \texttt{ResBos}\cite{Landry:2002ix,Guzey:2012jp} and \texttt{DYqT}/\texttt{DYRes} \cite{Catani:2007vq,Bozzi:2010xn,Catani:2015vma}) are following: (i) We extract the parameters related to individual TMDPDFs, which are suitable for phenomenological description of other TMD-related processes. (ii) We consistently include the perturbative ingredients, such as coefficient functions and anomalous dimensions, at the next-to-next-to-leading order (NNLO), introducing and using the $\zeta$-prescription to solve the problem of perturbative convergence at large-$b$ (where $b$ is the transverse distance). (iii) The TMDPDF parameterization is based on and is consistent with the theory expectation on the TMD behavior with $b$. To our knowledge this is the first attempt to include in a fit both high and low energy data at NNLO precision. The extraction of TMDs takes into account (for the first time to our knowledge) also LHC data. All this represents for us a clear improvement with respect to the more classical analyses.

In a modern view, a TMD distribution is a cumbersome function of many factors, which mix up perturbative and non-perturbative information. In this context, the issue of the separation of perturbative and non-perturbative physics requires a fine analysis and it is open to different solutions. The $\zeta$-prescription proposed in this work, is an attempt to consider the perturbative input to a TMD distribution as it is, without artificial regulators. The $\zeta$-prescription is founded on the fact that the TMD factorization introduces two factorization scales, one for the collinear and one for the soft exchanges. These scales are usually fixed to the same point, while in the $\zeta$-prescription they are chosen to eliminate the problematic double-logarithmic contributions. In other words, the $\zeta$-prescription is based on the freedom to select the normalization and factorization scales, which is guaranteed by the structure of the perturbative theory. The $\zeta$-prescription is essentially different from other used schemes. In particular, it does not strictly solve the problem of the large logarithmic contributions at large-$b$. It only decreases the power of the logarithmic contributions. However, the $x$-dependence of the remaining logarithmic terms has a form which prevents the blow up of the perturbative series, which is not accidental, but the result of the charge conservation. In this way, the $\zeta$-prescription postpones the large logarithm problem to the very far domain of $b$-space, where other factors suppress a TMD distribution. The practical implementation of the $\zeta$-prescription shows that it is efficacious and it allows a very accurate and sound  description of the data.

The description of the non-perturbative parts of TMD distributions is the most interesting task.  It is highly non-trivial because the definition of the non-perturbative part is strongly affected by schemes and prescriptions used in the perturbative implementation. In this respect a full NNLO can be clarifying. As an example, we recall that the non-perturbative behavior of the TMDPDFs is often assumed to have a Gaussian shape (see e.g. discussions in \cite{Landry:2002ix,Guzzi:2013aja,Schweitzer:2010tt,Bacchetta:2017gcc}). Although the Gaussian ansatz is widely used, it comes in the conflict with the usual picture of long-distance strong interaction fueled by light-meson exchanges. The typically expected behavior at long distances is exponential, which is confirmed also by model calculations \cite{Schweitzer:2012hh}.  However, the Gaussian shape is often introduced together with the $b^*$-prescription \cite{Collins:1981va}. Notwithstanding many positive features, the $b^*$-prescription has a serious issue: it introduces undesired $b$-even power corrections. In turn, these power correction introduced by $b^*$ can easily simulate the Gaussian behavior (see also discussion in \cite{Collins:2014jpa}). Once the $b^*$-prescription is removed the Gaussian ansatz for the TMD shape is no more essential, according to what we find.

An additional remarkable point of the present study is the wide range of energies covered by the data that we have analyzed. The lowest energy measurements included in the fits  have $(Q,\sqrt{s})=(4,19.4)\;\GeV$ (E288 experiment \cite{Ito:1980ev}), while the most energetic have $(Q,\sqrt{s})=(116-150,8\cdot 10^3)\; \GeV$ (ATLAS collaboration \cite{Aad:2015auj}). Typically, the low- and high-energy data are considered separately. The main reason for a separate scan is the assumed physical picture of strong interactions, which describes different energies. The description of the high-energy data requires a precise perturbative input and it is expected to be less sensitive to the fine non-perturbative dynamics. The situation is the opposite for the low-energy measurements. Our experience shows that the inclusion of data of different energies is not only possible within TMD formalism, but it is also desired because it cuts away inappropriate models very sharply. We find also that the precision achieved by LHC is already sensitive enough to the non-perturbative structure of TMDs.  We show that low and high energy data are sensitive to different regions of $b$-space, and consequently to different non-perturbative regimes of the TMDs: high energy data are better described by a Gaussian non-perturbative correction, while low energy data prefer an exponential type of non-perturbative models. The code  (\texttt{arTemiDe}) that we have prepared allows to test all these hypotheses, and  can be adaptded also to test different non-perturbative inputs for TMDs.

In order to extract the non-perturbative core of the TMDs, in the present study we choose a neutral tactic. We have scanned many possibilities such as a Gaussian/exponential behavior, with/without inclusion of power corrections, and so on. We have also studied the non-perturbative correction to the evolution kernel. During the examination of models we have prioritize the following criteria:
\begin{description}
\item{(i) \textit{Stability.}} The TMD factorization is valid at small-$q_T$ (the dilepton transverse momentum) up to a certain limit. Therefore, an acceptable model should produce a stable and good description within the allowed $q_T$-range. In other words, the value of $\chi^2$ should be valuable and values of parameters are stable independently on the number of included data points, as far as  the points belong to the  allowed range.
\item{(ii) \textit{Convergence.}} The agreement with data should improve with the increase of the perturbative order. Given the current state of the art of the theory, we can define four successive perturbative orders, which is enough to test the perturbative convergence. Also, the value of the phenomenological non-perturbative  constants that one extracts should converge to some central value.
\item{(iii) \textit{$\chi^2$ minimization.}} Naturally, among the models with similar behavior we select the model with the minimal $\chi^2$. We have found that it is difficult to find a model (with one or two parameters), which fulfills  the demands (i) and (ii), and that at the same time  provides  a good  $\chi^2$ value on the whole set of data points (although it is relatively easy to achieve this selecting a particular experiment). The models that we test consider a kind of minimal set of parameters which can be enlarged in future studies, refining the fitting hypotheses.
\end{description}
In the present fit, we have included the measurements of E288 at low-energies, Z-boson production at CDF, D0, ATLAS, CMS and LHCb, and Drell-Yan measurements from ATLAS. To our knowledge, this is the largest set of Drell-Yan data points ever simultaneously considered in a fit within the TMD formalism. We find also that the LHC data below the Z-boson peak and at small $q_T$ are very important for current/future TMD studies. In the article we present the most successful models that we have found, and discuss some popular models. 

In order to numerically evaluate the theoretical expressions, we have produced the package \texttt{arTeMiDe}. \texttt{arTeMiDe} has a flexible module structure and can be used at any level of TMD theory description, from the evaluation of a single TMDPDF or evolution factor to an evaluation of differential cross-section. The \texttt{arTeMiDe} code is available at \cite{web} and can be used to check  our statements or test a possible future/alternative ansatz (for instance \cite{Qiu:2000hf,Kang:2012am}). In \texttt{arTeMiDe} we have collected all recent achievements of TMD theory, including NNLO matching coefficient function, and N$^3$LO TMD anomalous dimensions. In the current version, \texttt{arTeMiDe} evaluates only unpolarized TMDPDFs and related cross-sections, however, we plan to extend it further. 

The body of the article is divided as in the following. In sec.~\ref{sec:theor} we review the theoretical construction of the Drell-Yan cross section and summarize the theoretical knowledge on unpolarized TMDPDFs. In this section, we also describe all the theoretical improvements which are original for this work. The main original point, namely $\zeta$-prescription is presented in sec. \ref{sec:scales} and appendix \ref{app:zeta_main}. The phenomenological studies are presented in sec.~\ref{sec:fit}. This  section includes also a dedicated discussion of the shape of the non-perturbative part of the TMD. The allowed range of validity of the TMD factorization is explored in sec.~\ref{sec:deltaT}, the presentation of theoretical uncertainties is given in sec.\ref{sec:HEE_unser}. The results of the final fit are presented in sec. \ref{sec:global}. A  final discussion and conclusions can be found in sec. \ref{sec:discussion}.

\section{Theoretical framework}
\label{sec:theor}

We consider the Drell-Yan reaction $h_1+h_2\to G(\to ll')+X$, where $G$ is the electroweak neutral gauge boson, $\gamma^*$ or $Z$. The incoming hadrons have momenta $p_1$ and $p_2$ with $(p_1+p_2)^2=s$. The gauge boson decays to the lepton pair with momenta $k_1$ and $k_2$. The momentum of the gauge boson or equivalently the invariant mass of lepton pair is $Q^2=q^2=(k_1+k_2)^2$. The differential cross-section for the Drell-Yan process can be written in the form \cite{Drell:1970wh,Altarelli:1978id}
\begin{eqnarray}\label{th:ds1}
d\sigma=\frac{d^4q}{2s}\sum_{G,G'=\gamma,Z}L_{GG'}^{\mu\nu}W_{\mu\nu}^{GG'}\Delta_G(q)\Delta_{G'}(q),
\end{eqnarray}
where $1/2s$ is the flux factor, $\Delta_G$ is the (Feynman) propagator for the gauge boson $G$. The hadron and lepton tensors are respectively
\begin{eqnarray}
W^{GG'}_{\mu\nu}&=&\int \frac{d^4 z}{(2\pi)^4} e^{-iqz}\langle h_1(p_1)h_2(p_2)|J^G_\mu(z)J^{G'}_\nu(0)|h_1(p_1)h_2(p_2)\rangle,
\\
\label{th:leptonTensor}
L_{\mu\nu}^{GG'}&=&\int\frac{d^3k_1}{(2\pi)^3 2E_1}\frac{d^3k_2}{(2\pi)^3 2E_2} (2\pi)^4 \delta^4(k_1+k_2-q) \langle l_1(k_1) l_2(k_2)|J_{\nu'}^G(0)|0\rangle\langle 0|J_{\mu'}^{G'}(0)|l_1(k_1) l_2(k_2)\rangle,
\nn \\ &&
\end{eqnarray}
where $J_\mu^G$ is the electroweak current.

The point of our interest is the $q_T$ dependence of the cross-section, where $q_T$ is the transverse component of the produced gauge boson in the center-of-mass frame. More precisely, we are interested in the regime $q_T\ll Q$, where the TMD factorization formalism can be applied. Within the TMD factorization, one obtains the following expression for the unpolarized hadron tensor (see e.g.~\cite{Tangerman:1994eh})
\begin{eqnarray}\label{th:hadron_tensor}
W_{\mu\nu}^{GG'}&=&\frac{-g_{T\mu\nu}}{\pi N_c}|C_V(q,\mu)|^2\sum_{f,f'}z_{ff'}^{GG'} \int \frac{d^2\vec b}{4\pi}e^{i(\vec q \vec b)}
F_{f\ot h_1}(x_1,\vec b;\mu,\zeta_1)F_{f'\ot h_2}(x_2,\vec b;\mu,\zeta_2)+Y_{\mu\nu},\nn
\\ &&
\end{eqnarray}
where $g_T$ is the transverse part of the metric tensor and the summation runs over the active quark flavors. The variable $\mu$ is the hard factorization scale.  The variables $\zeta_{1,2}$ are the scale of soft-gluons factorization, and the subject of relation $\zeta_1\zeta_2\simeq Q^4$. In the following, we consider the symmetric point $\zeta_1=\zeta_2=\zeta=Q^2$. The variables $x_{1,2}$ are the longitudinal parts of parton momenta
\begin{eqnarray}
x_1=\frac{\sqrt{Q^2+q_T^2}}{\sqrt{s}}e^y\simeq \frac{Q}{\sqrt{s}}e^y,
\qquad x_2=\frac{\sqrt{Q^2+q_T^2}}{\sqrt{s}}e^{-y}\simeq \frac{Q}{\sqrt{s}}e^{-y}.
\end{eqnarray} 
The factors $z_{ff'}^{GG'}$ are the electro-weak charges. The explicit form of factors $z$ is given in sec.~\ref{sec:z}. The factor $C_V$ is the matching coefficient of the QCD neutral current to the same current  expressed in terms of collinear quark fields. The explicit expressions for $C_V$ can be found in \cite{Kramer:1986sg,Matsuura:1988sm,Idilbi:2006dg}, and are also given in appendix \ref{app:CV}. The functions $F_{f\ot h}$ are the unpolarized TMDPDFs for quark $f$ in the hadron $h$. They are universal non-perturbative functions and the main objects of our study. The details of their definition and their parametrization are given in sec.~\ref{sec:TMDPDF}. Finally, the term $Y$ denotes the power corrections to the TMD factorization theorem (to be distinguished from the power corrections to the TMD operator product expansion). The $Y$-term is of the order $q_T/Q$ and composed of TMD distributions of the higher dynamical twist. In our study, we restrict ourself to the limit of low $q_T$ such that the $Y$-term can be dropped. 

Evaluating the lepton tensor, and combining together all factors one obtains the cross-section for the unpolarized Drell-Yan process at the leading order of TMD factorization in the form \cite{Collins:1984kg,Davies:1984sp,Ellis:1997ii,Becher:2010tm,GarciaEchevarria:2011rb,Collins:2011zzd}
\begin{eqnarray}\label{th:Xsec_gen}
\frac{d\sigma}{dQ^2dyd(q_T^2)}&=& \frac{4\pi}{3N_c}\frac{\mathcal{P}}{sQ^2}
\sum_{GG'}z_{ll'}^{GG'}(q)
\\\nn && \sum_{ff'}z_{ff'}^{GG'} |C_V(q,\mu)|^2\int \frac{d^2\vec b}{4\pi}e^{i(\vec b\vec q)}
F_{f\ot h_1}(x_1,\vec b;\mu,\zeta)F_{f'\ot h_2}(x_2,\vec b;\mu,\zeta)+Y,
\end{eqnarray}
where $y$ is the rapidity of the produced gauge boson.  The factor $\mathcal{P}$ is a part of the lepton tensor and contains information on the fiducial cuts. It is discussed in details in sec.~\ref{sec:cuts}. In the rest of this section a more detailed description of the particular components is presented.
 
\subsection{Expressions for cross-section for different produced bosons}
\label{sec:z}

In the case of neutral vector bosons production, the sum over $G$ and $G'$ in eq.~(\ref{th:Xsec_gen}) has three terms
\begin{eqnarray}\label{th:DY_all_terms}
\frac{d\sigma}{dQ^2dyd(q_T^2)}=\frac{d\sigma^{\gamma\gamma}}{dQ^2dyd(q_T^2)}+\frac{d\sigma^{ZZ}}{dQ^2dyd(q_T^2)}+\frac{d\sigma^{\gamma Z}}{dQ^2dyd(q_T^2)},
\end{eqnarray}
which correspond to $\gamma^*$-production, $Z$-production and interference of $\gamma^*$-$Z$ production amplitudes. These three terms of the cross-sections differ from each other only due to the  factors $z_{ff'}^{GG'}$ in eq.~(\ref{th:Xsec_gen}), which are
\begin{eqnarray}\label{th:z_gamma}
z^{\gamma\gamma}_{ll'}z^{\gamma\gamma}_{ff'}&=&\delta_{\bar ff'}\alpha^2_{\text{em}}(Q)e_f^2,
\nn
\\\label{th:z_Z}
z^{ZZ}_{ll'}z^{ZZ}_{ff'}&=&\frac{\delta_{\bar ff'}\alpha^2_{\text{em}}(Q)Q^4}{(Q^2-M_Z^2)^2+\Gamma_Z^2M_Z^2}\frac{1-4s_W^2+8s_W^4}{8s_W^2c_W^2}
\times 
\frac{1-4|e_f|s_W^2+8e_f^2s_W^4}{8s_W^2c_W^2}
\nn
\\\label{th:z_gammaZ}
z^{Z\gamma}_{ll'}z^{Z\gamma}_{ff'}&+&z^{\gamma Z}_{ll'}z^{\gamma Z}_{ff'}=\frac{\delta_{\bar ff'}\alpha^2_{\text{em}}(Q)2Q^2(Q^2-M_Z^2)}{(Q^2-M_Z^2)^2+\Gamma_Z^2M_Z^2}\frac{1-4s_W^2}{4s_Wc_W}
\times \frac{|e_f|(1-4|e_f|s_W^2)}{4s_Wc_W},
\end{eqnarray}
where $M_Z$ and $\Gamma_Z$ are the mass and the width of the Z-boson, $s_W$ and $c_W$ are sine and cosine of the Weinberg angle. We use the following of values \cite{Olive:2016xmw}
\begin{eqnarray}
M_Z=91.2~\GeV,\quad \Gamma_Z=2.5~\GeV,\quad s^2_W=0.2313.
\end{eqnarray}
In many studies (see e.g.\cite{DAlesio:2014mrz,Su:2014wpa,Landry:2002ix,Bacchetta:2017gcc,Bozzi:2008bb}) the contribution of $\gamma^*$ to the cross-section is neglected  in the vicinity of the Z-peak, i.e. the zero-width approximation is used.  Here, instead, we   include the $\gamma^*$ and interference terms in the evaluation of the  the cross-section. The inclusion of these terms is important for LHC  (in particular ATLAS experiment), where the measurement precision often exceeds the theory precision.

\subsection{TMD parton distributions: evolution}
\label{sec:evol}

\begin{figure}[t]
\centering
\includegraphics[width=0.45\textwidth]{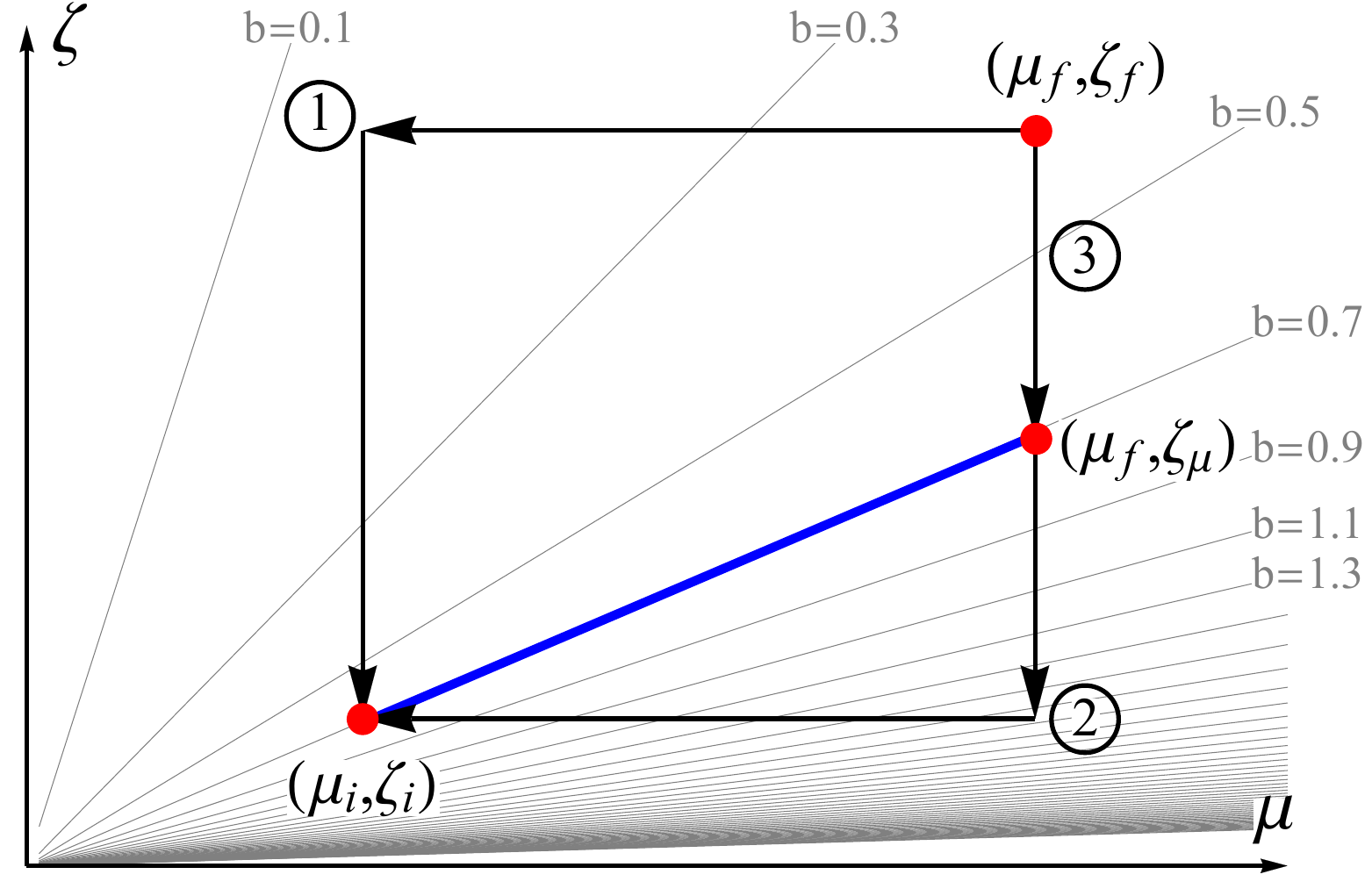}
\includegraphics[width=0.45\textwidth]{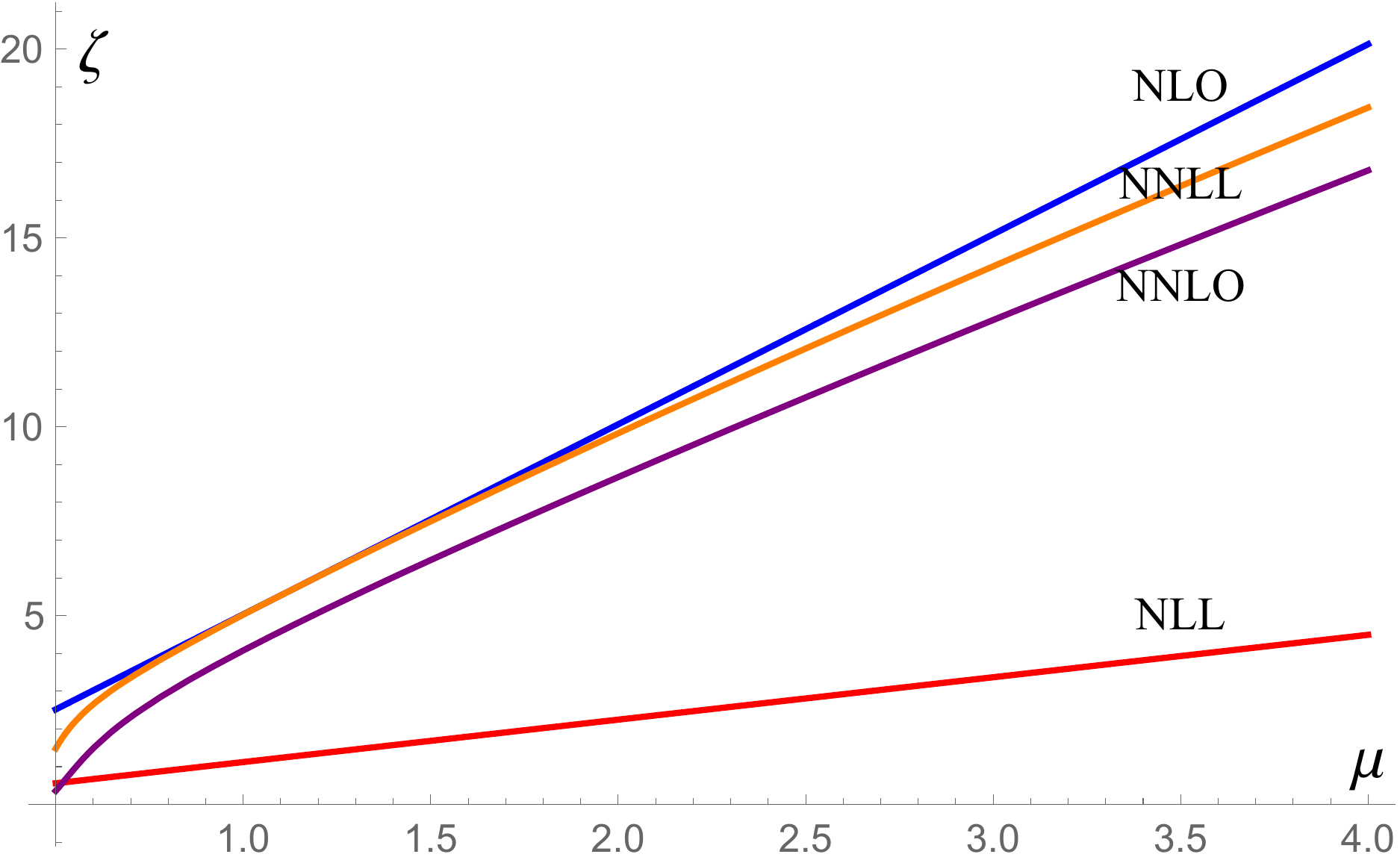}
\caption{(left) The evolution plane $(\mu,\zeta)$ and paths for the evolution integrals from $(\mu_f,\zeta_f)$ to $(\mu_i,\zeta_i)$. Gray lines are equi-evolution lines $\zeta_\mu$ at different $b$. Paths 1 and 2 reprent the solutions in eq.~(\ref{th:Rpath1}) and eq.~(\ref{th:Rpath2}), corespondingly. These solutions are equivalent to the evolution to the point $(\mu_f,\zeta_{\mu_f})$, which is shown by path 3, because there is no evolution along the blue segment (at $b=0.7$ GeV$^{-1}$).  (right) The plot of $\zeta_\mu$ at $b=1$ GeV$^{-1}$ for different orders.} \label{fig:EvolutionPaths}
\end{figure}

The quark unpolarized TMDPDFs are given by the matrix element \cite{Collins:2011zzd,GarciaEchevarria:2011rb,Echevarria:2016scs}
\begin{eqnarray}\label{def:TMDPDF}
&&F_{q\ot h}(x,\vec b;\zeta,\mu)= \frac{Z_q(\zeta,\mu) \mathcal{R}_q(\zeta,\mu)}{2}
\\\nn && \qquad \times
\sum_X\int \frac{d\xi^-}{2\pi}e^{-ix p^+\xi^-} \langle h|
\left\{T\[\bar q_i \,\tilde W_n^T\]_{a}\( \frac{\xi}{2} \) |X\rangle \gamma^+_{ij}\langle X|\bar T\[\tilde
W_n^{T\dagger}q_j\]_{a}\(-\frac{\xi}{2} \) \right\} | h\rangle,
\end{eqnarray}
where $n$ is the light-cone vector along the large component of the hadron momentum, $\xi=\{0^+,\xi^-,\vec b\}$, $Z$ and $\mathcal{R}$ are the ultraviolet and rapidity divergence renormalization factors. The Wilson lines $W_n$ pointing along the direction $n$ to the infinity. For the detailed definition of all constituents in this expression we refer to \cite{Echevarria:2016scs}.

The peculiar feature of the TMD operator is the presence of two types of divergences and, as a consequence, two renormalization factors $Z$ and $\mathcal{R}$. Firstly, we have ultraviolet divergences, which have their collinear counterpart in the coefficient function $C_V$. These divergences are the result of collinear factorization and give rise to the logarithms of the factorization scale $\mu$. Secondly, we have rapidity divergences, which arise in the factorization of the soft-gluon exchanges between partons. The singular soft-gluons exchanges can be collected into the soft factor, which in turn, can be written as a product of rapidity renormalization factors $\mathcal{R}$, see e.g. \cite{Echevarria:2015byo,Echevarria:2016scs,Vladimirov:2017ksc}. This procedure introduces the rapidity factorization scale $\zeta$. 

The dependence of TMDPDF on the factorization scales $\mu$ and $\zeta$ is given by the pair of evolution equations
\begin{eqnarray}\label{th:evol_mu}
\mu^2 \frac{d}{d\mu^2} F_{f\ot h}(x,\vec b;\mu,\zeta)&=&\frac{1}{2}\gamma^f_F(\mu,\zeta)F_{f\ot h}(x,\vec b;\mu,\zeta),
\\\label{th:evol_zeta}
\zeta \frac{d}{d\zeta} F_{f\ot h}(x,\vec b;\mu,\zeta)&=&-\mathcal{D}^f(\mu,\vec b)F_{f\ot h}(x,\vec b;\mu,\zeta).
\end{eqnarray}
The TMD anomalous dimensions $\gamma$ and $\mathcal{D}$ are known up to order $a_s^3$ (see \cite{Moch:2005id} for $\gamma_V$, and \cite{Vladimirov:2016dll,Li:2016ctv,Vladimirov:2017ksc} for $\mathcal{D}$). They satisfy the consistency condition (Cauchy-Riemann condition), which guaranties the existence of the common solution for equations (\ref{th:evol_mu}-\ref{th:evol_zeta}),
\begin{eqnarray}\label{th:consitency}
\zeta \frac{d}{d\zeta}\frac{\gamma^f_F(\mu,\zeta)}{2}=\mu^2\frac{d}{d\mu^2}(-\mathcal{D}^f(\mu,\vec b))=-\frac{\Gamma^f(\mu)}{2},
\end{eqnarray}
where $\Gamma^f$ is the cusp anomalous dimension. This equation fixes the logarithmic part of the anomalous dimensions. So, the anomalous dimension $\gamma$ is linear in logarithm at all orders, while the rapidity anomalous dimension $\mathcal{D}$ has all powers of logarithms,
\begin{eqnarray}\label{th:gammaV}
\gamma_F^f=\Gamma^f \mathbf{l}_\zeta-\gamma_V^f,\qquad\mathcal{D}^f=\sum_{n=1}^\infty a_s^n\sum_{k=0}^n \mathbf{L}_\mu^k d_f^{(n,k)}.
\end{eqnarray}
Here and in the following, we use the following notation for logarithms 
\begin{eqnarray}\label{th:log_notation}
\mathbf{L}_X=\ln\(\frac{\vec b^2 X}{4 e^{-2\gamma_E}}\),\qquad\mathbf{l}_X=\ln\frac{\mu^2}{X}.
\end{eqnarray}
The explicit expressions for the anomalous dimensions up to third-loop order can be found e.g. in the appendix of \cite{Echevarria:2016scs,Vladimirov:2017ksc}. 

The initial values of the factorization scales are dictated by the kinematics of the considered process. In particular, the scales $\zeta_{1,2}$ are related to the momentum of hard partons as
\begin{eqnarray}
\zeta_1\zeta_2=(2p_1^+p_2^-)^2=(Q^2+q_T^2)^2\simeq Q^4.
\end{eqnarray}
In the following, we use the symmetric normalization point, $\zeta_1=\zeta_2=\zeta=Q^2$. The $\mu$-dependence cancels between the parts of factorization formula, namely between hard coefficient function $|C_V|^2$ and the TMDPDFs. The natural choice of $\mu$ is such that logarithms appearing in $|C_V|^2$ are minimized, i.e. $\mu=Q$. Therefore, TMDPDFs enter in the cross-section in eq.~(\ref{th:Xsec_gen}) at the hard point $(\mu_f,\zeta_f)=(Q,Q^2)$. 

A typical construction of a model for a TMD distribution requires its evolution to a different set of scales. The evolution from $(\mu_f,\zeta_f)$ to $(\mu_i,\zeta_i)$ takes the form
\begin{eqnarray}\label{th:TMD_evol}
F_{f\ot h}(x,\vec b;\mu_f,\zeta_f)=R^f[\vec b;(\mu_f,\zeta_f)\to (\mu_i,\zeta_i)]F_{f\ot h}(x,\vec b;\mu_i,\zeta_i),
\end{eqnarray}
where
\begin{eqnarray}\label{th:TMD_R}
R^f[\vec b;(\mu_f,\zeta_f)\to (\mu_i,\zeta_i)]=\exp\[\int_P \(\gamma^f_F(\mu,\zeta)\frac{d\mu}{\mu} -\mathcal{D}^f(\mu,\vec b)\frac{d\zeta}{\zeta}\)\].
\end{eqnarray}
Here, the $\int_P$ denotes the integration along the path $P$ in the $(\mu,\zeta)$-plane from the point $(\mu_f,\zeta_f)$ to the point $(\mu_i,\zeta_i)$. The integration can be done on an arbitrary path $P$, and the solution is independent on it, thanks to the Cauchy-Riemann condition eq.~(\ref{th:consitency}). At a finite perturbative order, the condition eq.~(\ref{th:consitency}) is violated by the next perturbative order. As a consequence the expression for the evolution factor $R$ is dependent on the path of integration. The two simplest choices of integration paths are the  combinations of straight segments as
\begin{eqnarray*}
\text{path 1}&:& (\mu_f,\zeta_f)\to (\mu_i,\zeta_f)\to (\mu_i,\zeta_i),
\\
\text{path 2}&:& (\mu_f,\zeta_f)\to (\mu_f,\zeta_i)\to (\mu_i,\zeta_i).
\end{eqnarray*}
These paths are depicted in fig.~\ref{fig:EvolutionPaths}. The factor $R$ evaluated along these paths reads
\begin{eqnarray}\label{th:Rpath1}\nn
R^f[\vec b;(\mu_f,\zeta_f)\xrightarrow{1} (\mu_i,\zeta_i)]&=&\exp\[\int_{\mu_i}^{\mu_f}\frac{d\mu}{\mu}\gamma^f_F(\mu,\zeta_f)-\mathcal{D}^f(\mu_i,\vec b)\ln\(\frac{\zeta_f}{\zeta_i}\)\],
\\\label{th:Rpath2}
R^f[\vec b;(\mu_f,\zeta_f)\xrightarrow{2} (\mu_i,\zeta_i)]&=&\exp\[\int_{\mu_i}^{\mu_f}\frac{d\mu}{\mu}\gamma^f_F(\mu,\zeta_i)-\mathcal{D}^f(\mu_f,\vec b)\ln\(\frac{\zeta_f}{\zeta_i}\)\].
\end{eqnarray}
The numerical difference between these two expressions represents the value of the uncertainty at a given perturbative order.

The expressions for the evolution factor $R$ given in eqs.~(\ref{th:Rpath1}-\ref{th:Rpath2}), contain the rapidity anomalous dimension $\mathcal{D}(\mu,\vec b)$. The latter contains potentially large values of $\mathbf{L}_\mu$, which should be resummed with the help of eq.~(\ref{th:consitency}). Additionally, the rapidity anomalous dimension can acquire power corrections from the higher  orders in the power expansion   of the factorization theorem \cite{Becher:2013iya}. These power corrections can be also observed in the renormalon structure described in \cite{Scimemi:2016ffw}. The non-perturbative correction takes the form of a series of even powers of the transverse distance. Therefore, the practical expression for the rapidity anomalous $\mathcal{D}$ is
\begin{eqnarray}\label{th:D_evol}
\mathcal{D}^f(\mu,\vec b)=\int_{\mu_0}^\mu \frac{d\mu'}{\mu'}\Gamma^f+\mathcal{D}_{\text{pert}}^f(\mu_0,\vec b)+g_K \vec b^2,
\end{eqnarray}
where $g_K$ is an unknown parameter. Here, $\mathcal{D}_{\text{pert}}^f$ is the perturbative expression for $\mathcal{D}$. Correspondingly, the value $\mu_0$ should be chosen such that $\mathbf{L}_{\mu_0}$ is minimal in the perturbative region. Substituting this expression to the evolution factor, we obtain
\begin{eqnarray}\label{th:Rfactor}
&&R^f[\vec b;(\mu_f,\zeta_f)\to (\mu_i,\zeta_i);\mu_0]=\nn\\&&
\exp\[\int_{\mu_i}^{\mu_f}\frac{d\mu}{\mu}\gamma^f_F(\mu,\zeta_f)-\int_{\mu_0}^{\mu_i}\frac{d\mu}{\mu}\Gamma^f(\mu)\ln\(\frac{\zeta_f}{\zeta_i}\)\]\(\frac{\zeta_f}{\zeta_i}\)^{-\mathcal{D}^f_{\text{perp}}(\mu_0,\vec b)-g_K \vec b^2}.
\end{eqnarray}
In this form, the evolution factor $R$ is independent on the path of evolution, as can be checked explicitly. The perturbative uncertainty which previously has been given by the variation of evolution path, now is represented by the dependence on the parameter $\mu_0$. Thus, using eq.~(\ref{th:Rfactor}) the uncertainties of the perturbative calculation can be measured by varying the scale $\mu_0$. In the following, we use the evolution factor  as in 
eq.~(\ref{th:Rfactor}).

\subsection{TMD parton distributions: $b$-space behavior}
\label{sec:TMDPDF}

\begin{figure}[t]
\centering
\includegraphics[width=0.55\textwidth]{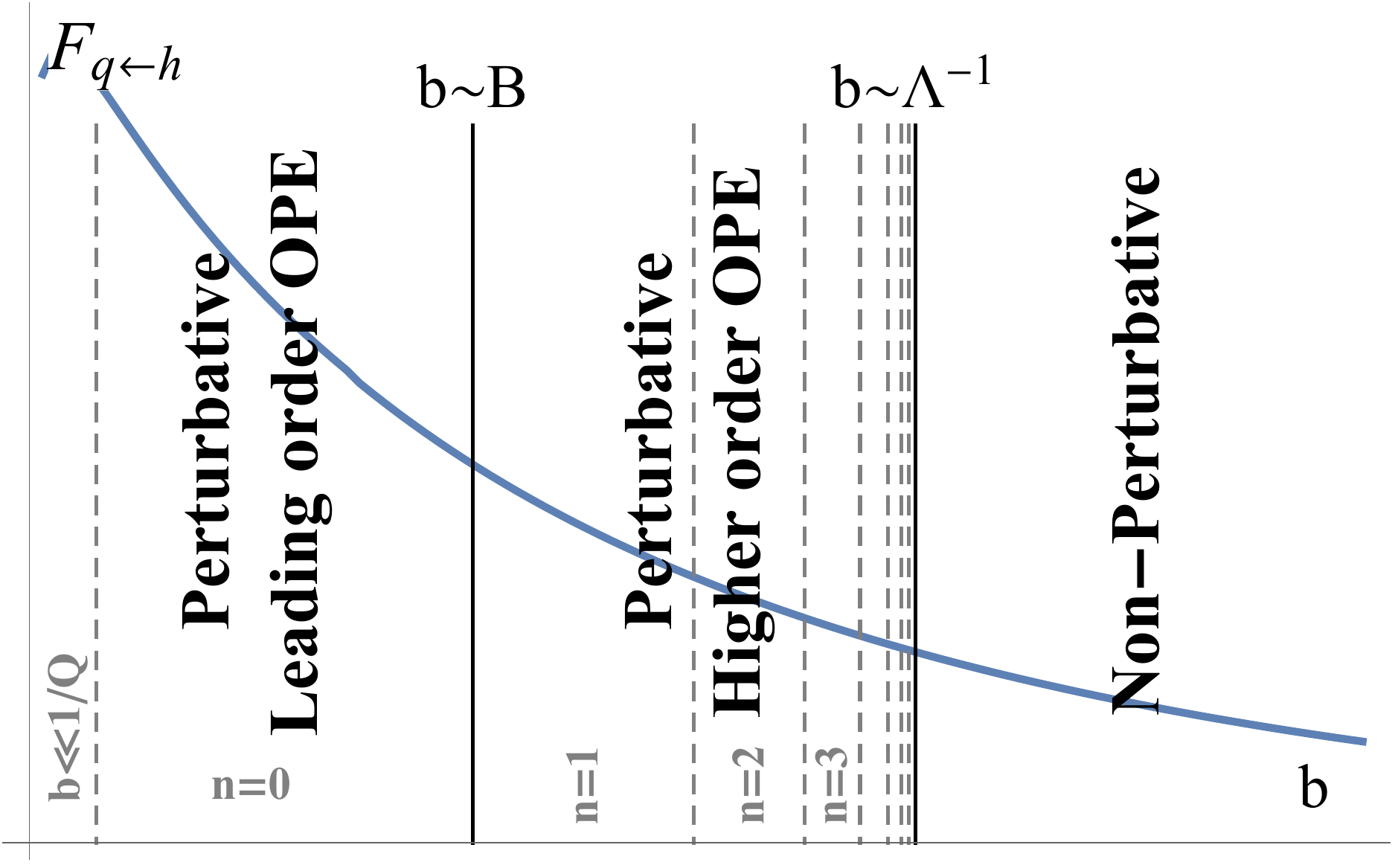}
\caption{Schematic depiction of the regions of the TMDPDF with the parameter $b$.} \label{fig:TMDRegions}
\end{figure}

The TMDPDF is a genuine non-perturbative function, which is to be fitted by a certain ansatz, which covers the whole domain in $b$-space. Different intervals of $b$-space describe different regimes of strong interactions. The schematic picture of $b$-regions is shown in fig.~\ref{fig:TMDRegions}.
 In order to construct an optimal and physically meaningful fitting ansatz, the behavior in every part of the $b$-space should be reproduced. In this section, we collect the main information on the $b$-dependence of TMDPDFs, as it is understood in the modern theoretical picture.   

The starting point of our description of a TMD distribution is the small-$b$ operator product expansion (OPE), which results in the  series
\begin{eqnarray}\label{th:OPE}
F_{q\ot h}(x,\vec b;\mu,\zeta)=\sum_{n=0}^\infty \(\frac{\vec b^2}{B^2}\)^n \sum_f \(C^{(n)}_{q\ot f}(\vec b;\mu,\zeta)\otimes f^{(n)}_{f\ot h}(\mu)\)(x),
\end{eqnarray}
where $f^{(n)}$ are PDFs of a $2(n+1)$-twist, $C^{(n)}$ are coefficient functions of OPE and the symbol $\otimes$ represents the convolution in momentum fractions of partons. The parameter $B$ is an unknown non-perturbative parameter which represents an intrinsic hadron scale.

\textit{Region 1:} In the range $b\ll B$ the TMDPDF is dominated by the $n=0$ term of OPE eq.~(\ref{th:OPE}). The leading term is  represented by the usual matching onto twist-2 PDFs and reads
\begin{eqnarray}\label{th:OPE_LO}
b\ll B &:&F_{q\ot h}(x,\vec b;\mu,\zeta)=\sum_f \int_x^1\frac{dz}{z} C_{q\ot f}(z,\mathbf{L}_\mu;\mu,\zeta) f_{f\ot h}\(\frac{x}{z},\mu\),
\end{eqnarray}
where $C$ is known up to two-loop order \cite{Echevarria:2016scs,Gehrmann:2014yya}. 

There is a subregion $b\ll 1/Q$, which should be considered specially. While the TMD distribution is completely perturbative within this region, the contributions of this region to the cross-section strongly overlaps with the $Y$-term, eq.~(\ref{th:hadron_tensor}), which is formally $\mathcal{O}(1/(bQ))$. The behavior of TMD distributions within this tiny range together with the Y-term dictates the asymptotics of the cross-section at large $q_T$. As a  consequence, it has a significant influence on the value of the total cross-section. In our current evaluation we restrict ourself to the range of small-$q_T$ (for a dedicated study of the applicability of this approximation in practice, see sec.~\ref{sec:deltaT}). Therefore, we drop the $Y$-term and do not need any special treatment of $b\ll Q^{-1}$ region.

\textit{Region 2:} In the range $b\lesssim B$ the OPE is still valid. However, one has to include the higher order terms in addition to the leading one. Very little is known about power suppressed terms of the small-$b$ OPE. Our recent study of the renormalon singularities~\cite{Scimemi:2016ffw} suggests several hints that can be used to model this region:
\begin{itemize}
\item[(i)] The OPE contains only even powers of $b$. Moreover, the coefficient function of $n$'th order has a prefactor $x^n$. In other word, the natural scale of OPE is $x\vec b^2/B^2$ rather then just $\vec b^2/B^2$.
\item[(ii)] The higher order OPE contributions induced by renormalons, can be summed together to some effective non-perturbative function \textit{under} the convolution integral.
\end{itemize}
Therefore, in this region the TMDPDF can be approximated by the form
\begin{eqnarray}\label{th:OPE_inter}
b\sim B &:&F_{q\ot h}(x,\vec b;\mu,\zeta)=\sum_f \int_x^1\frac{dz}{z} G_{q\ot f}\(z,\frac{z\vec b^2}{B^2},\mathbf{L}_\mu;\mu,\zeta\) f_{f\ot h}\(\frac{x}{z},\mu\),
\end{eqnarray}
where the leading term of the power series in $b/B$ of $G$ is given by $C$. 
As the power  $n$ grows,  the sub-leading terms of OPE switch on, which is schematically presented by gray lines in fig.~\ref{fig:TMDRegions}. The particular contributions at higher $n$ are not so important in the continuous TMD picture. However,
\begin{itemize}
\item[(iii)] The $n=1$ contribution to OPE can be estimated by the leading renormalon contribution of order $\sim x\vec b^2$ \cite{Scimemi:2016ffw}. It has the form
\begin{eqnarray}\label{th:ren}
C^{\text{ren}}_{q\ot q}(x,\vec b;\mu,\zeta)&=&2\bar x+\frac{2x}{(1-x)_+}-\delta(\bar x)\(\mathbf{L}_\Lambda-\mathbf{L}_{\sqrt{\zeta}}+\frac{2}{3}\),
\end{eqnarray}
where $\Lambda=\Lambda_{QCD}$ is the position of the Landau pole.
\end{itemize}

\textit{Region 3:} At $b\gg B$ the small-$b$ OPE cannot be considered as a source of information, and the TMD is completely non-perturbative. Luckily, this region is suppressed by the evolution factor. As a consequence, the cross-section is not very sensitive to the fine structure of TMD distribution in this region, but the general behavior is important. We have tested several asymptotical forms of the TMDPDF, including Gaussian, exponential and power-like and found that the best agreement with the experimental data is achieved with exponential behavior. This observation is in agreement with the general physical intuition, that at high distances the strong forces are dominated by meson exchange, while the Gaussian and power-like asymptotics can not be produced in any simple way.

We should mention that the size of the parameter $B$, as well as, the order of convergence of the small-$b$ OPE, which influences the size of the intermediate region 2, are not known. Our estimations of these characteristic sizes are presented in sec.~\ref{sec:discussion}.

\subsection{Definition of scaling parameters}
\label{sec:scales}

The small-$b$ matching is the starting point for the construction of the majority of phenomenological ansatzes for TMD distributions. It can be considered as an additional collinear factorization, which is performed at some convenient set of scales $(\mu_i,\zeta_i)$. The difference of $(\mu_i,\zeta_i)$ from the initial (defined by process kinematic) scales of TMD distribution is compensated by the evolution factor in  eq.~(\ref{th:TMD_evol}). As usual, the all-order expression is independent on $(\mu_i,\zeta_i)$, but in practice, these scales are to be chosen such that the coefficient function $C_{f\ot f'}$ has good perturbative convergence. This procedure is alike the choice of hard-factorization scale, with one \textit{essential} difference: the parameter $b$, which accompanies $\mu_i$ and $\zeta_i$ in the logarithms, has no fixed value. It varies from zero to infinity within the Fourier integral.

The choice of scales $(\mu_i,\zeta_i)$ is one of the central point of the TMD phenomenology. To make the discussion clearer, let us recall the expression for the coefficient function at NLO. It reads
\begin{eqnarray}\label{th:C_1loop}
&& C_{q\ot q}(x,\mathbf{L}_\mu;\mu,\zeta)=\delta(\bar x)+
\\\nn &&\qquad a_s(\mu)C_F\[-2\mathbf{L}_\mu \(\frac{2}{(1-x)_+}-1-x\)+2\bar x+\delta(\bar x)\(-\mathbf{L}_\mu^2+ 2\mathbf{L}_\mu \mathbf{l}_\zeta-\frac{\pi^2}{6}\)\],
\end{eqnarray}
where the notation for the  logarithms is defined in eq.~(\ref{th:log_notation}). Ideally, the scales $\mu$ and $\zeta$ should be chosen such that no large perturbative contributions appear in the coefficient function. Clearly, it cannot be done at arbitrary $b$ due to the presence of $\mu$ in the coupling constant and in $\mathbf{L}_\mu$. However, such a strict choice is not required. The only requirement for scales is to keep the perturbative ansatz stable, i.e. to prevent its blowing up. There are several solutions of this problem. The most famous is the $b^*$-prescription \cite{Collins:1981va}. Within the $b^*$-prescription the logarithms $\mathbf{L}_\mu$ are absent, an this fact allows the control of the perturbative series in the full region of $b$. However, the $b^*$-prescription introduces artificial power corrections to the small-$b$ OPE, which washes out any theoretical intuition. Another popular scheme~\cite{Laenen:2000de,Kulesza:2002rh} is based on the re-expression of Hankel-integral as an integral in the complex $b$-plane. In this way, the logarithms $\mathbf{L}_\mu$ can be minimized by $\mu\sim b^{-1}$ and the Landau pole at large-$b$ is by-passed in the complex plane. The drawback of this scheme is the necessity of the analytical continuation into the complex $b$-plane, and the restriction to NNLO (since the analytical solution for running coupling at N$^3$LO is unknown).  

In this work we use another scheme which we call $\zeta$-prescription. It is a novel one (to our best knowledge), and  it is described in the following.

The $\zeta$-prescription uses the fact that the TMD operator and hence its small-$b$ OPE depends on two scales $\mu$ and $\zeta$, which are entirely  independent. This simple fact has been overlooked so far. Indeed, the first typical step  is to fix $\zeta=C_0^2/\vec b^2$, or $\zeta=\mu^2$ \cite{Aybat:2011zv,Collins:2011zzd,Scimemi:2016ffw}. It reduces the problem to a single variable problem, which looks simpler, but finally, it does not provide  a simple solution for the appearance of large logarithms in the OPE.

The initial point of the $\zeta$-prescription is the observation that not all logarithms in the coefficient function are dangerous. So, the terms $\mathbf{L}^2_\mu$ and $\mathbf{L}_\mu \mathbf{l}_\zeta$ in eq.~(\ref{th:C_1loop})  are problematic, while the logarithm in first term is not. There are several reasons for it. First, the double logarithm contributions violate the normal perturbative counting and at large-$b$ grows faster than the single logarithms. Second, the first term of eq.~(\ref{th:C_1loop}) comes together with the DGLAP kernel, and thus, it preserves the area (say, the integral over $x$) of the TMDPDF, due to the conservation of the electromagnetic charge. We remind that logarithms accompanying the DGLAP kernel are related to PDF evolution, while the rest of logarithms are related to the TMD evolution. For this reason, the main problem of convergence is represented by the logarithms that are related to the TMD evolution. The logarithms related to the PDF evolution come with a particular $x$-dependent function. The probabilistic interpretation of PDF ensures their minimal contribution in the very large domain of $b$. Practically, this fact has been already demonstrated although not entirely realized in the fit~\cite{DAlesio:2014mrz}. In the realization of ref. \cite{DAlesio:2014mrz}, the DGLAP logarithms were left unregulated and they do not influence the convergence of the fit. 

The logarithms related to the TMD evolution can be eliminated completely by a particular choice of $\zeta=\zeta_\mu$. Along the curve $\zeta_\mu$, the TMD distributions are independent on $\mu$. In other words, the curve $\zeta_\mu$ is an equi-evolution curve in the plane $(\mu,\zeta)$. Such a curve satisfies the equation
\begin{eqnarray}
\mu^2 \frac{d F(x,\vec b;\mu,\zeta_\mu)}{d\mu^2}=0.
\end{eqnarray}
Using the definition of anomalous dimensions in eq.~(\ref{th:evol_mu}) we rewrite this equation as
\begin{eqnarray}\label{th:zeta_def}
\mathcal{D}(\mathbf{L}_\mu)f'(\mathbf{L}_\mu)+\frac{\Gamma}{2}f(\mathbf{L}_\mu)-\mathcal{D}(\mathbf{L}_\mu)-\frac{\gamma_V}{2}=0,
\end{eqnarray}
where $f(\mathbf{L}_\mu)=\mathbf{l}_{\zeta_\mu}$. The perturbative solution is discussed and presented in the appendix \ref{app:zeta}. The curve $\zeta_\mu$ is different for  quark and for  gluon TMDs, and it is expressed in  terms of the TMD anomalous dimensions eq.~(\ref{app:zeta_complete}). In our analysis, we need only the quark case. Up to NNLO it reads
\begin{eqnarray}\label{th:l_zeta}
\mathbf{l}_{\zeta_\mu}&=&\frac{\mathbf{L}_\mu}{2}-\frac{3}{2}+a_s\Big[\frac{11 C_A-4 T_F N_f}{36}\mathbf{L}_\mu^2
\\\nn&&
+C_F\(-\frac{3}{4}+\pi^2-12\zeta_3\)+C_A\(\frac{649}{108}-\frac{17\pi^2}{12}+\frac{19}{2}\zeta_3\)+T_FN_f\(-\frac{53}{27}+\frac{\pi^2}{3}\)\Big]+\mathcal{O}(a_s^2).
\end{eqnarray} 
Note, that in eq.~(\ref{th:l_zeta}) we have set the boundary condition such that no terms singular at $\mathbf{L}_\mu\to 0$ appear in $\mathbf{l}_\zeta$ (see appendix \ref{app:zeta}, for details). Also, in the current work we drop the power contributions to the rapidity anomalous dimension $\mathcal{D}$. The influence of these decisions should be investigated later. One can check that the leading term of $\zeta_\mu$ (i.e. $\mathbf{l}_\zeta=\mathbf{L}_\mu/2$) cancels leading powers of logarithms at \textit{all orders} in perturbation theory  (i.e. all terms $a_s^n \mathbf{L}_\mu^{2n}$). Then, including the next correction ($a_s\beta_0\mathbf{L}_\mu^2/12 $) cancels subleading powers of logarithms at \textit{all orders} of the perturbation theory (i.e. all terms $a_s^n \mathbf{L}_\mu^{2n-1}$) , and so on. 

Substituting the leading term of the solution in eq.~(\ref{th:l_zeta}) to the quark small-$b$ coefficient function, we obtain
\begin{eqnarray}\label{th:C_1loop_zeta-prescription}
&& C_{q\ot q}(x,\mathbf{L}_\mu;\mu,\zeta_\mu)=\delta(\bar x)+
\\\nn &&\qquad a_s(\mu)C_F\[-2\mathbf{L}_\mu \(\frac{2}{(1-x)_+}-1-x\)+2\bar x+\delta(\bar x)\(-3\mathbf{L}_\mu-\frac{\pi^2}{6}\)\].
\end{eqnarray}
This coefficient function is stable for any value of $\mathbf{L}_\mu$, which can be seen by considering its integral
\begin{eqnarray}
\int_0^1 dx C_{q\ot q}(x,\mathbf{L}_\mu;\mu,\zeta_\mu)=1+a_s(\mu)C_F\(1-\frac{\pi^2}{6}\),
\end{eqnarray}
which is independent on $\mathbf{L}_\mu$.

The general expression for the coefficient of arbitrary flavour at NNLO has the form
\begin{eqnarray}\label{th:coeffF_NNLO}
C_{f\ot f'}(x,\vec b;\mu,\zeta_\mu)&=&\delta_{ff'}\delta(\bar x)+a_s\(-\mathbf{L}_\mu P^{(1)}_{f\ot f'}+C^{(1,0)}_{f\ot f'}\)+
\\\nn && a_s^2\Big[\mathbf{L}_\mu^2\frac{P_{f\ot k}^{(1)}\otimes P_{k\ot f'}^{(1)}-\beta_0 P^{(1)}_{f\ot f'}}{2}
 -\mathbf{L}_\mu\(P^{(2)}_{f\ot f'}+C^{(1,0)}_{f\ot k}\otimes P^{(1)}_{k\ot f'}-\beta_0 C_{f\ot f'}^{(1,0)}\)
 \\\nn && \qquad+\frac{d^{(2,0)}_f\gamma_V^{f(1)}}{\Gamma_0^f}\delta(\bar x)+C^{(2,0)}_{f\ot f'}\Big]+\mathcal{O}(a_s^3),
\end{eqnarray}
where $C^{(n,0)}$ is the finite part of the coefficient function at $n$'th perturbative order, and $P(x)=\sum a_s^n P^{(n)}$ is the DGLAP kernel. The detailed derivation of eq.~(\ref{th:coeffF_NNLO}) is presented in the appendix (\ref{app:zeta_in_coeff}). Eq.~(\ref{th:coeffF_NNLO}) has the form of the usual coefficient function for an object without external evolution (e.g. coefficient function for DIS). In other words, it is straightforward to check that
\begin{eqnarray}
\mu^2\frac{d}{d\mu^2}C_{f\ot f'}(x,\vec b;\mu,\zeta_\mu)\otimes f_{f'\ot h}(x,\mu)=0,
\end{eqnarray}
by direct differentiation of eq.~(\ref{th:coeffF_NNLO}). The integral of this function over $x$  is independent on $\mathbf{L}_\mu$  due to the charge conservation, and thus at least the area of TMDPDF is stable at large $b$. 

A further positive point of the $\zeta$-prescription is that the scale $\mu$ remains unconstrained.  Often, the scale $\mu$ is selected such that it behaves as $\sim 1/b$ at $b\to 0$. Such a choice minimizes the small-$b$ logarithms in small-$b$ OPE and in the evolution exponent. At large-$b$ the scale $\mu$ should be frozen to some fixed value (of the order of a few GeV's), in order to avoid the Landau pole. We use the simplest function which satisfies these criteria
\begin{eqnarray}\label{th:mub}
\mu=\mu_b=\frac{C_0}{b}+2\; {\rm GeV},
\end{eqnarray}
where the low-energy fixed scale $2$ GeV is chosen due to the fact that it is the standard scale of PDF extractions.

Finally, we should also select the value for the parameter $\mu_0$ that enters expression for the evolution factor eq.~(\ref{th:Rfactor}). To keep our discussion simple, we set $\mu_0=\mu_b$.

\subsection{Theoretical uncertainties and perturbative ordering}
\label{sec:th_uncertain}

\begin{table}[t]
\begin{center}
\begin{tabular}{||c||c|c||c|c|c||c|c||c||}
Name & $|C_V|^2$ & $C_{f\ot f'}$ & $\Gamma$ & $\gamma_V$&$\mathcal{D}$ & PDF set & $a_s$(run) & $\zeta_\mu$
\\\hline\hline 
NLL/LO & $a_s^0$ & $a_s^0$ & $a_s^2$ & $a_s^1$& $a_s^2$ & nlo & nlo & NLL
\\\hline
NLL/NLO & $a_s^1$ & $a_s^1$ & $a_s^2$ & $a_s^1$& $a_s^2$ & nlo & nlo & NLO
\\\hline
NNLL/NLO & $a_s^1$ & $a_s^1$ & $a_s^3$ & $a_s^2$& $a_s^3$ & nlo & nlo & NNLL
\\\hline
NNLL/NNLO & $a_s^2$ & $a_s^2$ & $a_s^3$ & $a_s^2$& $a_s^3$ & nnlo & nnlo & NNLO
\end{tabular}
\end{center}
\caption{\label{tab:orders}The perturbative  orders studied in the  fit. For each order we indicate the  power of $a_s$ of each piece that enters in the TMDs.
 Note, that the order of $a_s$ and PDF set are related, since the values of $a_s$ are taken from the PDF set.}
\end{table}

In the construction of the cross section, one finds several sources of perturbative uncertainties. The size of these uncertanties can be estimated by the variation of associated scales. We list here the ones that we have considered in the present work.

\begin{itemize}

\item \textit{Uncertainty associated with the perturbative matching of rapidity anomalous dimension :} This uncertainty arises from the dependence (at the fixed perturbative order) on $\mu_0$, which should be compensated between the Sudakov factor and the boundary term $\mathcal{D}(\mu_0)$ in the TMD evolution factor eq.~(\ref{th:Rfactor}). This uncertainty can be tested by changing $\mu_0\to c_1\mu_0$ and varying $c_1\in [0.5,2]$.
\item \textit{Uncertainty associated with the hard factorization scale:} This uncertainty arises from the dependence (at the fixed perturbative order) on the scale $\mu_f(\sim Q)$ which is to be compensated between the hard coefficient function $|C_V|^2$ and the TMD evolution factor. This uncertainty can be tested by changing $\mu_f\to c_2\mu_f$ and varying $c_2\in [0.5,2]$.
\item \textit{Uncertainty associated with the TMD evolution factor:} This uncertainty arises from the dependence (at the fixed perturbative order) on initial scale of TMD evolution $\mu_i$, which is to be compensated between the evolution integral and the $\mu$-dependence of $\zeta_i$ in eq.~(\ref{th:Rfactor}). This uncertainty can be tested by changing $\mu_i\to c_3\mu_i$ and varying $c_3\in [0.5,2]$.
\item \textit{Uncertainty associated with the small-$b$ matching:} This uncertainty arises from the dependence (at the fixed perturbative order) on the scale of the small-$b$ matching $\mu_{\text{OPE}}$ which is to be compensated between the small-$b$ coefficient function $C_{f\ot f'}$ and evolution of PDF. This uncertainty can be tested by changing $\mu_{\text{OPE}}\to c_4\mu_{\text{OPE}}$ and varying $c_4\in [0.5,2]$.
\end{itemize}

We remark that our definition of perturbative uncertainties $c_{1,2}$ is commonly used in the literature (as far as it can be compared among different schemes of calculation), see e.g.~\cite{Catani:2015vma,Bozzi:2008bb}. The uncertainties $c_{3,4}$ are usually non distinguished and 
they are commonly 
varied simultaneously i.e. in the literature one finds discussions of errors for the case $c_4=c_3$. To our best knowledge, the distinction of the matching and evolution uncertainties is made here for the first time.

In this way, the general expression for the  cross-section in eq.~(\ref{th:Xsec_gen}) with our choice of scales reads
\begin{eqnarray}\label{th:Xsec_gen_with_uncert}
\frac{d\sigma}{dQ^2dyd(q_T^2)}&=& \frac{4\pi}{3N_c}\frac{\mathcal{P}}{sQ^2}
\sum_{GG'}z_{ll'}^{GG'}(q)\sum_{ff'}z_{ff'}^{GG'}
\\\nn &&\times \int \frac{d^2\vec b}{4\pi}e^{i(\vec b\vec q)}|C_V(Q,c_2Q)|^2
\Big\{R^f[\vec b;(c_2Q,Q^2)\to (c_3\mu_i,\zeta_{c_3\mu_i});c_1\mu_i]\Big\}^2
\\\nn &&  \qquad\qquad\qquad\qquad\times
F_{f\ot h_1}(x,\vec b;c_4\mu_{\text{OPE}},\zeta_{c_4\mu_{\text{OPE}}})F_{f'\ot h_2}(x,\vec b;c_4\mu_{\text{OPE}},\zeta_{c_4\mu_{\text{OPE}}}),
\end{eqnarray}
where the evolution factor $R$ is given in eq.~(\ref{th:Rfactor}) and the explicit expression for the $\zeta_\mu$ is given in eq.~(\ref{th:l_zeta}). The low-normalization point $\mu_i$ and the scale of small-b operator product expansion $\mu_{\text{OPE}}$ are fixed  at the same point (\ref{th:mub})
\begin{eqnarray}
\label{eq:q0}
\mu_i=\mu_{\text{OPE}}=\frac{C_0}{b}+2\;{\rm GeV}.
\end{eqnarray}
The central value of the  constants  $c_{1,2,3,4}$ is $1$ and they are varied in order to estimate the theoretical uncertainties in the usual range $(0.5,2)$. 

The perturbative orders of each constituent are to be combined  consistently. Having at our disposal the  NNLO expressions for coefficient function and N$^3$LO expressions for anomalous dimensions, we can define four successive self-contained sets of ordering. This is reported in table \ref{tab:orders}. In our definition of orders we use the following logic: (i) The order of the $a_s$-running should be the same as the order of PDF set, since their extraction are correlated. (ii) The order of $\mathcal{D}$ should be the same as the order of $\Gamma$ since they enter the evolution kernel $R$ with the same counting of logarithms (i.e. $a_s^n \ln^{n+1}\mu$), and one-order higher then the order of $\gamma_V$, since it has counting $a_s^n \ln^n\mu$. (iii) The order of small-$b$ matching coefficient should be the same as the order of evolution of a PDF, because large logarithms of $b$ are to be compensated by the PDF evolution. (iv) The order of $\zeta_\mu$ should be such that no logarithms appear in the coefficient function, and the general logarithm counting coincides with the counting of the evolution factor. In  table \ref{tab:orders} the order of the $\zeta_\mu$ is defined as following: NLL is $\mathbf{l}_\zeta=\mathbf{L}_\mu/2$, NLO has in addition finite part at order $a_s^0$ (i.e. two first terms of eq.~(\ref{th:l_zeta})), NNLL has in addition logarithmic part at order $a_s^1$ (i.e. the first line of eq.~(\ref{th:l_zeta})), and NNLO is given by  whole expression eq.~(\ref{th:l_zeta}).

To label the orders we use the nomenclature where the part with 'LO suffix designates the order of coefficient functions, and the part with 'LL suffix designates the order of the evolution factor in the $a_s \ln\mu \sim 1$ scheme. So, our highest order is NNLL/NNLO, which at the moment the highest available combination of the perturbative series. The order NLL/LO appears to be  barely inconsistent, because it requires the LO PDF evolution to match the trivial coefficient function. Therefore, we \textit{exclude} the NLL/LO from our phenomenological studies.

\subsection{Implementation of lepton cuts}  
\label{sec:cuts}

In modern experiments, the cross-section is often evaluated with the fiducial cuts on the dilepton momenta. That is, the lepton tensor in eq.~(\ref{th:leptonTensor}) should be evaluated  taking into account the experimental cut phase-space. At the leading order the lepton tensor takes the form
\begin{eqnarray}
(-g_T^{\mu\nu})L_{\mu\nu}^{GG'}&=&32z^{GG'}_{ll'}\int\frac{d^3k_1}{2E_1}\frac{d^3k_2}{2E_2}((k_1\cdot k_2)+(\vec k_1\cdot \vec k_2))\theta(k_{1,2}\in \text{cuts})\delta^4(k_1+k_2-q),
\end{eqnarray}
where $\theta$-function restricts the lepton momenta to the allowed region. 

In the limit $Q\to \infty$ and no restriction on the lepton pair  phase space we obtain
\begin{eqnarray}
\lim_{Q\to \infty}(-g_T^{\mu\nu})L_{\mu\nu}^{GG'}&=&\frac{16\pi}{3}z^{GG'}_{ll'}Q^2.
\end{eqnarray}
Substituting this expression to the cross-section we obtain the standard formula to the Drell-Yan cross-section within TMD factorization \cite{Collins:1984kg,Davies:1984sp,Ellis:1997ii,Becher:2010tm,GarciaEchevarria:2011rb,Collins:2011zzd}. In order  to  include the  corrections due to a finite $Q$  and   experimental cuts let us  introduce a factor $\mathcal{P}$, i.e.
\begin{eqnarray}
(-g_T^{\mu\nu})L_{\mu\nu}^{GG'}&=&\frac{16\pi}{3}z^{GG'}_{ll'}Q^2\mathcal{P},
\end{eqnarray}
which is consistent with  the cross section expression presented in eq.~(\ref{th:Xsec_gen}). The function $\mathcal{P}$ in the absence of cuts reads
\begin{eqnarray}
\mathcal{P}(\text{no cuts})=1+\frac{q_T^2}{2Q^2}.
\end{eqnarray}
In the presence of cuts the expression for $\mathcal{P}$ is involved. For example, at $q_T=0$ and $y=0$ it reads
\begin{eqnarray}
\mathcal{P}_{q_T=0,y=0}(|k_{1,2}|>p_T;|\eta_{1,2}|<\eta)&=&\left\{
\begin{array}{lc}
0, & Q<2p_T \\
\(1-\frac{p_T^2}{Q^2}\)\sqrt{1-\frac{4p_T^2}{Q^2}},&2p_T<Q<2p_T \cosh\eta
\\
\(1-\frac{1}{4\cosh^2\eta}\)\th \eta, & 2p_T\cosh\eta<Q.
\end{array}
\right.
\end{eqnarray}
Generally, $\mathcal{P}$ cannot be evaluated analytically, but it is rather easy to evaluate numerically.

\section{Comparison with experiment}
\label{sec:fit}
\subsection{Review of experimental data}
\label{sec:datareview}

In this section we present the experimental data that have been included in our fit.  We have splitted the data into two large data sets with respect to a general energy scaling. They include the measurements from the following experiments:
\begin{itemize}
\item Low-energy data set
\begin{itemize}
\item E288: Drell-Yan process, at $4<Q<14$ GeV.
\end{itemize}
\item High energy data set:
\begin{itemize}
\item CDF/D0: Z-boson production at $s=1.8,~1.96$ TeV.
\item ATLAS/CMS/LHCb: Z-boson production at $s=7,8,13$ TeV.
\item ATLAS: Vector boson production outside the Z-peak ($46<Q<66$ and $116<Q<150$ GeV) at $s=8$ TeV.
\end{itemize}
\end{itemize}
In the present  study, we have  not included the data of other experiments, such as E605, or R209. The main reason is that these data require some special discussion, due to possible internal inconsistencies (see e.g.  \cite{DAlesio:2014mrz}), or due to low precision. We expect  that the  results of the present work are not  sensibly affected by this omission.

In the following, we present each included measurement in more detail.
\begin{table}[t]\begin{center}
\begin{tabular}{|c||c|c|c|c|}
 & E288 200 & E288 300 & E288 400 
\\\hline\hline
$\sqrt{s}$ & 19.4 GeV & 23.8 GeV & 27.4 GeV
\\\hline
 process & p+Cu$\to \gamma\to \mu^+\mu^-$ & p+Cu$\to \gamma\to \mu^+\mu^-$&p+Cu$\to \gamma\to \mu^+\mu^-$
\\\hline
$Q$ range & 4-9 GeV & 4-9 GeV & 5-14 GeV 
\\\hline
$\Delta Q$-bin & 1 GeV & 1 GeV & 1 GeV 
\\\hline 
y & y=0.4 & y=0.21 & y=0.03 
\\\hline
 Observable & $E\frac{d^3\sigma}{d^3q}$ &$E\frac{d^3\sigma}{d^3q}$ &$E\frac{d^3\sigma}{d^3q}$  
\\\hline
Ref. & \cite{Ito:1980ev} & \cite{Ito:1980ev} & \cite{Ito:1980ev} 
\end{tabular}
\caption{\label{tab:E288} The characteristics of the data measured at E288 experiment.}
\end{center}
\end{table}
\paragraph{E288.}
The E288 experiment \cite{Ito:1980ev} presents a large number of low energy points. So, the total number of low-energy point is nearly equal the total number of points of high energy experiments. For convenience we have splitted this data set into three subsets with different center of mass energy $s$. The characteristics of the measurements are shown in table \ref{tab:E288}. Concerning these data we can comment the following:
\begin{itemize}
\item We exclude the data points in the range $9<Q<11$ GeV, because these data are dominated by the $\Upsilon$-resonance ($M_\Upsilon\simeq 9.5$ GeV). The description of $\Upsilon$-resonance production is beyond the scope of current TMD factorization approach.
\item The E288 experiment is made on a copper target. To simulate the effects of copper nuclei we replace the proton PDFs by the following combinations
\begin{eqnarray}\label{pdf:cuprum}
u_{Cu}(x)=\frac{Z u(x)+N d(x)}{A},\qquad d_{Cu}(x)=\frac{Z d(x)+N u(x)}{A},\qquad s_{Cu}(x)=s(x),
\end{eqnarray}
where $Z=29$, $A=63$ and $N=A-Z=34$, are charge, atomic number and the number of neutrons in copper correspondingly.
\item The absolute normalization of the E288 $p_T$-cross-section is unknown. Typically, one includes an additional normalization factor $N_{E288}$, as a parameter of the fit, see e.g. \cite{DAlesio:2014mrz,Su:2014wpa,Landry:1999an,Landry:2002ix}. There is no agreement on this factor values, it varies from $\sim 0.8$ \cite{DAlesio:2014mrz,Landry:1999an,Su:2014wpa} to $\sim 1.2$ \cite{Landry:2002ix}. In our analysis we fix $\mathcal{N}_{E288}=0.8$. 

The theoretical uncertainties for  low energy experiments are large, of the order $\pm 10 \%$ at the best (see 
sec.~\ref{sec:HEE_unser}). As a consequence, the value of the cross-section is very sensitive to the choice of the PDF set and the overall normalization factor. For example, we have checked that the E288 data can be fitted also with $N_{E288}=0.9$ with the same (or better) value of $\chi^2$ by an additional variation of $\mu_b$. However, we consider  this  as a bad practice and restrict ourself to $\mathcal{N}_{E288}=0.8$, as the most conventional solution.

\item The data are splitted into  different bins with different dilepton invariant mass.
For each bin we  evaluate the cross-section eq.~(\ref{th:Xsec_gen}) as 
\begin{eqnarray}
E\frac{d \sigma}{dq^3}=\int_{Q_{\text{min}}}^{Q_{\text{max}}}dQ\, \frac{2Q}{\pi} \frac{d\sigma}{dQ^2 dy d(q_T)^2},
\end{eqnarray}
where $Q_{\max,\min}$ are the boundary of the $Q$-bin.
\end{itemize}

\begin{table}[t]
\begin{center}
\begin{tabular}{|c||c|c|}
 & CDF run I & D0 run I 
\\\hline\hline
$\sqrt{s}$ & 1.8 TeV & 1.8 TeV 
\\\hline
 process & $p+\bar p\to Z\to e^+e^-$ & $p+\bar p\to Z\to e^+e^-$ 
\\\hline
$M_{ll}$ range & 66-116 GeV & 75-105 GeV 
\\\hline 
y & y-integrated & y-integrated 
\\\hline
Observable & $\frac{d\sigma}{dq_T}$ &$\frac{d\sigma}{dq_T}$ 
\\\hline
Exp. $\sigma_{\text{tot}}$ [pb] & $248\pm17$ & $\sigma=221\pm 11$
\\\hline
$\sigma_{\text{tot}}$[pb] \begin{tabular}{@{}c@{}}\cite{Catani:2007vq,Catani:2009sm}$_\text{NLO}$\\ \cite{Catani:2007vq,Catani:2009sm}$_\text{NNLO}$:\end{tabular}&
\begin{tabular}{@{}c@{}}$223.8\pm0.05$\\ $237.63 \pm 0.18$\end{tabular}
&
\begin{tabular}{@{}c@{}}$223.8\pm0.05$\\ $237.63 \pm 0.18$\end{tabular}
\\\hline
Ref. & \cite{Affolder:1999jh} & \cite{Abbott:1999wk,Abbott:1999yd} 
\end{tabular}
\caption{\label{tab:Run1} The characteristics of the data measured at CDF and D0 collaborations at run 1.}
\end{center}
\end{table}
\begin{table}[t]
\begin{center}
\begin{tabular}{|c||c|c|}
 & CDF run II & D0 run II
\\\hline\hline
$\sqrt{s}$ & 1.96 TeV & 1.96 GeV
\\\hline
 process & $p+\bar p\to Z\to e^+e^-$& $p+\bar p\to Z\to e^+e^-$
\\\hline
$M_{ll}$ range & 66-116 GeV & 70-110 GeV 
\\\hline 
y & y-integrated & y-integrated  
\\\hline
 Observable & $\frac{d\sigma}{dq_T}$ & $\frac{1}{\sigma}\frac{d\sigma}{dq_T}$
 \\\hline
Exp. $\sigma_{\text{tot}}$ [pb] & $256 \pm 2.91 $ & $\sigma=255$
\\\hline
$\sigma_{\text{tot}}$[pb] \begin{tabular}{@{}c@{}}\cite{Catani:2007vq,Catani:2009sm}$_\text{NLO}$\\ \cite{Catani:2007vq,Catani:2009sm}$_\text{NNLO}$:\end{tabular}&
\begin{tabular}{@{}c@{}}$245.0\pm0.06$\\ $259.77 \pm 0.22$\end{tabular} 
&
\begin{tabular}{@{}c@{}}$245.0\pm0.06$\\ $259.77 \pm 0.22$\end{tabular} 
\\\hline
Ref.& \cite{Aaltonen:2012fi} & \cite{Abazov:2007ac}
\end{tabular}
\caption{\label{tab:Run2} The characteristics of the data measured at CDF and D0 collaborations at run 2.}
\end{center}
\end{table}

\paragraph{CDF and D0.}
The data on the Z-boson production measured by CDF and D0 collaborations at Tevatron Run 1 and Run 2 \cite{Affolder:1999jh,Abbott:1999wk,Abbott:1999yd,Aaltonen:2012fi,Abazov:2007ac} have been used nearly in every fit of unpolarized TMDPDFs. They are summarized in tables~\ref{tab:Run1}-\ref{tab:Run2}. Concerning these data we can comment the following:
\begin{itemize}
\item There is a known tension between the values of total cross-section at run I of CDF and D0. Here we restrict ourself to the fit of the shape of the cross-section and normalize the theoretical points on the bin-by-bin integrals in the allowed range of $q_T$. I.e. we multiply the theoretical cross-section by the factor
\begin{eqnarray}\label{LHC_norm}
\mathcal{N}=\frac{\sum_{\substack{\text{included}\\{\text{bins}}}} \Delta q_T \frac{d\sigma_{\text{exp.}}}{dq_T}}{\sum_{\substack{\text{included}\\{\text{bins}}}} \Delta q_T \frac{d\sigma_{\text{th.}}}{dq_T}},
\end{eqnarray}
where $\Delta q_T$ is the size of $q_T$-bins. As we show in sec.~\ref{sec:LHC_norm} the obtained normalization factors are very close to one (at NNLO), and the values of partial cross-sections are in  agreement with the experimental ones within error-bars. In the tables~\ref{tab:Run1}-\ref{tab:Run2}, we also present the values of the total cross-sections evaluated by DYNNLO code \cite{Catani:2007vq,Catani:2009sm}. In this calculation of the total-cross-section, we have used the same inputs as in the TMD fits, i.e. the PDF are taken from MMHT2014 set~\cite{Harland-Lang:2014zoa}.
\item The experimental values for cross-section points are obtained by integrating over all values of $y$, integrating over measure range of $Q$ and averaging in $q_T$. Consequently, we have used the following expression for the cross-section
\begin{eqnarray}\label{eq:X_TeV}
\frac{d\sigma}{dq_T}=\frac{1}{\Delta q_T}\int_{q_{T,\text{low}}}^{q_{T,\text{high}}}2 q'_T dq'_T \int_{-y_0}^{y_0}dy \int_{M_{ll,\min}}^{M_{ll,\max}} 2 Q dQ~\frac{d\sigma}{dy d({q'_T}^2)dQ^2},
\end{eqnarray}
where $y_0=\frac{1}{2}\ln(s/Q^2)$, $q_{T,\text{low}}$ and $q_{T,\text{high}}$ are boundaries of $q_T$-bin, and $\Delta q_T$ is the size of the $q_T$-bin.
\end{itemize}

\begin{table}[t]
\begin{center}
\begin{tabular}{|c||c|c|}
 & ATLAS & ATLAS 
\\\hline\hline
$\sqrt{s}$ & 7 TeV & 8 TeV 
\\\hline
 process & $pp\to Z\to ee+\mu\mu$ & $pp\to Z\to \mu\mu$ 
\\\hline
$M_{ll}$ range & 66 - 116 GeV & 66 - 116 GeV 
\\\hline
lepton cuts &
\begin{tabular}{@{}c@{}}$p_T>20$ GeV\\ $|\eta|<2.4$\end{tabular}&
\begin{tabular}{@{}c@{}}$p_T>20$ GeV\\ $|\eta|<2.4$\end{tabular} 
\\\hline 
$y$ & $-2.4<y<2.4$ & $-2.4<y<2.4$
\\\hline
Observable & $\frac{1}{\sigma}\frac{d\sigma}{dq_T}$ &$\frac{1}{\sigma}\frac{d\sigma}{dq_T}$
\\\hline
Exp.$\sigma_{\text{fid}}$[pb] & - & $537.1\pm 0.63(\pm 2.8\%) $ 
\\\hline
Theor.$\sigma_{\text{fid}}$[pb] &\begin{tabular}{@{}c@{}}
\cite{Catani:2007vq,Catani:2009sm}$_\text{NLO}$: $448.56\pm0.19$\\ 
\cite{Catani:2007vq,Catani:2009sm}$_\text{NNLO}$: $471.53\pm0.94$\end{tabular}&
 \begin{tabular}{@{}c@{}}
\cite{Catani:2007vq,Catani:2009sm}$_\text{NLO}$: $505.53\pm0. 21$\\ 
\cite{Catani:2007vq,Catani:2009sm}$_\text{NNLO}$: $531.39\pm 0.93$ \\ 
\cite{Ridder:2016nkl}: $507.9^{+2.4}_{-0.7}$ \end{tabular} 
\\\hline
Ref. & \cite{Aad:2014xaa} & \cite{Aad:2015auj} 
\end{tabular}
\caption{\label{tab:Z_ATLAS} The characteristics of the Z-boson production data measured by ATLAS collaborations.}
\end{center}
\end{table}

\begin{table}[t]
\begin{center}
\begin{tabular}{|c||c|c|}
 & ATLAS & ATLAS 
\\\hline\hline
$\sqrt{s}$ &  8 TeV & 8 TeV 
\\\hline
 process & $pp\to Z/\gamma^*\to \mu\mu$ & $pp\to Z/\gamma^*\to \mu\mu$ 
\\\hline
$M_{ll}$ range & 46 - 66 GeV & 116 - 150 GeV 
\\\hline
lepton cuts &
\begin{tabular}{@{}c@{}}$p_T>20$ GeV\\ $|\eta|<2.4$\end{tabular} 
&\begin{tabular}{@{}c@{}}$p_T>20$ GeV\\ $|\eta|<2.4$\end{tabular} 
\\\hline 
$y$ & $-2.4<y<2.4$ & $-2.4<y<2.4$ 
\\\hline
Observable & $\frac{1}{\sigma}\frac{d\sigma}{dq_T}$& $\frac{1}{\sigma}\frac{d\sigma}{dq_T}$ 
\\\hline
Exp.$\sigma_{\text{fid}}$[pb] &  $14.96\pm2.62(\pm 2.8\%) $ & $5.59\pm 1.52(\pm 2.8\%)$
\\\hline
Theor.$\sigma_{\text{fid}}$[pb] & - & -
\\\hline
Ref. & \cite{Aad:2015auj} & \cite{Aad:2015auj} 
\end{tabular}
\caption{\label{tab:DY_ATLAS} The characteristics of the  data for the vector boson production off the Z-peak measured by ATLAS collaborations.}
\end{center}
\end{table}

\paragraph{ATLAS.}
The data by ATLAS collaboration in \cite{Aad:2014xaa,Aad:2015auj} cover a broad range of dilepton invariant masses for the Drell-Yan process with small experimental error-bands. So, this set provide the biggest constraints on TMD extraction coming from  high energy data points. The characteristics of the measurements are resumed in 
tables~\ref{tab:Z_ATLAS}-\ref{tab:DY_ATLAS} and here we comment the following:
\begin{itemize}
\item The data from the ATLAS detector at 8 TeV run are presented in several sets \cite{Aad:2015auj}, which corresponds to  different treatment of final-state photon radiation. We have considered the "dressed" set of the data.
\item The values of cross-section have been calculated by the expression in eq.~(\ref{eq:X_TeV}), where $y_0=2.4$, as it is presented in the tables \ref{tab:Z_ATLAS}-\ref{tab:DY_ATLAS}.
\item There is a known tension between the theoretical calculation of the integrated cross-section and the measured one, see e.g.~\cite{Aad:2015auj,Ridder:2016nkl}. Moreover the theoretical cross-section for vector boson production is not known (at least, the DYNNLO package \cite{Catani:2007vq,Catani:2009sm}, which we have used for evaluation of total cross-sections, does not guarantee the accurate calculation outside the Z-peak). Therefore, we normalize the calculated cross-sections by a factor, as explained in more detail  in the text  around eq.~(\ref{LHC_norm1}). In sec. \ref{sec:LHC_norm}, we compare the obtained values of normalization to the total cross-section. We have found that the values of obtained normalization are practically independent on the non-perturbative input of the TMD model, and at NNLL/NNLO correctly reproduce (within the error-bars) the measured total cross-sectio.
\item All data sets from LHC are presented within fiducial cross-sections. Therefore, we have implemented the cut leptonic tensor as it is discussed in sec.~\ref{sec:cuts}.
\end{itemize}

\begin{table}[t]
\begin{center}
\begin{tabular}{|c||c|c|}
 & CMS & CMS 
\\\hline\hline
$\sqrt{s}$ & 7 TeV & 8 TeV 
\\\hline
 process & $pp\to Z\to ee+\mu\mu$ & $pp\to Z\to \mu\mu$ 
\\\hline
$M_{ll}$ range & 60-120 GeV & 60-120 GeV 
\\\hline
lepton cuts &
\begin{tabular}{@{}c@{}}$p_T>20$ GeV\\ $|\eta|<2.1$\end{tabular}
&
\begin{tabular}{@{}c@{}}$p_T>15$ GeV\\ $|\eta|<2.1$\end{tabular} 
\\\hline 
y & $|y|<2.1$ & $|y|<2.1$ 
\\\hline
Observable & $\frac{1}{\sigma}\frac{d\sigma}{dq_T}$ &$\frac{1}{\sigma}\frac{d\sigma}{dq_T}$
\\\hline
Norm. exp. & - & - 
\\\hline
$\sigma_{\text{fid}}$[pb] \begin{tabular}{@{}c@{}}\cite{Catani:2007vq,Catani:2009sm}$_\text{NLO}$\\ \cite{Catani:2007vq,Catani:2009sm}$_\text{NNLO}$ 
\end{tabular}& 
\begin{tabular}{@{}c@{}}$379.43\pm 0.16$\\ $398.27\pm 0.71$\end{tabular}&
\begin{tabular}{@{}c@{}}$427.32\pm0.53$\\ $448.04\pm 0.83$\end{tabular}
\\\hline
Ref. & \cite{Chatrchyan:2011wt} & \cite{Khachatryan:2016nbe} 
\end{tabular}
\caption{\label{tab:Z_CMS} The characteristics of the Z-boson production data measured by CMS collaborations.}
\end{center}
\end{table}

\begin{table}[t]
\begin{center}
\begin{tabular}{|c||c|c|c|}
 &  LHCb  & LHCb & LHCb
\\\hline\hline
$\sqrt{s}$ &  7 TeV & 8 TeV & 13 TeV
\\\hline
 process &  $pp\to Z\to \mu\mu$ & $pp\to Z\to \mu\mu$ & $pp\to Z\to \mu\mu$
\\\hline
$M_{ll}$ range &  60-120 GeV & 60-120 GeV & 60-120 GeV
\\\hline
lepton cuts &
\begin{tabular}{@{}c@{}}$p_T>20$ GeV\\ $2<\eta<4.5$\end{tabular}  
&
\begin{tabular}{@{}c@{}}$p_T>20$ GeV\\ $2<\eta<4.5$\end{tabular}  
&
\begin{tabular}{@{}c@{}}$p_T>20$ GeV\\ $2<\eta<4.5$\end{tabular}  
\\\hline 
y &  $2<y<4.5$ & $2<y<4.5$ & $2<y<4.5$
\\\hline
Observable & $d\sigma(q_T)$  & $d\sigma(q_T)$ & $\frac{d\sigma}{dq_T}$
\\\hline
Norm. exp. & $\sigma=76.0\pm 3.1$ pb & $\sigma=95.0\pm 3.2$ pb & $\sigma=198.0\pm 13.3$ pb
\\\hline
$\sigma_{\text{fid}}$[pb] \begin{tabular}{@{}c@{}}\cite{Catani:2007vq,Catani:2009sm}$_\text{NLO}$\\ \cite{Catani:2007vq,Catani:2009sm}$_\text{NNLO}$ 
\end{tabular} & 
\begin{tabular}{@{}c@{}}$69.85\pm 0.3$\\ $74.30\pm 0.21$\end{tabular}&
\begin{tabular}{@{}c@{}}$88.98\pm 0.397$\\ $93.50\pm 0.3$\end{tabular}&
\begin{tabular}{@{}c@{}}$185.0\pm 0.09$\\ $192.78 \pm 0.82$\end{tabular}
\\\hline
Ref. &  \cite{Aaij:2015gna}& \cite{Aaij:2015zlq} & \cite{Aaij:2016mgv}
\end{tabular}
\caption{\label{tab:Z_LHCb} The characteristics of the Z-boson production data measured by LHCb collaborations.}
\end{center}
\end{table}

\paragraph{CMS and LHCb.}
The CMS and LHCb collaborations provide data  around the Z-boson  peak in \cite{Chatrchyan:2011wt,Khachatryan:2016nbe,Aaij:2015gna,Aaij:2015zlq,Aaij:2016mgv}, see tables \ref{tab:Z_CMS},\ref{tab:Z_LHCb}. The treatment of these data is similar to the case  of ATLAS data:
\begin{itemize}
\item The values of cross-section have been calculated by the expression in eq.~(\ref{eq:X_TeV}), where the limits for $y$-integration $y_0$ are taken in accordance to the tables  \ref{tab:Z_CMS}-\ref{tab:Z_LHCb}.
\item Just as in the case of ATLAS data we have normalized the calculated cross-sections by the factor provided in eq.~(\ref{LHC_norm}) discussed in sec.\ref{sec:LHC_norm}.
We have found a good agreement between the theoretical and experimental values for total cross-section for LHCb data.
\item All data sets from LHC are fiducial cross-sections. Therefore, we have implemented the cut leptonic tensor as it is discussed in sec.~\ref{sec:cuts}.
\end{itemize}

Finally, we have considered only points which allow a consistent TMD treatment. I.e. the points with the value of $q_T<\delta_T Q$, where $\delta_T$ is sufficiently small. In the literature we have not found any special study on this limiting ratio. So, we present our study in sec.~\ref{sec:deltaT}, and conclude that TMD factorization range is $q_T/Q<0.2$.

\subsection{\texttt{arTeMiDe}}

In order to evaluate the cross-sections we have prepared  the program package \texttt{arTeMiDe}. The \texttt{arTeMiDe} package is a collection of FORTRAN modules that evaluates individual terms of the TMD factorization formalism, such as TMD evolution factors, TMDPDFs, and combines them into the differential cross-sections. \texttt{arTeMiDe} forms a flexible package for TMDPDF phenomenology based on the $\zeta$-prescription, as described in this article. It is publicly available at the web-page \cite{web}. 

\texttt{arTeMiDe} version 1.1  evaluates the quark and gluon unpolarized TMDPDFs (although in the discussed fit the gluon TMDPDFs are not necessary) for any given function $f_{NP}$, at any composition of perturbative orders from LO to NNLO,  with or without renormalon-induced power corrections. For the current study,  the input PDFs are taken from the MMHT2014 PDF set~\cite{Harland-Lang:2014zoa}.

The most time-consuming part of the numerical evaluation of the TMDPDFs, is the convolution integral in eq.~(\ref{eq:cfnp}), which is especially expensive at NNLL/NNLO. Within the \texttt{arTeMiDe} package the convolution integral is optimized using an approximate expression for NNLO coefficient functions. The approximate form of the NNLO coefficient function is (note, that NLO and renormalon coefficient functions can be presented in this form without approximation)
\begin{eqnarray}\nn
C(\mathbf{L}_{\mu},x)&=&A_1 \delta(\bar x)+A_2\(\frac{1}{1-x}\)_++A_3\(\frac{\ln \bar x}{1-x}\)_++A_4 \ln \bar x+A_5 \ln^2 \bar x++A_6 \ln^3 \bar x
\\ && \label{th:aprox_coef}
+B_1\ln x+B_2 \ln^2 x+B_3 \ln^3 x+B_4\frac{1}{x}+B_5\frac{\ln x}{x}
\\\nn && +c_1+c_2 x+c_3 x^2 +c_4 x^3+c_5 \ln \bar x \ln x+c_6 \ln \bar x \ln^2 x,
\end{eqnarray}
where coefficients $A$, $B$ and $c$ are functions of $\mathbf{L}_\mu$. Such an approximate form is widely used in NNLO+ phenomenology of PDFs, see e.g.~\cite{vanNeerven:1999ca}. Here, the coefficients $A$ and $B$ represent the singular at $x\to 1$  and $x\to 0$ terms, and are evaluated exactly. The coefficients $c$ represent the smooth part of the coefficient function, which is reconstructed by appropriate values of $c_i$ with  better then $1\%$ accuracy. The values of constants $A$, $B$ and $c$ are presented in the appendix \ref{app:coef}. In the convolution integral the main numerical contribution comes from the singular terms proportional to $A$ and $B$, which are exact. The relative difference between the convolution with exact coefficient function and approximate expression in eq.~(\ref{th:aprox_coef}) is of order $10^{-6}$. This numerical error is compatible with the numerical error of integration procedure and far inside the theoretical error-bands. 

The evaluation of the Hankel-type integral over $b$ is one of the main source of numerical errors. Typically, in order to obtain sufficient precision one should include a large number of points into the integral, which is very costly especially at NNLL/NNLO. \texttt{arTeMiDe} evaluates this integral with the Ogata quadratures \cite{Ogata_quadrature}. The Ogata quadrature is a double exponential quadrature, whose nodes are the zeros of the Bessel function. It provides a fast and precise evaluation of Hankel-type integrals with the minimal number of integrand calls.

The fitting procedure has been performed by minimizing the $\chi^2$-function. The minimization of the $\chi^2$ distribution have been done using the MINUIT package from the CERN library \cite{James:1975dr,James:1994vla}. The estimation of the statistical uncertainties for non-perturbative parameters is made with the  MINOS procedure, performing the variation of parameters in the range $\chi^2\pm \Delta\chi^2$, with $\Delta\chi^2$ corresponding to the 68\% confidence level (i.e. $\Delta \chi^2\simeq\{1.03,2.32,3.55\}$ for 1,2,3 fitting parameters, correspondingly.) The sources of  theoretical uncertainties have been pointed in  sec \ref{sec:th_uncertain}, and parameterized by the constants $c_{1,2,3}$. The variation of these constants in the region $(0.5,2)$ produces the error-bands. The discussion on the individual contributions of theoretical uncertainties associated with different scales is given in sec. \ref{sec:HEE_unser}. 

\subsection{Models for non-perturbative part of TMDPDFs}
\label{sec:models}

The non-perturbative part of the  TMDPDF in general needs some ansatz, the parameters of which are to be extracted from data. In our study we have tested different ansatzes of the following general form
\begin{eqnarray}
\label{eq:cfnp}
F_{q\ot h}(x,\vec b)=\int_x^1 \frac{dz}{z} \sum_{f}C_{q\ot f}\(z,\vec b;\mu,\zeta_\mu\)f_{f\ot h}\(\frac{x}{z},\mu\)f_{NP}\(z,\vec b\),
\end{eqnarray}
where $f_{f\ot h}$ is PDF of the parton with the flavour $f$. The non-perturbative information of the TMDPDF, which is unreachable from the PDFs, is contained in $f_{NP}$. In order to match the perturbative regime, the function $f_{NP}$ should approach $1$ for $b\to0$. Instead, the behavior of  $f_{NP}$ for $b\to \infty$  is not so well established, which requests a  test of different models.  In the current study, we restrict ourself to flavor independent $f_{NP}$, i.e. $f_{NP}$ is common for TMDPDFs of different flavours. The flavour-dependence of TMDPDFs enters only through PDFs and coefficient functions, i.e. it is completely determined.

The large-$b$ behavior of TMD distributions is the key point of TMD parametrization and extraction. There is no common agreement on this behavior. Clearly, such an agreement cannot be achieved in general, since the $b$-shape of a TMD distribution is strongly dependent on the large-$b$ prescription. For example, the Gaussian behavior is typically observed in the models based on $b^*$-prescription.  Moreover, the classical fits by \texttt{ResBos} package \cite{Landry:2002ix} disfavor other non-perturbative behaviors, such an exponential one (for more recent discussion, see \cite{Guzzi:2013aja}). Also the Gaussian shape is used in \texttt{DYRes} code \cite{Catani:2015vma} (together with $b^*$-prescription) and in \texttt{DYqT} code \cite{Bozzi:2010xn} (together with the minimal prescription). Contrary, the fit made in ref.~\cite{DAlesio:2014mrz},  which does not employ the $b^*$-prescription,  uses an exponential shape of $f_{NP}$ and also obtains an agreement with data. We point out  that the use of LHC data for TMD extraction is made here for the first time (to our knowledge). Given the precision of LHC data, the consistency and/or  goodness of all previous hypotheses have to be rediscussed.

In order to decide the best shape of $f_{NP}$ within $\zeta$-prescription, we have considered several subsets of the data. It appears important to include simultaneously both high-energy and low-energy data because they are sensitive to different parts of the $b$-space spectrum. We have found that the most optimal data subset is given by the  E288 data and the ATLAS Z-boson production data, see tables \ref{tab:E288} and \ref{tab:Z_ATLAS}. In this subset, the very small error-bands of ATLAS data are compensated by a large number of points in E288 data, and as a result,  we have a certain equilibrium between  low and high-energy inputs.

Using the E288/ATLAS subset we have performed multiple fits using several different functional forms of $f_{NP}$. Probably, the most informative preliminary test is the comparison of the pure Gaussian and exponential behavior for separate/joint  low and high energy data points. In table \ref{tab:diff_fNP} we demonstrate results of fit with some simple single-parameter models. According to this table, although the quality of the fit is still not optimal, the high-energy data clearly favor the Gaussian shape of $f_{NP}$, while the low-energy data favor the exponential behavior of $f_{NP}$. This difference is simply explained if we recall that at higher energies (and thus at generally higher $q_T$) the Fourier integral in eq.~(\ref{th:hadron_tensor}) is saturated by small values of $b$. At lower energies (and thus at generally smaller $q_T$) the Fourier integral in eq.~(\ref{th:hadron_tensor}) is affected by a wider interval of values of $b$. Therefore, the results presented in the table~\ref{tab:diff_fNP}, suggest that $f_{NP}$ should be Gaussian at small-$b$ and exponential at large-$b$. This is in complete agreement with the theory expectations discussed in sec.~(\ref{sec:TMDPDF}). The expected $f_{NP}$ should be a function with a Taylor series expansion (around $b=0$) of even powers of $b$, with an exponential decay at $b\to \infty$. A simple representative of such functions is $\cosh^{-1}(\lambda b)$. The test of this $f_{NP}$ is given in the last columns of the table~\ref{tab:diff_fNP} which  clearly shows that this function alone, although it works much better than a Gaussian or an exponential, is not able to describe both low and high energy data, and thus we need extra parameters.

\begin{table}[t]
\begin{center}
\begin{tabular}{||c||c|c|c||}
data & $f_{NP}=e^{-\lambda b}$ & $f_{NP}=e^{-\lambda b^2}$ & $f_{NP}=\cosh^{-1}(\lambda b)$ 
\\\hline\hline
ATLAS & \specialcellcenter{4.78} & \specialcellcenter{1.43} & \specialcellcenter{1.42} 
\\
\hline
E288 & \specialcellcenter{2.70} & \specialcellcenter{5.68} & \specialcellcenter{3.64} 
\\
\hline
E288+ATLAS & \specialcellcenter{8.18} & \specialcellcenter{5.77} & \specialcellcenter{3.72} 
\end{tabular}
\end{center}
\caption{\label{tab:diff_fNP} The values of $\chi^2/d.o.f$ for different single-parameter non-perturbative functions $f_{NP}$, minimized on  different data sets. The  $\chi^2/d.o.f$  values correspond to $\delta_T=0.2$ and NNLL/NNLO.}
\end{table}

The preliminary tests with simple one-parameter dependence for the $f_{NP}$ shape can  be summerized by the following:
\begin{itemize}
\setlength\itemsep{0pt}
\item[(i)] The high and low energy data should be considered altogether, because they test different intervals of the $b$-space spectrum of  $f_{NP}$.

\item[(ii)] The subset of data points E288 and ATLAS Z-boson, is very selective for the $f_{NP}$. A good fit of this subset guaranties the good fit for  the whole set of data points. Nevertheless, in the following sections, we include all experiments, for consistency.

\item[(iii)] Both theoretically and phenomenologically,  we argue  that $f_{NP}$ should be a function of even powers of  $b$ with an exponential asymptotic behavior at $b\to \infty$.  Using a minimal set of  two parameters (and the evolution parameter $g_K$) we find that one can easily fit the data with a $\chi^2/d.o.f\sim 1.2$-1.3. The addition of more parameters (say for the control of $b^4$ correction and/or flavor dependence) has the possibility to increase the quality of the fit. However, in this work, we do not consider extra parameters, since the current quality of the fit is already typical and reasonable for the modern TMD extraction (compare e.g. with \cite{Bacchetta:2017gcc}).

\item[(iv)] One needs at least two parameters (one to control $\sim b^2$ behavior at $b\to 0$ and another to control the asymptotics) to fit simultaneously low and high-energy data. However, the multiplication by polynomials (e.g. $f_{NP}\sim (1+\lambda b^2)/\cosh(b)$) does not work well, which suggests that the asymptotic terms $\sim b^2 e^{-b}$ are disfavored.
\end{itemize}

Based on this experience we have formulated some simple ansatzes for $f_{NP}$.
\begin{itemize}
\item \textit{Model 1}: This ansatz uses the fact that the simplest even-$b$ function with exponent asymptotics is the hyperbolic cosine. The model reads
\begin{eqnarray}\label{MODEL1}
f_{NP}(b)=\frac{\cosh\(\(\frac{\lambda_2}{\lambda_1}-\frac{\lambda_1}{2}\)b\)}{\cosh\(\(\frac{\lambda_2}{\lambda_1}+\frac{\lambda_1}{2}\)b\)},
\end{eqnarray}
where $\lambda_1[\GeV]>0$ and $\lambda_2[\GeV^2]>0$ are free parameters.  The advantage of this model is its simplicity and independence of the Bjorken variable. The model 1 has a quadratic (Gaussian) behavior at small-$b$ $f_{NP}\sim e^{-\lambda_2 b^2}$ and exponential behavior at large-$b$ $f_{NP}\sim e^{-\lambda_1 b}$.

\item \textit{Model 2}: The model 2 reads
\begin{eqnarray}\label{MODEL2}
f_{NP}(z,b)=\exp\(\frac{-\lambda_2 z b^2}{\sqrt{1+z^2 b^2\frac{\lambda_2^2}{\lambda_1^2}}}\),
\end{eqnarray}
where $\lambda_1[\GeV]>0$ and $\lambda_2[\GeV^2]>0$ are free parameters. In this model we attempt to incorporate the theoretical expectations on the $z$-dependence of $f_{NP}$. So,  the model 2 has a $zb^2$-behavior at small-$b$ $f_{NP}\sim e^{-\lambda_2 zb^2}$ and exponential behavior at large-$b$ $f_{NP}\sim e^{-\lambda_1 b}$.
\end{itemize}

Both models have two parameters, which we include in the parameterization such that the parameter $\lambda_1[\GeV]$ dictates the asymptotical behavior at large $b$. and the parameter $\lambda_2[\GeV^2]$ gives the quadratic term. A priory, the parameter $\lambda_1$ should be of  order of $m_\pi\sim 0.14\;\GeV$, since it is the only natural scale of strong forces at large distances. The parameter $\lambda_2[\GeV^2]$ roughly corresponds to the size of the leading power correction to small-$b$ OPE, see sec.~\ref{sec:TMDPDF}. We can associate $\lambda_2$ with the scale $B$ as $\lambda_2\sim B^{-2}$. In ref.~\cite{Scimemi:2016ffw} we have estimated the size of this parameter in the large-$\beta_0$ approximation as $\lambda_2\sim 0.075\;\GeV^2$.

Additionally, to the parameters $\lambda_{1,2}$ we have studied the parameter $g_K[\GeV^2]>0$, which parametrizes the non-perturbative contribution to the rapidity evolution kernel $\mathcal{D}$ (see eq.~(\ref{th:D_evol})). The importance of this parameter is not clear from the literature. In ref.~\cite{Scimemi:2016ffw} we have estimated its size in the large-$\beta_0$ approximation as $0.01\pm 0.03\;\GeV^2$, i.e. consistent with zero. Also, the fit of~\cite{DAlesio:2014mrz} shows a negligible influence of this parameter on the final results. Therefore, in the following we consider both possibilities $g_K=0$ and $g_K\neq 0$. In section~\ref{sec:global}, we demonstrate that the parameter $g_K$ is important at lower perturbative order, but its influence is negligible at NNLL/NNLO.

\subsection{The domain of TMD factorization}
\label{sec:deltaT}

\setlength{\tabcolsep}{5pt}
\begin{table}[t]
\begin{tabular}{|c||cccc|c|cccc|}
$\delta_T$ & 0.1 & 0.125 & 0.15 & 0.175 & 0.2 & 0.225 & 0.25 & 0.275 & 0.3
\\\hline\hline
CDF+D0 run1					 	& 27 & 34 & 38 & 41 & 44 & 47 & 49 & 51 & 52
\\
CDF+D0 run2 					& 22 & 28 & 32 & 38 & 43 & 49 & 55 & 60 & 63
\\
ATLAS Z-production (7+8 TeV) 	& 10 & 12 & 14 & 16 & 18 & 20 & 21 & 23 & 24
\\
ATLAS DY (8 TeV) 				& 9  & 11 & 12 & 14 & 14 & 16 & 16 & 18 & 18
\\
CMS (7+8 TeV) 					& 8  & 10 & 10 & 12 & 14 & 16 & 16 & 18 & 18
\\
LHCb (7+8+13 TeV) 				& 18 & 21 & 24 & 27 & 30 & 30 & 33 & 33 & 33
\\\hline
High energy total 				& 94 & 116& 130& 148& 163& 178& 190& 203& 208
\\\hline
E288 200 GeV 					& 16 & 20 & 24 & 29 & 35 & 36 & 41 & 44 & 47
\\
E288 300 GeV 					& 22 & 27 & 33 & 38 & 45 & 46 & 51 & 55 & 59
\\
E288 400 GeV 					& 33 & 40 & 49 & 57 & 66 & 69 & 76 & 82 & 85
\\\hline
Low energy total 				& 71 & 87 & 106& 124& 146& 151& 168& 181& 191
\\\hline\hline
Total 							& 165& 203& 236& 272& 309& 329& 358& 384& 399
\end{tabular}
\caption{\label{tab:number_of_points} The number of points  with $q_T<\delta_T Q$ for each data set. In the  majority of fits we use $\delta_T=0.2$, see explanation in the text.}
\end{table}
\setlength{\tabcolsep}{6pt}

\begin{figure}[t]
\centering
\includegraphics[width=0.45\textwidth]{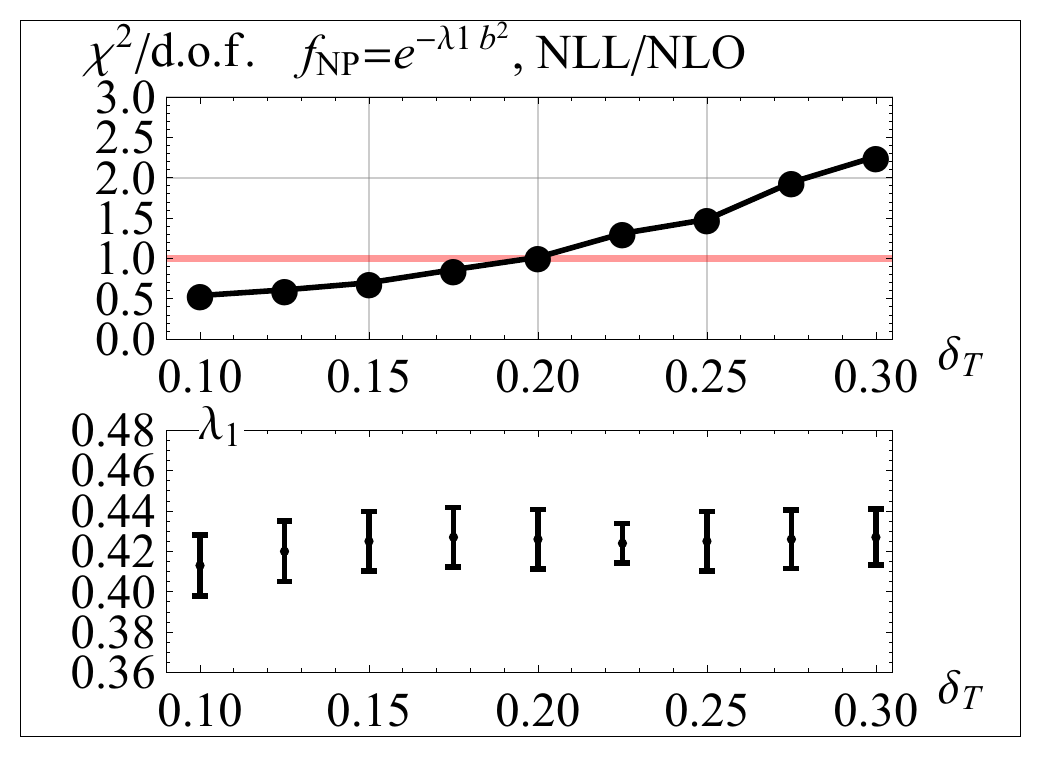}
\includegraphics[width=0.45\textwidth]{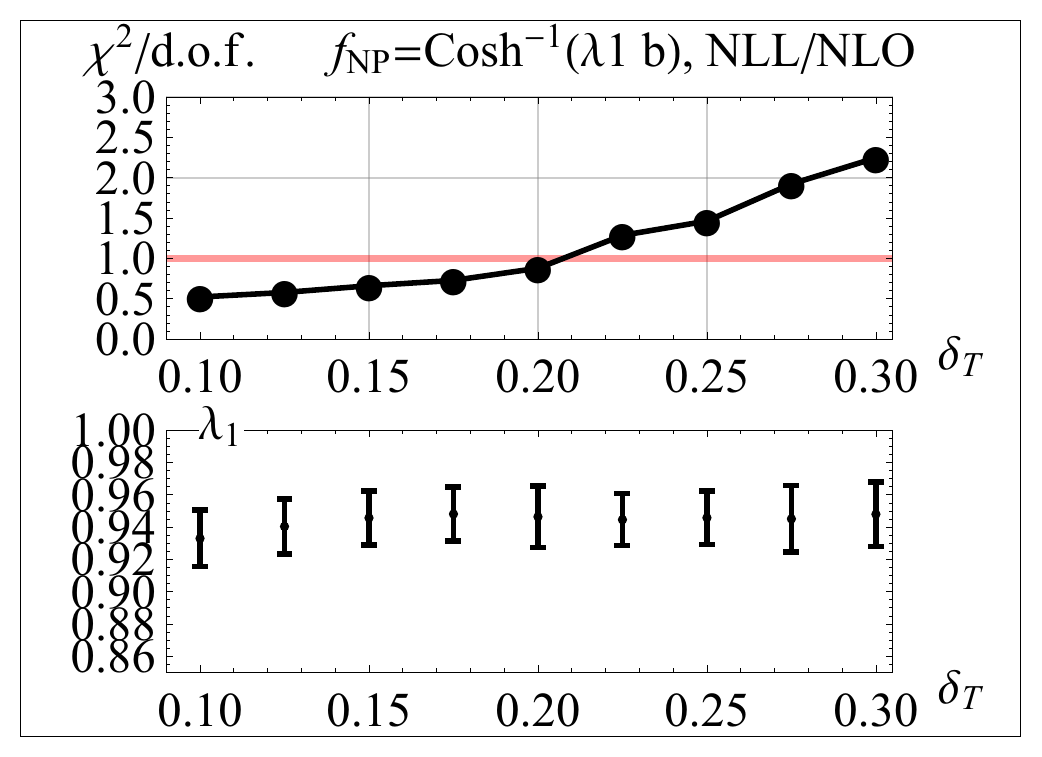}

\includegraphics[width=0.45\textwidth]{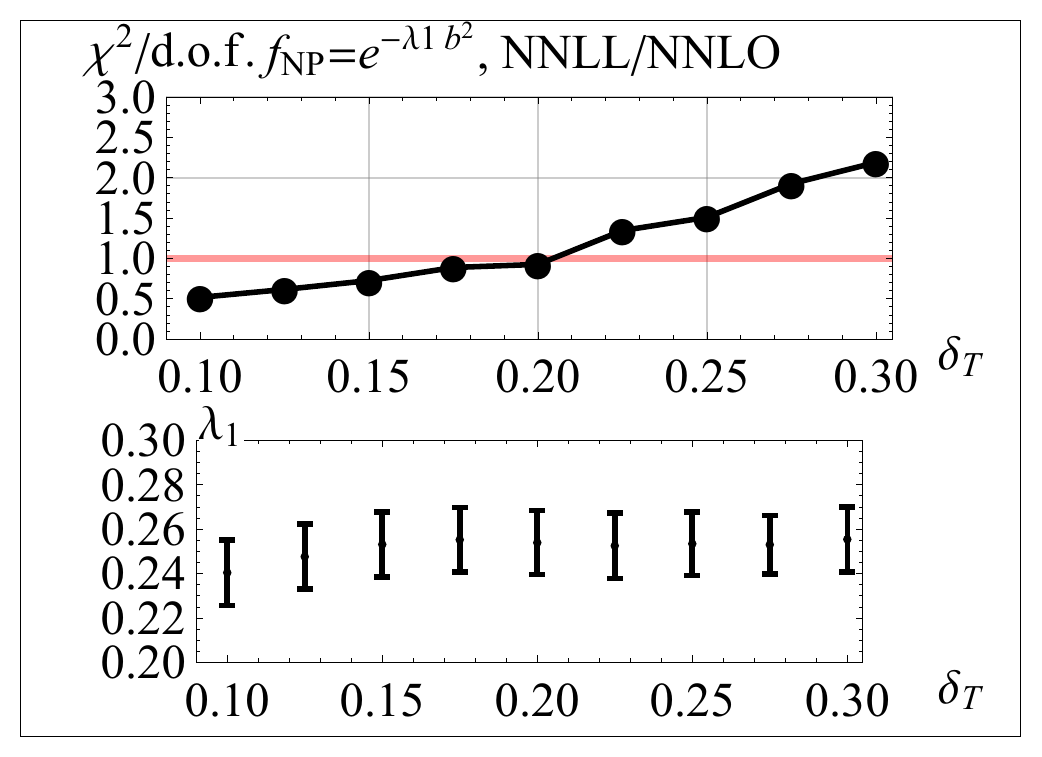}
\includegraphics[width=0.45\textwidth]{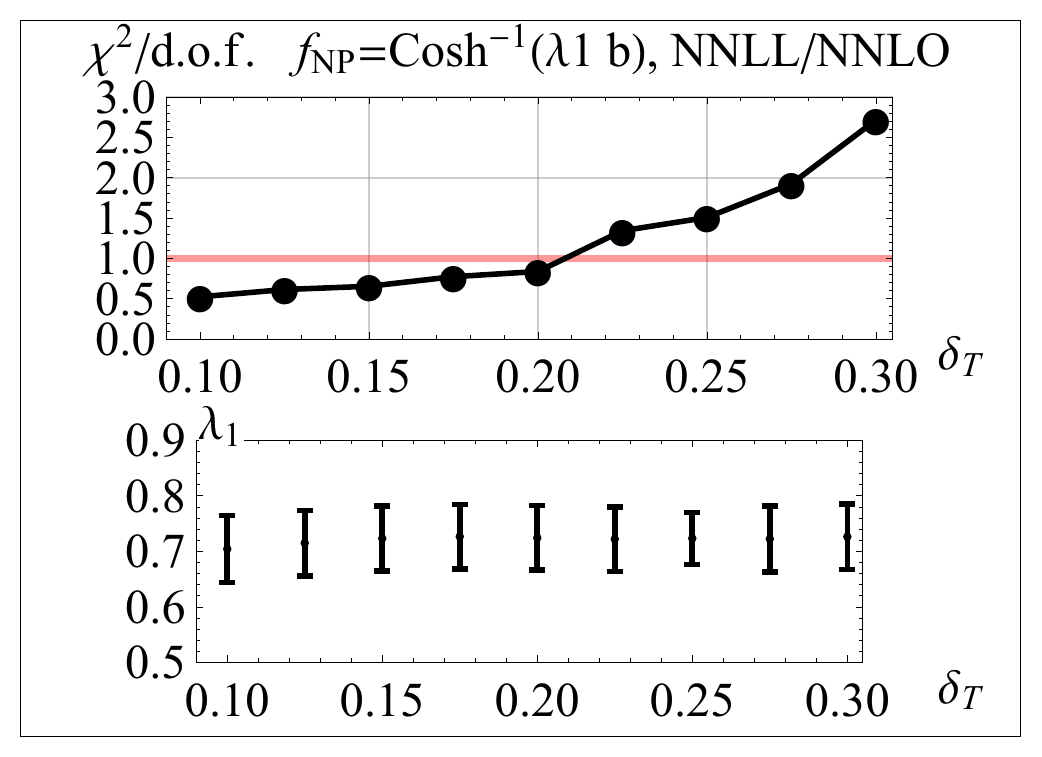}
\caption{\label{fig:scan_dT} The $\delta_T$ dependence of the value of $\chi^2/d.o.f.$ for some one-parameter models. The value of the parameter coming from the fit is also shown together with systematic uncertainties.} 
\end{figure}

The TMD factorization is restricted to the small-$q_T$ range. The size of the allowed $q_T$-region is \textit{a priory} unknown. We have not found any phenomenological studies on this point but only some statement on the strong dependence of the fit on the $q_T$-window. A specific study on TeVatron Z-boson production data in ref.~\cite{Becher:2011xn} shows that the Y-term contribution is extremely marginal for  $q_T<30$ GeV. 


In order to make a qualitative study, we introduce the parameter $\delta_T$ and we consider all data points with $q_T<\delta_T Q$. The amount of data points which are allowed by such a restriction are shown in the table~\ref{tab:number_of_points}. In order to estimate the maximum value of $\delta_T$ we perform a series of fits with increasing values of $\delta_T$. Ideally, the $\chi^2/d.o.f.$ and the fitting parameters should be stable within and unstable outside of the allowed $\delta_T$ interval. In this way, considering the dependence on $\delta_T$ one should find an interval  of $\delta_T$ for which the fit is not sensitive to  the $Y$-term. This point indicates the region of TMD-factorization, and should not depend of the perturbative order.

We have performed such a test for high-energy data set with different one-parameter forms of $f_{NP}$. We have especially used the one parameter models to guarantee the absence of fine-tuning of the cross-section. For this reason we also exclude the E288 data, because it is impossible to describe high- and low-energy data with a single non-perturbative parameter. The result of the fits practically agrees for all tested models and orders. In fig.~\ref{fig:scan_dT}, we present some typical outcome of the fits. 

In plots \ref{fig:scan_dT} one can see that all models reproduce the data very-well at very small $\delta_T$, which is expected since the TMD factorization is only valid at $q_T\ll Q$. Then the value of $\chi^2$ slightly grows but keeps less then one until $\delta_T=0.2$ and  after this threshold it jumps to higher values. The next jump is seen at $\delta_T=0.25$. After  $\delta_T=0.25$ the value of $\chi^2$ increases rapidly. 
We interpret this fact saying 
 that at $\delta_T=0.2$ we become sensitive to $Y$-term, and at $\delta_T=0.25$ the $Y$-term starts to dominate the cross-section, i.e. we leave the domain of TMD factorization. We have found that the presented plots rather strongly depend on the set of pertubative scales. 
For
 some choice of these scales, one can obtain an ideally flat plateau of $\chi^2$ for $\delta_T\leqslant 0.2$. However, the values of the two important thresholds, namely, $\delta_T=0.2$ (where deviation form TMD factorization appears) and $\delta_T=0.25$ (where the TMD factorization is completely broken), are stable with perturbative scales.

As a result of  these tests, in the following we use the data points with $q_T\lesssim 0.2\; Q$, or say $\delta_T=0.2$. The choice of $\delta_T$ that we make is consistent with~\cite{Becher:2011xn}. This range includes $163$ high-energy and $146$ low-energy data points (in total 309 data points). Comparing this number of points with the literature, we observe that, it is the largest set of points for  Drell-Yan/Z-boson production used up to present in a simultaneous fit of TMDPDF (to our knowledge), which also has the largest considered range of energies from $(Q,\sqrt{s})=(4,19.4)\;\GeV$ (from the E288 experiment) to $(Q,\sqrt{s})=(150,8000)\;\GeV$ (from the ATLAS experiment).


\subsection{Scale variations and theoretical uncertainties}
\label{sec:HEE_unser}

\begin{figure}[t]
\centering
\includegraphics[width=0.95\textwidth]{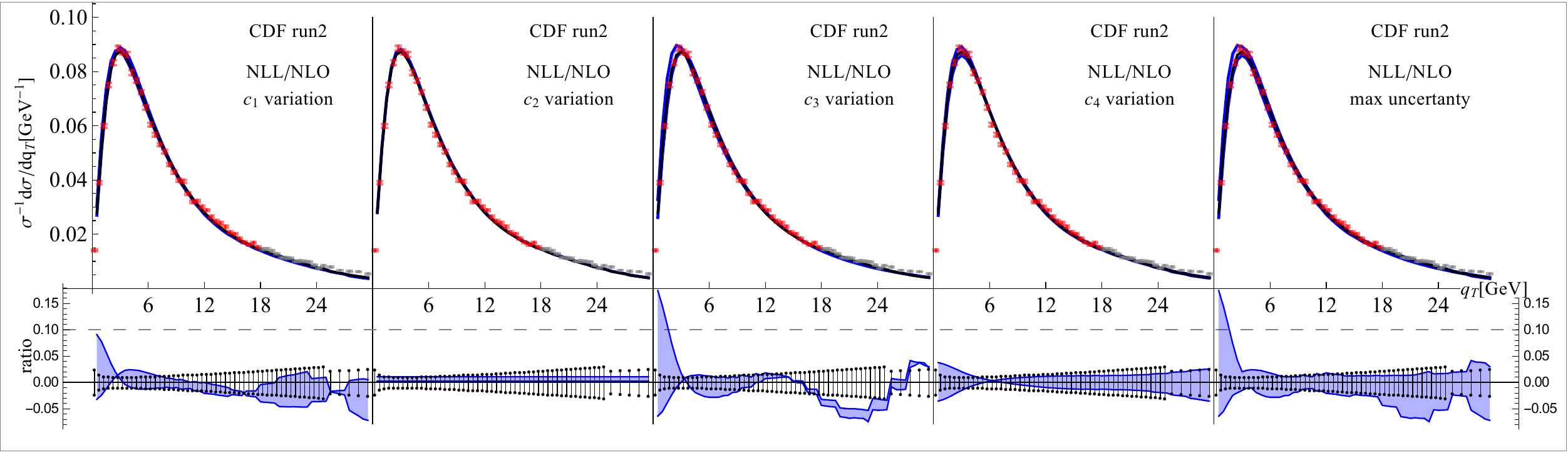}

\vspace{-3.5pt}
\includegraphics[width=0.95\textwidth]{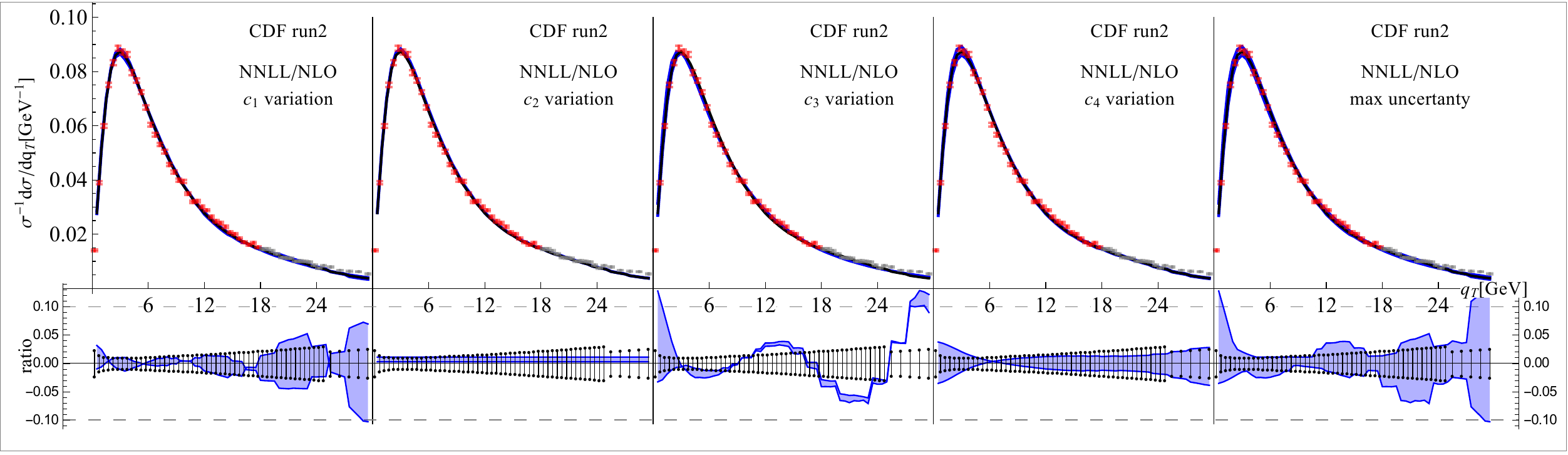}

\vspace{-3.5pt}
\includegraphics[width=0.95\textwidth]{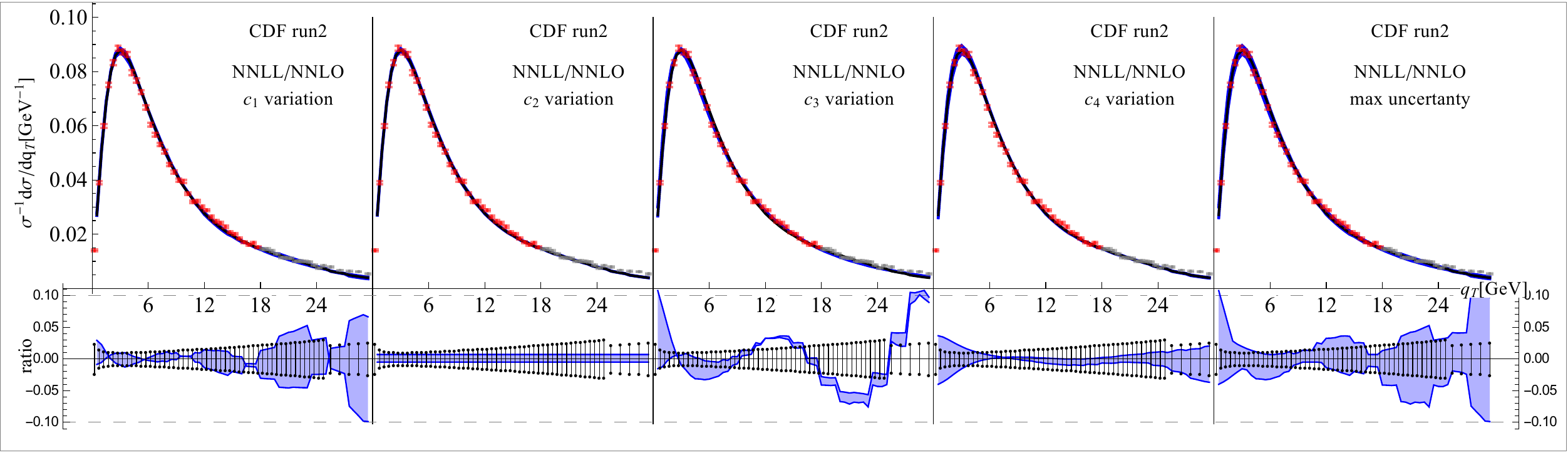}
\caption{\label{fig:scale_vary1} 
Theoretical error-bands and  experimental data points for CDF-Run 2  experiment.
The theoretical error is estimated changing  $c_{1,2,3,4}$ in the range $(0.5,2)$ at each perturbative order.
The nonpertubative input is provided by  \textit{model 2}.
The sub-panels show the relative size of error-band for theory and experiment.} 
\end{figure}

\begin{figure}[h]
\centering
\includegraphics[width=0.95\textwidth]{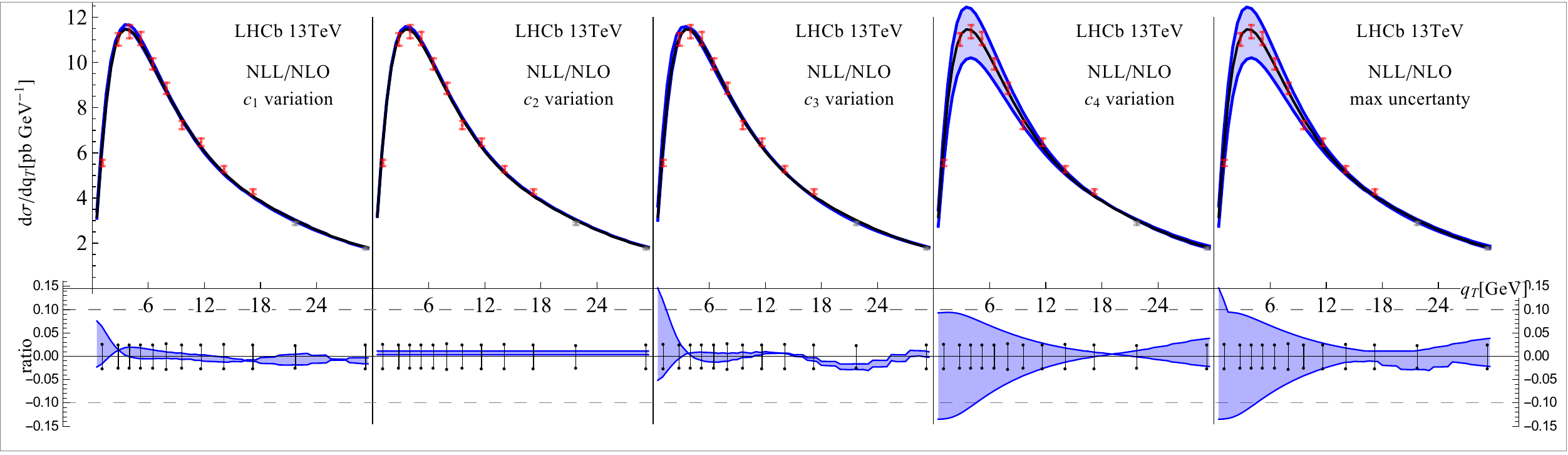}

\vspace{-4pt}
\includegraphics[width=0.95\textwidth]{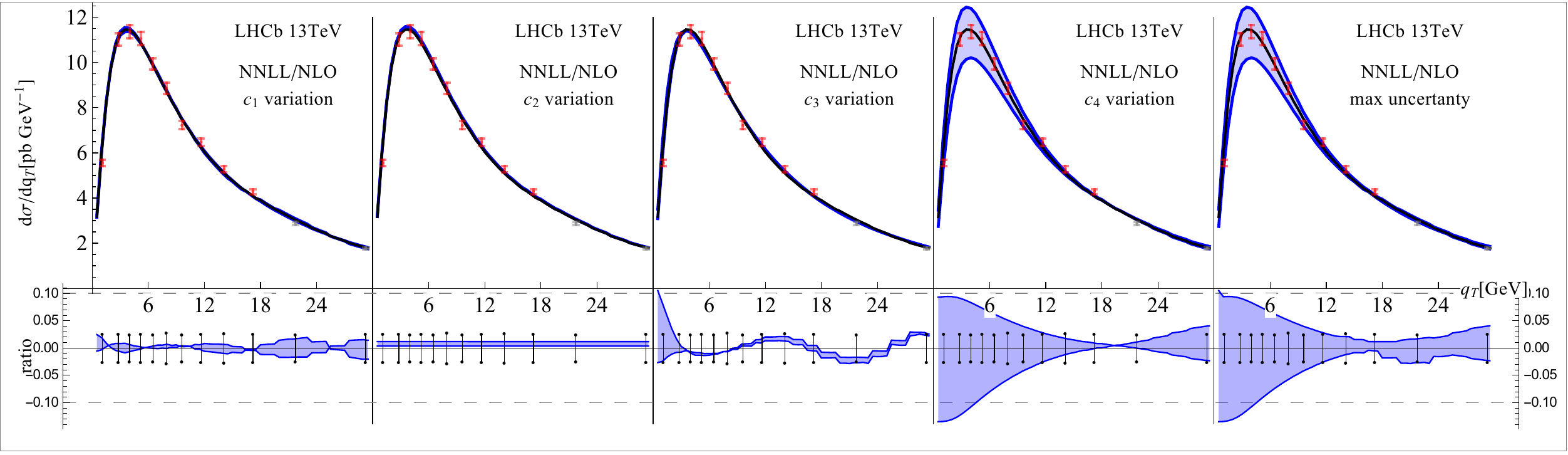}

\vspace{-4pt}
\includegraphics[width=0.95\textwidth]{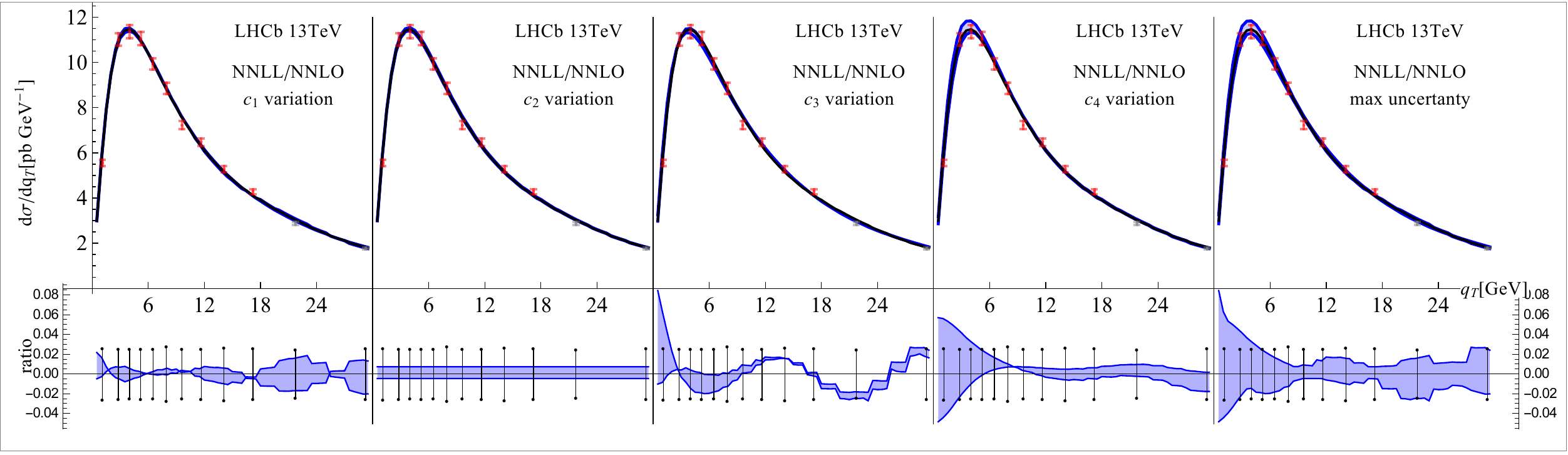}
\caption{\label{fig:scale_vary2}Theoretical error-bands and  experimental data points for LHCb  (13 TeV)  experiment at $13-14$ GeV.
The theoretical error is estimated changing  $c_{1,2,3,4}$ in the range $(0.5,2)$ at each perturbative order.
The nonpertubative input is provided by  \textit{model 2}.
 The sub-panels show the relative size of error-band for theory and experiment.} 
\end{figure}

\begin{figure}[t]
\centering

\includegraphics[width=0.95\textwidth]{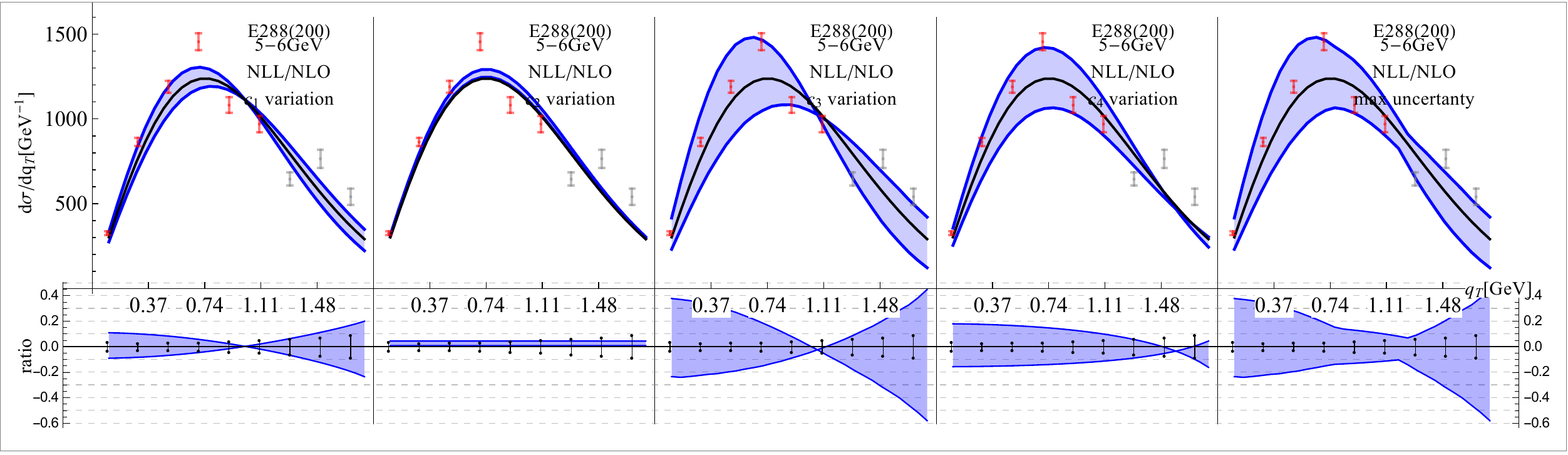}

\includegraphics[width=0.95\textwidth]{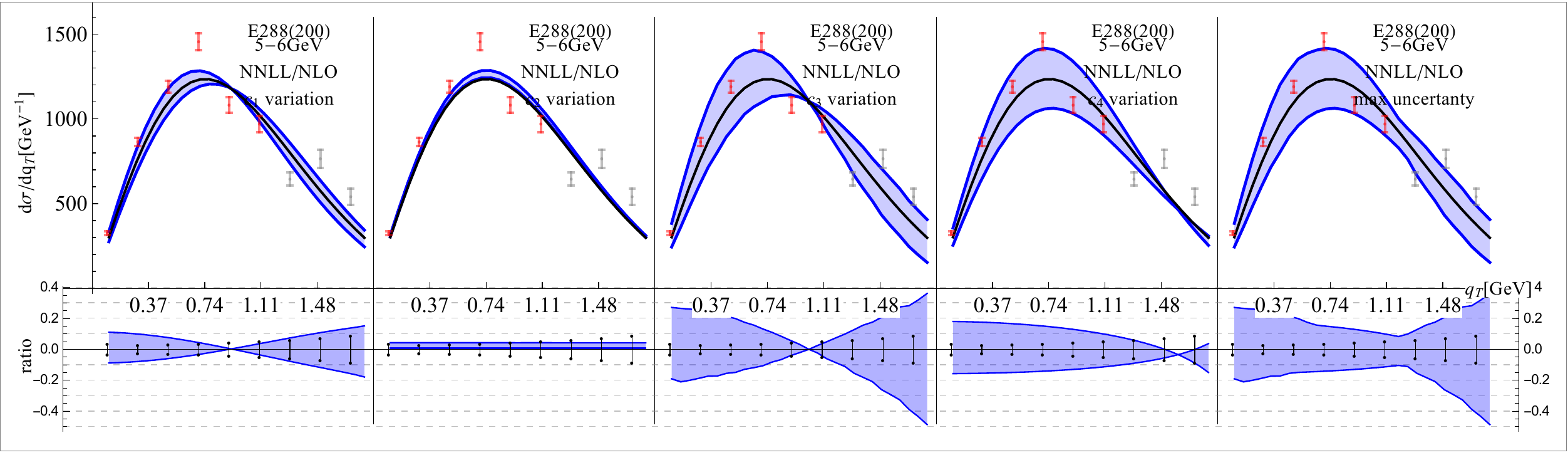}

\includegraphics[width=0.95\textwidth]{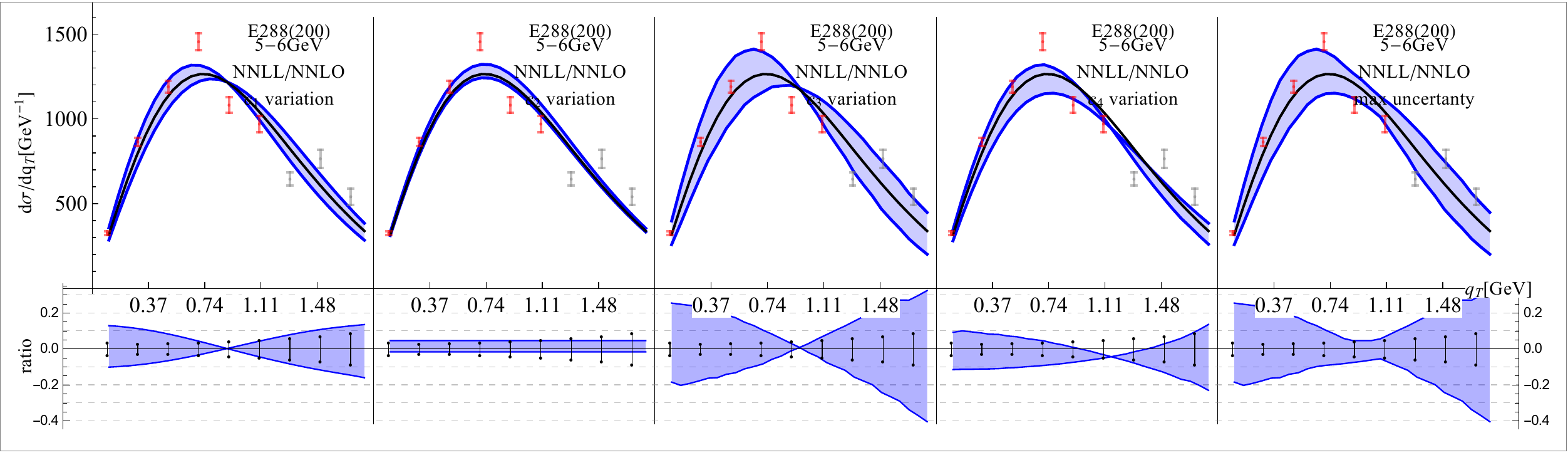}
\caption{\label{fig:scale_vary3}
Theoretical error-bands and  experimental data points for E288  (200 GeV)  experiment at 5-6 GeV.
The theoretical error is estimated changing  $c_{1,2,3,4}$ in the range $(0.5,2)$ at each perturbative order.
 The nonpertubative input is provided by  \textit{model 2}. The sub-panels show the ratio of deviation to the central line (with $c_i=1$).} 
\end{figure}

\begin{figure}[h]
\centering

\includegraphics[width=0.95\textwidth]{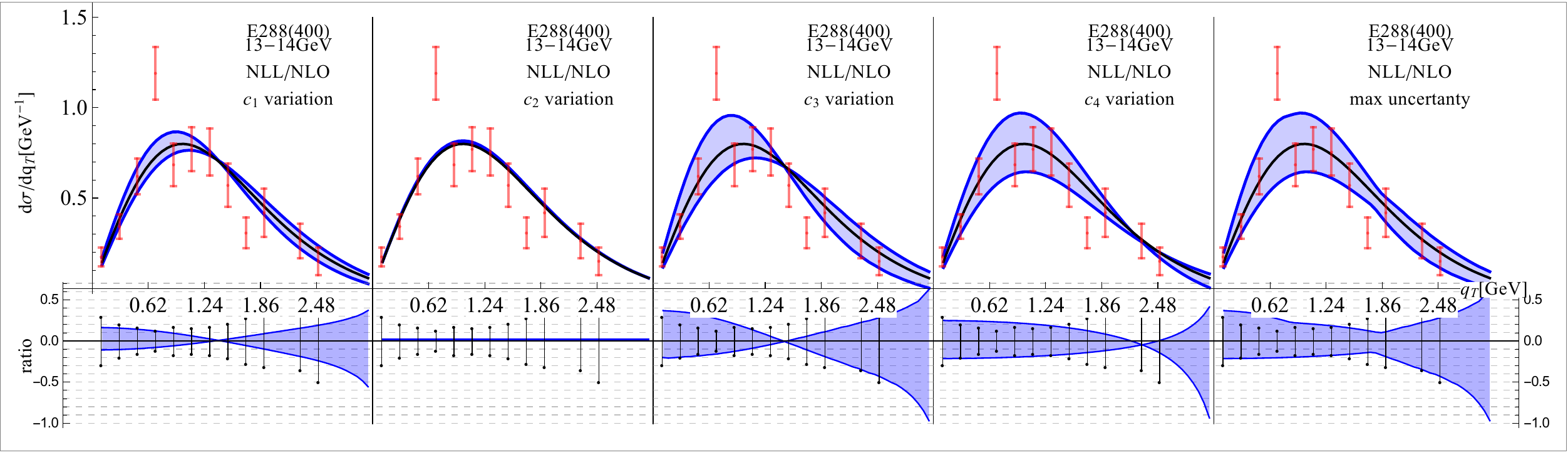}

\includegraphics[width=0.95\textwidth]{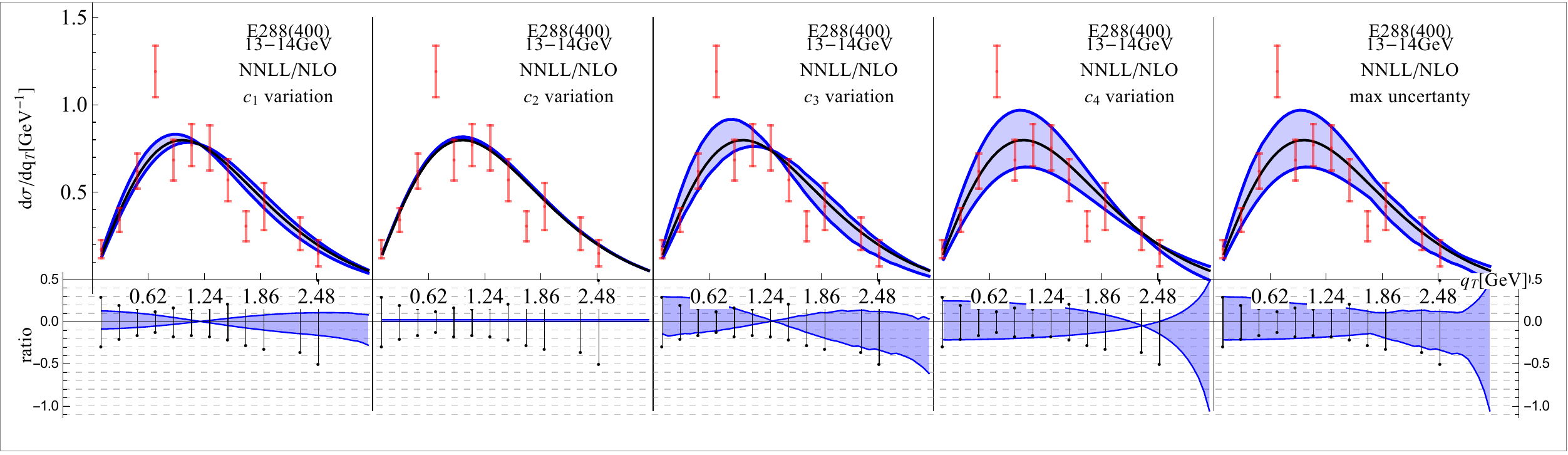}

\includegraphics[width=0.95\textwidth]{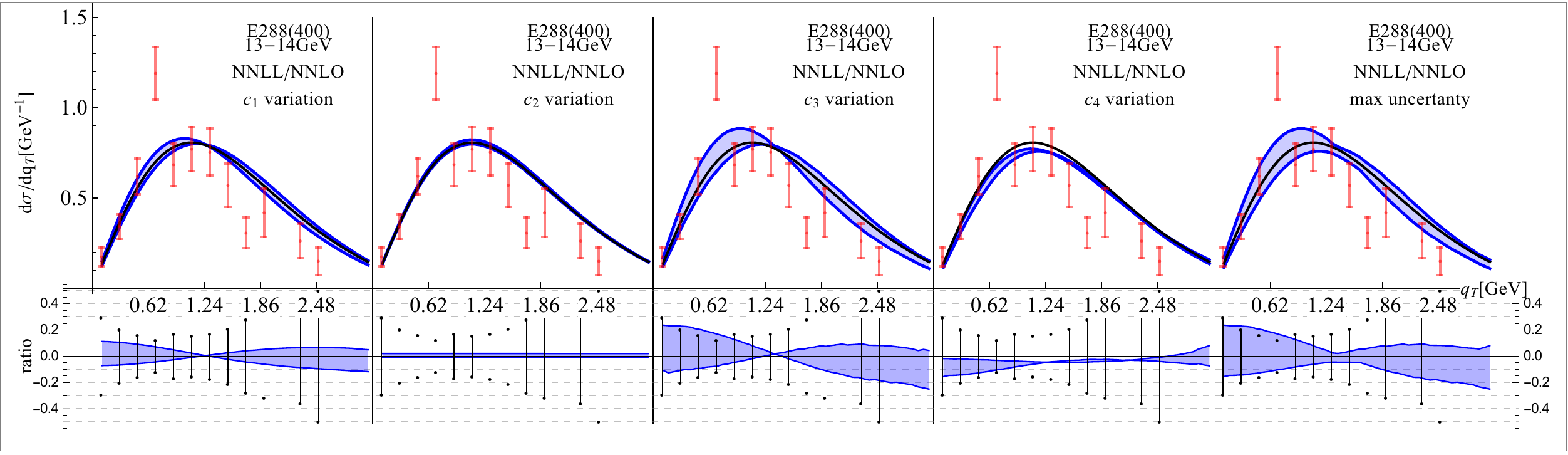}
\caption{\label{fig:scale_vary4}
Theoretical error-bands and  experimental data points for E288  (400 GeV)  experiment at 13-14 GeV.
The theoretical error is estimated changing  $c_{1,2,3,4}$ in the range $(0.5,2)$ at each perturbative order.
The nonpertubative input is provided by  \textit{model 2}.
The sub-panels show the ratio of deviation to the central line (with $c_i=1$).} 
\end{figure}

The theoretical uncertainties of the perturbative inputs are  tested by varying the perturbative scales around their central values, as it is discussed in  sec.~\ref{sec:scales}. The distribution of uncertainties through orders for a typical high energy experiment is shown in figs.~\ref{fig:scale_vary1} and~\ref{fig:scale_vary2}, and for a typical low-energy experiments in fig.~\ref{fig:scale_vary3}-\ref{fig:scale_vary4}. The complete set of plots for every included experiment can be found in \cite{web}.

The uncertainty associated with the TMD evolution factor is parameterized  by  the $c_1$-variation. This uncertainty drops down  between NLL/NLO and NNLL/NLO orders, that is together with the increase of the perturbative order for $\mathcal{D}$  (see table \ref{tab:orders}). The size of the band is correlated with the energy of the process, that is,  it is less significant for higher-energy experiments.

The uncertainty associated with the hard scale depends on the  $c_2$-variation. This band is independent on $q_T$. This error is the main one at NLL/LO (which we do not present here), but becomes negligible at higher orders.

The uncertainty associated with the low-energy behavior of the evolution factor is parameterized by the $c_3$-variation. We have found that it significantly influences the shape of the cross-section and also it is  rather large at small-$q_T$. As expected it is decreases going from NLL/NLO to NNLL/NNLO. At NNLL/NNLO it gives the main contribution to the uncertainty band for the cross-section.

The uncertainty associated with the small-$b$ matching of coefficients and PDFs is represented by the $c_4$-variation. It is the most interesting error because it checks the convergences of the $\zeta$-prescription. The corresponding error-band is larger at $q_T\to 0$, which corresponds to the contribution of large $\mathbf{L}_\mu$ (we remind that in $\zeta$-prescription, $\mathbf{L}_\mu$ grows unrestrictedly). The important observation is that the large uncertainty area significantly shrinks between NLL/NLO and NNLL/NNLO, although the NNLL/NNLO contains a higher power of $\mathbf{L}_\mu$. This shows a very good behavior of the $\zeta$-prescription. In total this error is dominant at NLL/NLO, but becomes smaller  (although compatible) to the one coming from the $c_3$ variation  at NNLL/NNLO.

The size of the theoretical error-band is significantly bigger  at small-$Q$, as can be visually checked comparing figs.~\ref{fig:scale_vary1}-\ref{fig:scale_vary2} to  figs.~\ref{fig:scale_vary3}-\ref{fig:scale_vary4}. The uncertainties reduces when one increases the perturbative orders, both in high and low energy cases. However  for  the low energy case the error remains of order  $\sim 20\%$ or higher even at NNLL/NNLO, which can be problematic for a precise description of  these experiments. We additionally stress that at NLL/LO the uncertainties range from $50\%$ to $100\%$ and higher. This shows that this particular order has no prediction power, and should not be considered any serious for a well based extraction of TMDs. This is the main reason for excluding NLL/LO order from our analysis.

In order to provide a  final definition of the theoretical error, we use all  scale variations and we take the maximum deviation among them. We have found that our definition of uncertainties is close, as far as one can compare different theoretical expressions, to the common definition used e.g. in \cite{Catani:2015vma,Bozzi:2008bb}.  In total, for the high-energy experiments we find that the theoretical uncertainty (at NNLO) is of the order $2-3$\% at the peak. It grows to $\sim 5-6\%$ at maximum allowed $q_T$, and to $\sim 10\%$ at $q_T\to 0$. This value seems to be  smaller (but comparable) to the typical values of uncertainties presented \texttt{ResBos} or \texttt{DYRes}. This is a definite positive point of the $\zeta$-prescription. Indeed, the main contribution (at high energies) to it comes from the $c_3$- and $c_4$-error-bands, which are controlled by $\zeta$-prescription. The $c_4$-band would be significantly larger in the presence of double-logarithms, which are absent due to the $\zeta$-prescription.

\subsection{Normalization}
\label{sec:LHC_norm}

\setlength{\tabcolsep}{4pt}
\begin{table}[t]\begin{center}
\begin{tabular}{|c||cc|cc|cc|ccc|}
order & \specialcellcenter{ATLAS \\ Z-boson\\7TeV} & \specialcellcenter{ATLAS\\Z-boson \\ 8TeV} & 
\specialcellcenter{ATLAS \\ 46-66 \\8TeV} & \specialcellcenter{ATLAS \\ 116-150 \\8TeV} & 
\specialcellcenter{CMS \\7TeV} & \specialcellcenter{CMS \\8TeV} & 
\specialcellcenter{LHCb \\7TeV} & \specialcellcenter{LHCb \\8TeV} & 
\specialcellcenter{LHCb \\13TeV}
\\\hline\hline
NLL/NLO & 438 pb & 0.92 & \bf{1.01} & \bf{0.93} & 369 pb & 407 pb & 0.92 & 0.93 & 0.93
\\\hline
NNLL/NLO & 438 pb & 0.92 & \bf{1.01} & \bf{0.93} & 368 pb & 407 pb & 0.92 & 0.93 & 0.93
\\\hline
NNLL/NNLO & 461 pb & \bf{0.97} &  \bf{1.08} & \bf{0.98} & 387 pb & 429 pb & \bf{0.97} & \bf{0.99} & \bf{0.98}
\end{tabular}
\end{center}
\caption{\label{tab:LHC_norm} The normalization factors for the cross-section for each experiment. The dimensional-less numbers are ratios of partially integrated cross section over $q_T$ (\ref{LHC_norm1})(theory/experiment, i.e. $\mathcal{N}^{-1}$), for the data with the published value of total cross-section. For the data sets with unpublished values of total cross-section, the value of the total cross-section used for normalization is presented. The numbers are given for the model 1. The variation of the scales and models gives the change of numbers in the unrepresented digits. The numbers shown in bold are those which agree with the measured cross-section within the error bars. 
}
\end{table}
\setlength{\tabcolsep}{6pt}

As the TMD factorization approach describes the shape of the  differential cross section only in a limited range of $q_T$, we need some extra input to normalize the curves. In order to compare with the data, we weight the differential cross-section by the total (or fiducial) cross-section. The values of the theory predictions for total cross-sections can be obtained from the studies of other groups. For example, one can use the DYNNLO code \cite{Catani:2007vq,Catani:2009sm}. Its predictions for the total cross-sections are presented in the tables \ref{tab:Run1}-\ref{tab:Run2}. However, we found that such a strategy is unreliable, because even tiny disagreement in the normalization leads to huge effects in the $\chi^2$-minimization. 
 This is especially important for LHC data sets, which have very small error-bands. Additionally, as we demonstrate later, the DYNNLO predictions are worse than that obtained using our normalization factors. 

Therefore, to fit the high energy data set we introduce a normalization factor for each data set. This factor equals the partial integral over $q_T$ for experimental and theoretical cross-sections, and reads
\begin{eqnarray}\label{LHC_norm1}
\mathcal{N}=\frac{\sum_{\substack{\text{included}\\{\text{bins}}}} \Delta q_T \frac{d\sigma_{\text{exp.}}}{dq_T}}{\sum_{\substack{\text{included}\\{\text{bins}}}} \Delta q_T \frac{d\sigma_{\text{th.}}}{dq_T}},
\end{eqnarray}
where $\Delta q_T$ is the size of the $q_T$ bin. In this way, we fit only the $q_T$-shape of cross-section, which is already very restricting, as we discussed in the previous section. 

The values of $\mathcal{N}^{-1}$ resulting from the calculations are presented in  table \ref{tab:LHC_norm}. It is clear that the deviation between the theory and experiment decreases with perturbative orders. For the majority of experiments (excluding the Z-boson production measured by ATLAS), we find a good agreement for the absolute value of the differential cross-section obtained from the data points and the TMD factorization. It is important that the values of $\mathcal{N}$ are very stable with respect to the change of non-perturbative model and to the scale variation. In particular, we do not present the error-band on the normalization values in the table \ref{tab:LHC_norm}, because they are smaller then the present precision. 

\renewcommand{\arraystretch}{1.1}
\begin{table}[t]
\begin{center}
\begin{tabular}{||c||c||c|c|c||}
Order & $~~\frac{\chi^2}{d.o.f.}~~$ & $~~~~~~~\lambda_1~~~~~~~$ & $~~~~~\lambda_2~~~~~$ 
\\\hline \hline
\multicolumn{4}{|c|}{Model 1}
\\\hline 
NLL/NLO &
$2.33~^{+2.76}_{-0.68}$ &
$0.321^{+0.008}_{-0.007}~^{+0.095}_{-0.100}$ &
$0.271^{+0.014}_{-0.013}~^{+0.155}_{-0.063}$ 
\\ \hline
NNLL/NLO &
$1.76~^{+1.25}_{-0.48}$ &
$0.289^{+0.004}_{-0.004}~^{+0.007}_{-0.121}$ &
$0.424^{+0.051}_{-0.045}~^{+0.673}_{-0.139}$
\\ \hline
NNLL/NNLO &
$1.34~^{+0.44}_{-0.20} $ &
$0.271^{+0.007}_{-0.006}~^{+0.076}_{-0.073}$ &
$0.277^{+0.015}_{-0.012} ~^{+0.081}_{-0.042}$ 
\\\hline \hline
\multicolumn{4}{|c|}{Model 2}
\\\hline 
NLL/NLO &
$2.19~_{-0.64}^{+2.34}$ &
$0.329^{+0.008}_{-0.008}~^{+0.047}_{-0.101}$ &
$0.289^{+0.019}_{-0.017}~^{+0.276}_{-0.008}$
\\ \hline
NNLL/NLO &
$1.65~^{+1.32}_{-0.39}$ &
$0.236^{+0.005}_{-0.004}~^{+0.070}_{-0.064}$ &
$0.440^{+0.049}_{-0.044}~^{+0.573}_{-0.126}$
\\ \hline
NNLL/NNLO &
$1.36~^{+0.35}_{-0.18}$ &
$0.284^{+0.007}_{-0.006}~^{+0.074}_{-0.079}$ &
$0.280^{+0.019}_{-0.017}~^{+0.086}_{-0.034}$ 
\end{tabular}
\end{center}
\caption{\label{tab:results} The results of the $\chi^2$-minimization procedure with $g_K=0$. The values of $\chi^2$ are given including the  theoretical error-band. The values of extracted parameters are given with statistical error-band (the first pair of numbers) and the theoretical error-band (the second pair of numbers). The visual presentation of this table is given in fig.\ref{fig:parameters_gK}.}
\end{table}

\begin{figure}[h]
\centering
\includegraphics[width=0.75\textwidth]{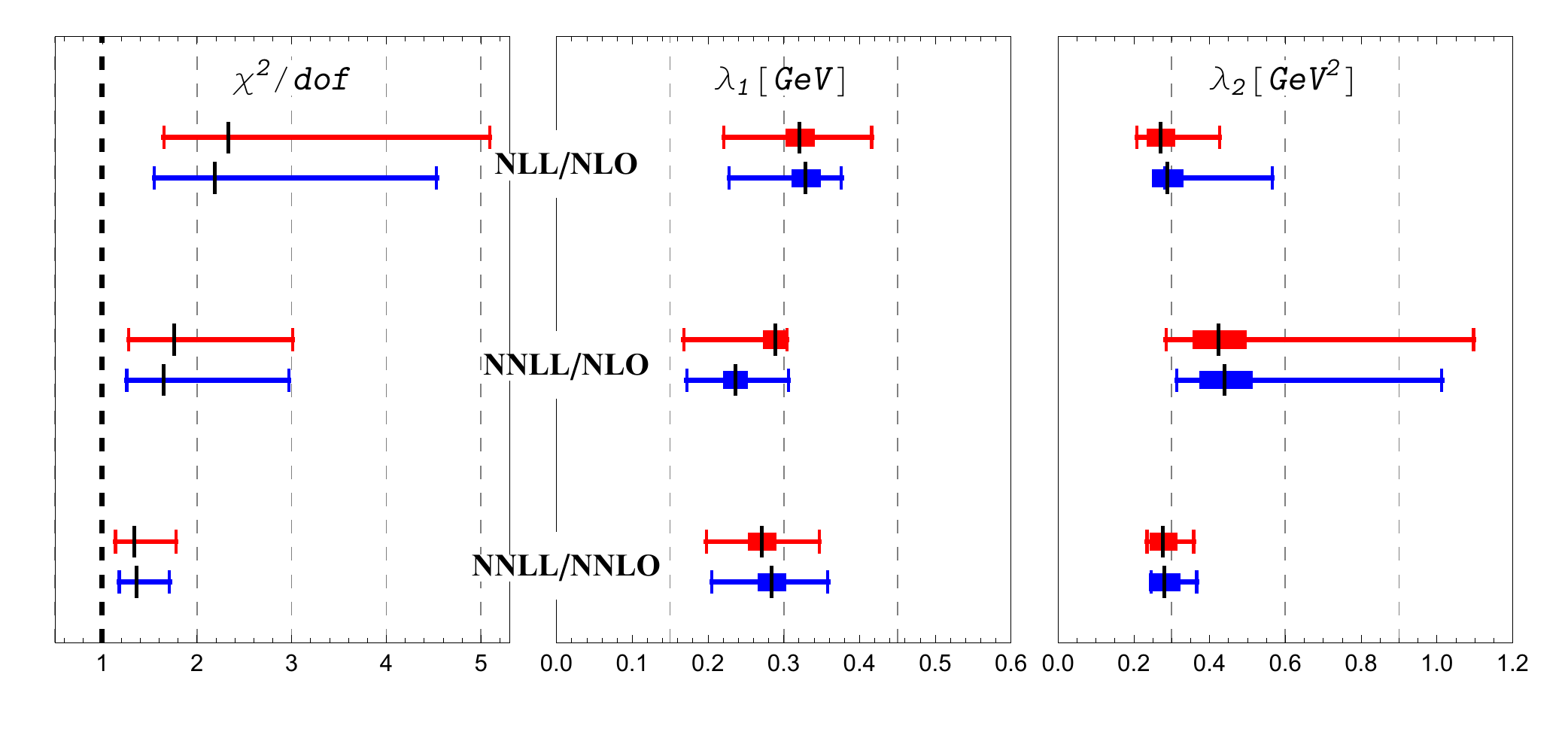}
\caption{\label{fig:parameters_nogK} The values of parameters for $f_{NP}$ extracted from the global fit with $g_K=0$. Red marks represent the extraction with model 1. Blue marks represent the extraction with model 2. The black marks show the values of parameters extracted at $c_{1,2,3,4}=1$. The thick bands represent the statistical errors of parameter determination. The thin error-bands represent the theoretical error on extracted parameters due to variation of $c_{1,2,3,4}\in[0.5,2]$. The numerical values of parameters are given in the table \ref{tab:results}.} 
\end{figure}

The normalization of the data from E288 experiment is generally unknown. Most probably, the main source of discrepancy comes from the fiducial cuts made for E288 experiment, which cannot be restored nowadays. The small fiducial cuts do not seriously influence the $q_T$-shape of the differential cross-section, but can sizably decrease the total normalization. In our analysis, we change the common normalization of all E288 data points as
\begin{eqnarray}
\mathcal{N}_{E288}=0.8.
\end{eqnarray}
This or close values have been used in different fits, see e.g. \cite{Landry:2002ix,DAlesio:2014mrz}. However, we do not seriously ground on it, e.g. we can switch to $0.85$ or $0.9$ without significant loss in $\chi^2$ (however, the value $1$ produces serious disagreement with our current input). One should take into account that the theoretical uncertainty at small$-Q$ is very large, see figures \ref{fig:scale_vary3} and \ref{fig:scale_vary4}. It also implies that low-energy cross-sections are very sensitive to the choice of PDF set (in particular, our approximation of eq.~(\ref{pdf:cuprum}) for nuclei PDF could be too crude). We have checked that the E288 data can be also fitted with $N_{E288}=1$ to the same values (or better) of $\chi^2$ by additional variation of $Q_0$ (similar to the fit made in \cite{Guzzi:2013aja}). Such an ambiguity represents a problem in the analysis of the low-energy data. 
 
\subsection{Results of the fits and TMD extraction}
\label{sec:global}

\begin{figure}[t]
\centering
\includegraphics[width=0.95\textwidth]{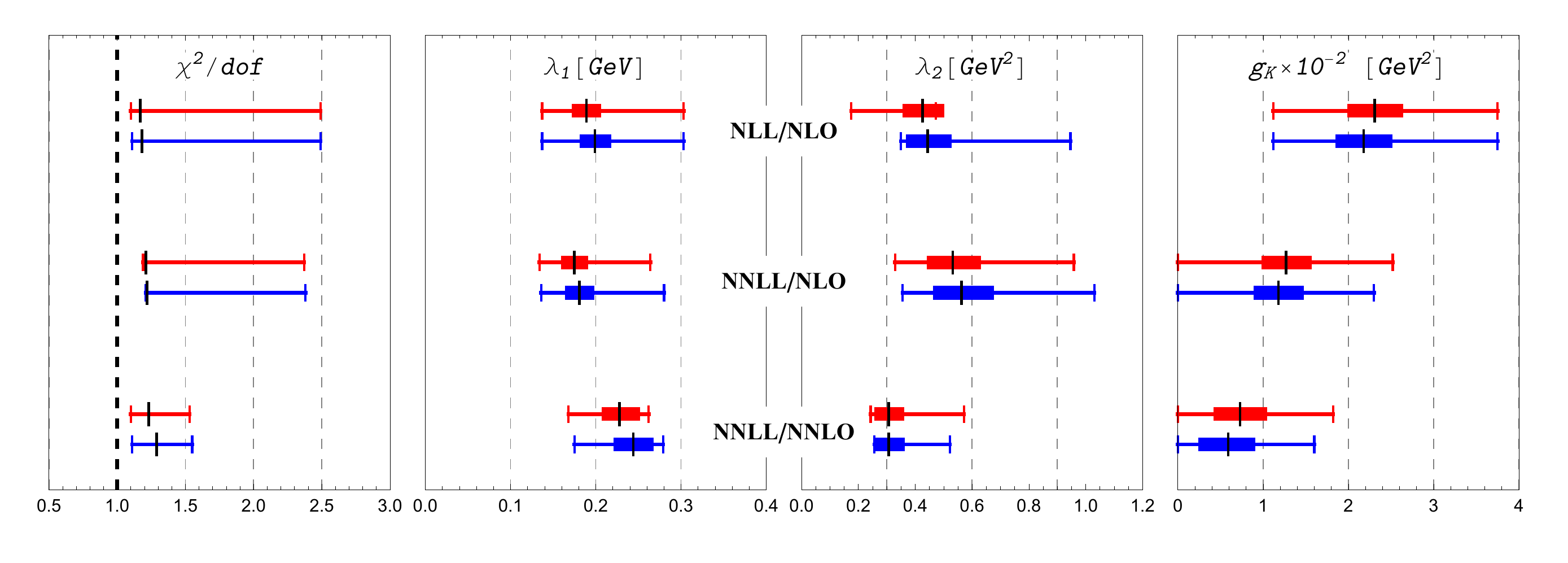}
\caption{\label{fig:parameters_gK}The values of parameters for $f_{NP}$ extracted from the global fit with $g_K\neq 0$. Red marks represent the extraction with model 1. Blue marks represent the extraction with model 2. The black marks show the values of parameters extracted at $c_{1,2,3,4}=1$. The thick bands represent the statistical errors of parameter determination. The thin error-bands represent the theoretical error on extracted parameters due to variation of $c_{1,2,3,4}\in[0.5,2]$. The numerical values of parameters are given in the table \ref{tab:results_gK}.} 
\end{figure}

In this section, we present the results of the global fit for the complete data sets presented in sec.~\ref{sec:datareview}, which allows the extraction of the unpolarized TMDPDF. We have made two independent fits, with $g_K=0$ and with $g_K\neq 0$. The results of the $\chi^2$ minimization  and the values of the extracted parameters are presented in tab.~\ref{tab:results} and \ref{tab:results_gK}. The visual presentation is given in fig.~\ref{fig:parameters_nogK} and~\ref{fig:parameters_gK}.

\begin{table}[t]
\begin{center}
\begin{tabular}{||c||c||c|c|c||}
Order & $~~\frac{\chi^2}{d.o.f.}~~$ & $~~~~~~~\lambda_1~~~~~~~$ & $~~~~~\lambda_2~~~~~$  
& $~~~~~g_K~\times 10^{-2}~~~$ 
\\\hline \hline
\multicolumn{5}{|c|}{Model 1}
\\\hline 
NLL/NLO &
$1.17~^{+1.32}_{-0.07} $ &
$0.189^{+0.009}_{-0.009}~^{+0.114}_{-0.052}$ &
$0.425^{+0.054}_{-0.045}~^{+0.047}_{-0.250}$ &
$2.31^{+0.25}_{-0.24}~^{+1.44}_{-1.19}$
\\ \hline
NNLL/NLO &
$1.21~^{+1.16}_{-0.02}$ &
$0.175^{+0.008}_{-0.008}~^{+0.089}_{-0.041}$ &
$0.532^{+0.076}_{-0.067}~^{+0.426}_{-0.203}$ &
$1.27^{+0.22}_{-0.21}~^{+1.19}_{-1.27}$
\\ \hline
NNLL/NNLO &
$1.23~^{+0.30}_{-0.13} $ &
$0.228^{+0.016}_{-0.013}~^{+0.034}_{-0.060}$ &
$0.306^{+0.031}_{-0.026}~^{+0.265}_{-0.063}$ &
$0.73^{+0.24}_{-0.23}~_{-0.73}^{+1.09}$ 
\\\hline \hline
\multicolumn{5}{|c|}{Model 2}
\\\hline 
NLL/NLO &
$1.18~^{+1.31}_{-0.07}$ &
$0.199^{+0.011}_{-0.010}~^{+0.104}_{-0.062}$ &
$0.443^{+0.061}_{-0.052}~^{+0.503}_{-0.093}$&
$2.18^{+0.26}_{-0.25}~^{+1.57}_{-1.06}$
\\ \hline
NNLL/NLO &
$1.22~^{+1.16}_{-0.01}$ &
$0.181^{+0.009}_{-0.009}~^{+0.099}_{-0.045}$ &
$0.562^{+0.092}_{-0.075}~^{+0.468}_{-0.206}$ &
$1.18^{+0.22}_{-0.21}~^{+1.12}_{-1.18}$
\\ \hline
NNLL/NNLO &
$1.29 ~^{+0.26}_{-0.18}$ &
$0.244^{+0.016}_{-0.015}~^{+0.035}_{-0.069}$ &
$0.306^{+0.034}_{-0.029}~^{+0.216}_{-0.050}$ &
$0.59^{+0.24}_{-0.27}~^{+1.01}_{-0.59}$
\end{tabular}
\end{center}
\caption{\label{tab:results_gK} The results of the $\chi^2$-minimization procedure with non-zero $g_K$. The values of $\chi^2$ are given with theoretical error-band. The values of extracted parameters are given with statistical error-band (the first pair of numbers) and the theoretical error-band (the second pair of numbers). The visual presentation of this table is given in fig.\ref{fig:parameters_gK}.}
\end{table}

We have estimated both statistical and theoretical errors on the fit parameters. The statistical errors are related to the uncertainty of the $\chi^2$-minimization and are induced by the experimental error-bands. The statistical errors have been estimated by the MINOS procedure of MINUIT package \cite{James:1994vla}. The theoretical errors are related to the uncertainty of perturbation series.  There is no common procedure for the estimation of the theoretical error. Therefore, we propose the method presented in the following. 

The theoretical error is estimated by a set of independent fitting procedures for each variation of the scale constants $c_{1,2,3,4}\in [0.5,2]$, as discussed in sec.~\ref{sec:th_uncertain}. In other words, we set, say, $c_1=2$ and perform the minimization of $\chi^2$. In this way, we obtain a new set of model parameters (and a new value of $\chi^2$). In total, we have 8 independent variations and hence have 8 values of parameters. The final theoretical error-band is given by the maximal positive and minimal deviations from the central value and the results are reported in tab.~\ref{tab:parameter_errors}. A drawback of this procedure is the variation of a scale can lead to the serious increase in $\chi^2$. In other words changing the matching scales affects also the quality of the fit. In general, the size of the band for $\chi^2$ value represents the stability of the theoretical model, and they are also reported in tab.~\ref{tab:parameter_errors}. One can see that the error for $\chi^2$ significantly drops with orders.

\renewcommand{\arraystretch}{0.9}
\begin{table}[t]
\begin{center}
\begin{tabular}{||l||c||c|c|c||}
Variation & $~~\frac{\chi^2}{d.o.f.}~~$ & $~~~~~~~\lambda_1~~~~~~~$ & $~~~~~\lambda_2~~~~~$  
& $~~~~~g_K~\times 10^{-2}~~~$ 
\\\hline \hline
\multicolumn{5}{|c|}{Model 1 NNLL/NLO}
\\\hline 
$c_{1,2,3,4}=1$ & $1.17 $ & $0.189$ & $0.425$ & $2.31$
\\ \hline
$c_{1}=2$ 		& $1.31~(\mathbf{+0.14}) $ & $0.201 ~(\mathbf{+0.012})$ & $0.316 ~(\mathbf{-0.109})$ & $3.00 ~(\mathbf{+0.69})$
\\ \hline
$c_{1}=0.5$ 	& $1.10~(\mathbf{-0.07}) $ & $0.184 ~(\mathbf{-0.005})$ & $0.308 ~(\mathbf{-0.117})$ & $1.60 ~(\mathbf{-0.71})$
\\ \hline
$c_{2}=2$ 		& $1.19~(\mathbf{+0.02}) $ & $0.204 ~(\mathbf{+0.015})$ & $0.223 ~(\mathbf{-0.202})$ & $2.12 ~(\mathbf{-0.19})$
\\ \hline
$c_{2}=0.5$ 	& $1.20~(\mathbf{+0.03}) $ & $0.219 ~(\mathbf{+0.030})$ & $0.226 ~(\mathbf{-0.199})$ & $1.93 ~(\mathbf{-0.38})$
\\ \hline
$c_{3}=2$ 		& $1.23~(\mathbf{+0.06}) $ & $0.251 ~(\mathbf{+0.062})$ & $0.315 ~(\mathbf{-0.110})$ & $3.75 ~(\mathbf{+1.44})$
\\ \hline
$c_{3}=0.5$ 	& $1.13~(\mathbf{-0.04}) $ & $0.160 ~(\mathbf{-0.029})$ & $0.220 ~(\mathbf{-0.205})$ & $1.12 ~(\mathbf{-1.19})$
\\ \hline
$c_{4}=2$ 		& $1.76~(\mathbf{+0.59}) $ & $0.137 ~(\mathbf{-0.052})$ & $0.473 ~(\mathbf{+0.046})$ & $2.71 ~(\mathbf{+0.40})$
\\ \hline
$c_{4}=0.5$ 	& $2.49~(\mathbf{+1.32}) $ & $0.303 ~(\mathbf{+0.114})$ & $0.175 ~(\mathbf{-0.250})$ & $1.15 ~(\mathbf{-1.16})$
\\ \hline & &&& \\[-10pt] 
Result			& $1.17^{+1.32}_{-0.07}$ & $0.189^{+0.114}_{-0.052}$ & $0.425^{+0.047}_{-0.250}$ & $2.31^{+1.44}_{-1.19}$
\\[2pt]\hline \hline
\multicolumn{5}{|c|}{Model 1 N$^3$LL/NNLO}
\\\hline 
$c_{1,2,3,4}=1$ & $1.23 $ & $0.228$ & $0.306$ & $0.73$
\\ \hline
$c_{1}=2$ 		& $1.40~(\mathbf{+0.17}) $ & $0.242 ~(\mathbf{+0.014})$ & $0.296 ~(\mathbf{-0.010})$ & $1.21 ~(\mathbf{+0.48})$
\\ \hline
$c_{1}=0.5$ 	& $1.14~(\mathbf{-0.09}) $ & $0.221 ~(\mathbf{-0.007})$ & $0.346 ~(\mathbf{+0.020})$ & $0.12 ~(\mathbf{-0.61})$
\\ \hline
$c_{2}=2$ 		& $1.22~(\mathbf{-0.01}) $ & $0.217 ~(\mathbf{-0.011})$ & $0.295 ~(\mathbf{-0.011})$ & $0.86 ~(\mathbf{+0.13})$
\\ \hline
$c_{2}=0.5$ 	& $1.26~(\mathbf{+0.03}) $ & $0.252 ~(\mathbf{+0.024})$ & $0.326 ~(\mathbf{+0.020})$ & $0.48 ~(\mathbf{-0.25})$
\\ \hline
$c_{3}=2$ 		& $1.27~(\mathbf{+0.04}) $ & $0.260 ~(\mathbf{+0.032})$ & $0.344 ~(\mathbf{+0.038})$ & $1.82 ~(\mathbf{+1.09})$
\\ \hline
$c_{3}=0.5$ 	& $1.31~(\mathbf{+0.08}) $ & $0.198 ~(\mathbf{-0.030})$ & $0.358 ~(\mathbf{+0.052})$ & $0.00 ~(\mathbf{-0.73})$
\\ \hline
$c_{4}=2$ 		& $1.10~(\mathbf{-0.13}) $ & $0.168 ~(\mathbf{-0.060})$ & $0.571 ~(\mathbf{+0.265})$ & $1.27 ~(\mathbf{+0.54})$
\\ \hline
$c_{4}=0.5$ 	& $1.53~(\mathbf{+0.30}) $ & $0.262 ~(\mathbf{+0.034})$ & $0.243 ~(\mathbf{-0.063})$ & $0.68 ~(\mathbf{-0.05})$
\\ \hline & &&& \\[-10pt] 
Result			& $1.23^{+0.30}_{-0.13}$ & $0.228^{+0.034}_{-0.060}$ & $0.306^{+0.265}_{-0.063}$ & $0.73^{+1.09}_{-0.73}$
\\[2pt]\hline \hline
\end{tabular}
\end{center}
\caption{\label{tab:parameter_errors} Example of parameter extraction with the variation of $c_{1,2,3,4}$ constants, and evaluation of the theoretical error. Bold numbers in brackets represent the deviation of the parameter from its central value.}
\end{table}

The values of the parameter $\lambda_1$, which parametrizes the asymptotics of TMDPDFs, extracted at different orders agree with each other within the error-band, that slightly reduces with the increase of the order. It has a natural size of the order of pion mass, $\lambda_1\sim 1.3 m_\pi - 2.3 m_\pi$. The values of the parameter $\lambda_2$, which parameterizes the quadratic correction to the small-$b$ regime, are not so stable although the values at different orders are compatible within the errors. In particular, they have large error-bars at NNLL/NLO order. The behavior of $g_K$ is the most peculiar. It decreases with the increase of the perturbation order. Moreover, at NNLL (both /NLO and /NNLO) its error-band touches the zero. It can be interpreted as following: the parameter $g_K$ is very small (or even zero) but within the fit, it tends to compensate the missing higher perturbative orders of evolution exponent. We also observe that all extractions of $g_K$ agrees with the theoretical estimation $g_K=0.01\pm 0.03\;\GeV^2$ made in \cite{Scimemi:2016ffw}. One can see that both models produce very similar results both for $\chi^2$ and the  parameters. 

\setlength{\tabcolsep}{3pt}
\begin{table}[t]
\begin{center}
\begin{tabular}{||c||c||c|c|c||c|c|c||}
\hline
Data set & point &\multicolumn{3}{|c||}{Model 1} &\multicolumn{3}{|c||}{Model 2}

\\ & & \specialcell{NLL/ \\ NLO} & \specialcell{NNLL/ \\ NLO} & \specialcell{NNLL/ \\ NNLO}   
& \specialcell{NLL/ \\ NLO} & \specialcell{NNLL/ \\ NLO} & \specialcell{NNLL/ \\ NNLO}
\\\hline\hline
CDF run1 & 30 					& 0.67& 0.68& 0.64& 	0.67& 0.67& 0.64
\\
D0 run1 & 14 					& 0.50& 0.52& 0.60&		0.49& 0.51& 0.62
\\
CDF run2 &36 					& 1.22& 1.36& 1.30&		1.17& 1.29& 1.33
\\
D0 run2 & 7 					& 2.52& 2.69& 2.75&		2.45& 2.64& 2.79
\\
ATLAS (7TeV) Z-boson  &9 		& 1.54& 1.55& 2.01&		1.60& 1.59& 2.27
\\
ATLAS (8TeV) Z-boson  & 9 		& 2.32& 2.48& 2.69& 	2.46& 2.70& 2.79
\\
ATLAS\,(8TeV)\,46-66\,GeV &5 	& 0.04& 0.05& 0.16&		0.05& 0.04& 0.20
\\
ATLAS\,(8TeV)\,116-150\,GeV &9	& 0.30& 0.35& 0.31&		0.30& 0.36& 0.30
\\
CMS (7 TeV) & 7					& 1.38& 1.39& 1.36&		1.38& 1.38& 1.36
\\
CMS (8 TeV) & 7					& 1.38& 1.38& 1.54&		1.38& 1.37& 1.58
\\
LHCb (7 TeV) & 10				& 0.26& 0.26& 0.31&		0.25& 0.26& 0.33
\\
LHCb (8 TeV) & 10				& 0.11& 0.12& 0.27&		0.11& 0.12& 0.32
\\
LHCb (13 TeV) &10				& 0.50& 0.50& 0.28&		0.50& 0.50& 0.27
\\\hline
High energy data & 163			& 0.95& 1.00& 0.94&		0.94& 1.00& 1.04
\\\hline\hline
E288(200) 4-5 GeV &5			& 3.86& 4.28& 3.86&		4.25& 4.59& 4.30
\\
E288(200) 5-6 GeV &6			& 3.00& 3.03& 1.92&		3.05& 3.07& 1.92
\\
E288(200) 6-7 GeV &7			& 1.68& 1.68& 0.84&		1.66& 1.67& 0.79
\\
E288(200) 7-8 GeV &8			& 1.10& 1.10& 0.93&		1.13& 1.11& 1.00
\\
E288(200) 8-9 GeV &9			& 1.83& 1.84& 0.78&		1.89& 1.87& 1.87
\\
E288(300) 4-5 GeV &5			& 1.93& 2.20& 4.09&		2.24& 2.44& 4.90
\\
E288(300) 5-6 GeV &6			& 1.15& 1.18& 1.15&		1.19& 1.21& 1.21
\\
E288(300) 6-7 GeV &7			& 0.84& 0.83& 0.66&		0.85& 0.83& 0.69
\\
E288(300) 7-8 GeV &8			& 1.18& 1.17& 0.90&		1.16& 1.17& 0.86
\\
E288(300) 8-9 GeV &9			& 1.13& 1.14& 1.13&		1.11& 1.36& 1.10
\\
E288(300) 11-12 GeV &12			& 1.08& 1.08& 1.00&		1.11& 1.10& 1.04
\\
E288(400) 5-6 GeV &6			& 2.11& 2.04& 1.12&		1.94& 1.92& 1.01
\\
E288(400) 6-7 GeV &7			& 2.59& 2.68& 2.55&		2.59& 2.64& 2.55
\\
E288(400) 7-8 GeV &8			& 0.83& 0.97& 2.02&		0.99& 1.07& 2.44
\\
E288(400) 8-9 GeV &9			& 1.36& 1.31& 1.37&		1.37& 1.32& 1.54
\\
E288(400) 11-12 GeV &12			& 1.08& 1.06& 1.25&		1.05& 1.05& 1.17
\\
E288(400) 12-13 GeV &12			& 0.88& 0.88& 1.10&		0.87& 0.88& 1.14
\\
E288(400) 13-14 GeV &12			& 0.39& 0.38& 0.72&		0.39& 0.39& 0.71
\\\hline
Low energy data &146 			& 1.38& 1.41& 1.35&		1.50& 1.48& 1.49
\\\hline\hline
Total & 309 					& 1.17& 1.21& 1.23&		1.18& 1.22& 1.29
\\\hline\hline
\end{tabular}
\end{center}
\caption{\label{tab:partial_chi} The values of $\chi^2/points$ for individual data sets. The boxes indicate the values of partial $\chi^2$ which are responsible for the increment of $\chi^2/d.o.f.$ from NLL/NLO to NNLL/NNLO.}
\end{table}
\setlength{\tabcolsep}{6pt}

As expected the theoretical error is reduced with the increase of the perturbative order. In particular, the band on the value of $\chi^2$ is significantly smaller at NNLL/NNLO. The distribution of parameter values over perturbative orders presented in table \ref{tab:parameter_errors} is typical. The variation of $c_1$ does not represent the main contribution to the error-band. It implies that the low-energy matching for the rapidity anomalous dimension is not so important (in comparison to other matchings), as  typically expected. 

The variation of $c_2$ is almost negligible. Here, however, we recall that $c_2$ influences only the common normalization factor, and thus the effect of its variation could be underestimated due to our fitting procedure. The variation of $c_3$ and $c_4$ produces the most part of the error-band and the strongest variation of $\chi^2$. At $g_K=0$ these variation are more-or-less equivalent. At $g_K\neq0$ there is a clear pattern. In this case, the variation of $c_3$ gives the main error-band on $g_K$, while the variation of $c_4$ gives the main error-band on parameters $\lambda_{1,2}$. It is very natural since the variation of $c_3$ tests the low-energy normalization point of the evolution factor, and $c_4$ tests the uncertainties of perturbation determination of the TMDPDF.

\begin{figure}[t]
\centering
\includegraphics[width=0.6\textwidth]{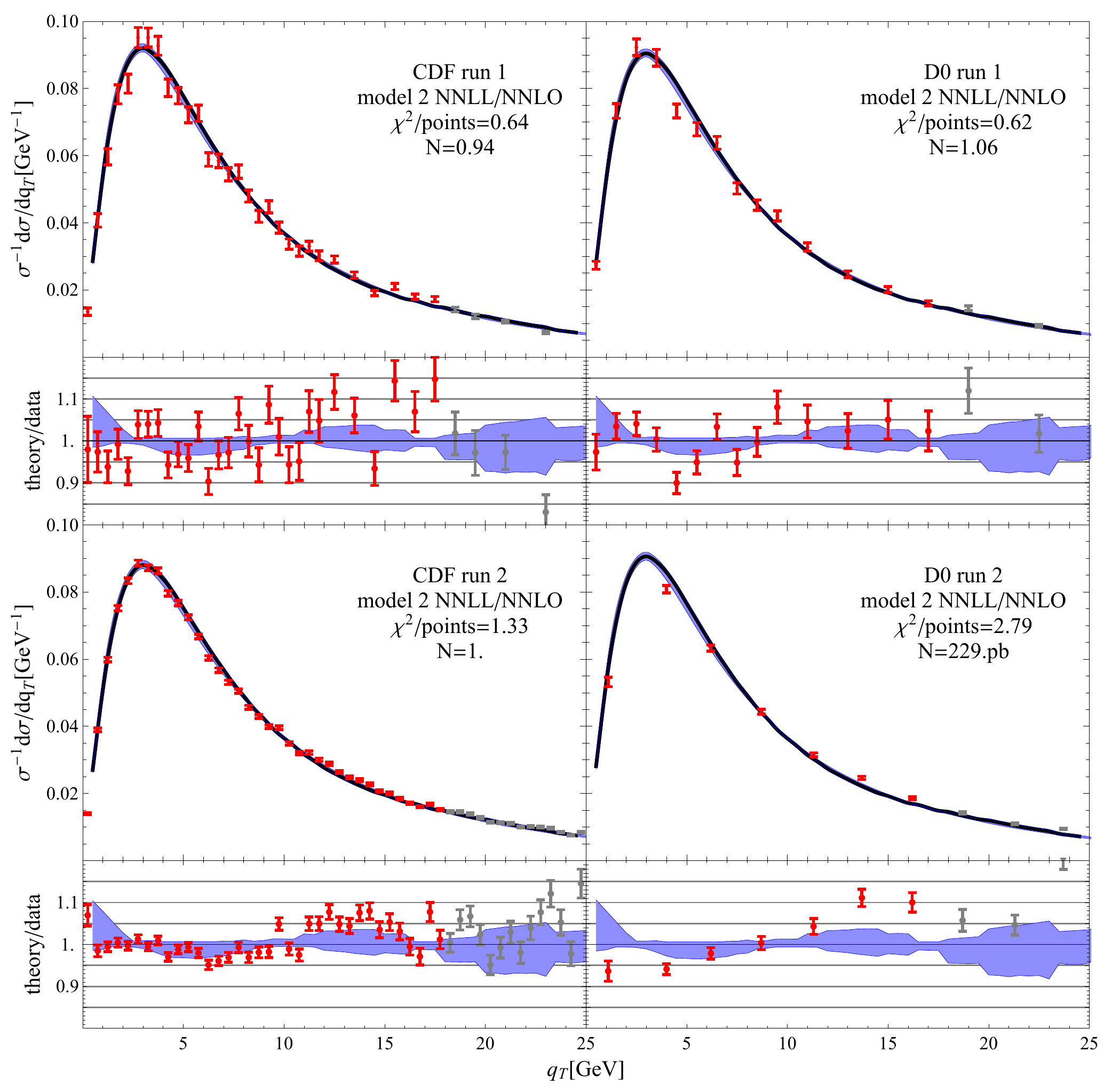}
\caption{\label{fig:Tevatron_M2_NNLO}The comparison of the data for Z-boson production collected at Tevatron experiments (run 1 and run 2) to the fit of \textit{model 2} at NNLL/NNLO. Red data points are those which included in the fit (i.e. with $\delta_T=0.2$). Gray data points are those which are not include in the fit (i.e. $\delta_T>0.2$). The blue band is the theoretical uncertainty obtained from the variation of scales.} 
\end{figure}

\begin{figure}[t]
\centering
\includegraphics[width=0.6\textwidth]{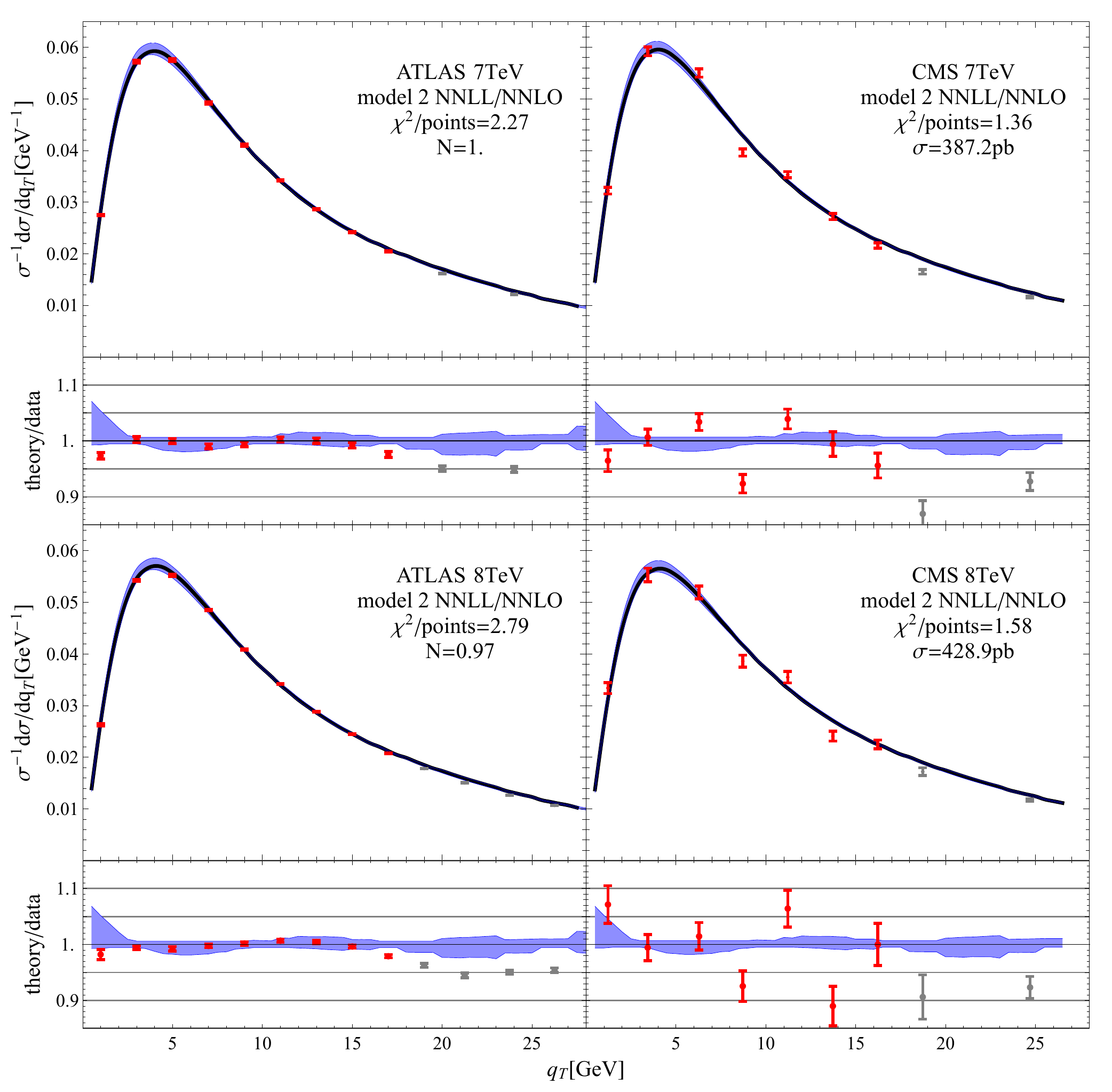}
\caption{\label{fig:ATLAS+CMS_M2_NNLO}The comparison of the data for Z-boson production collected at ATLAS and CMS experiments to the fit of \textit{model 2} at NNLL/NNLO. Red data points are those which included in the fit (i.e. with $\delta_T=0.2$). Gray data points are those which are not include in the fit (i.e. $\delta_T>0.2$). The blue band is the theoretical uncertainty obtained from the variation of scales.} 
\end{figure}

\begin{figure}[t]
\centering
\includegraphics[width=0.9\textwidth]{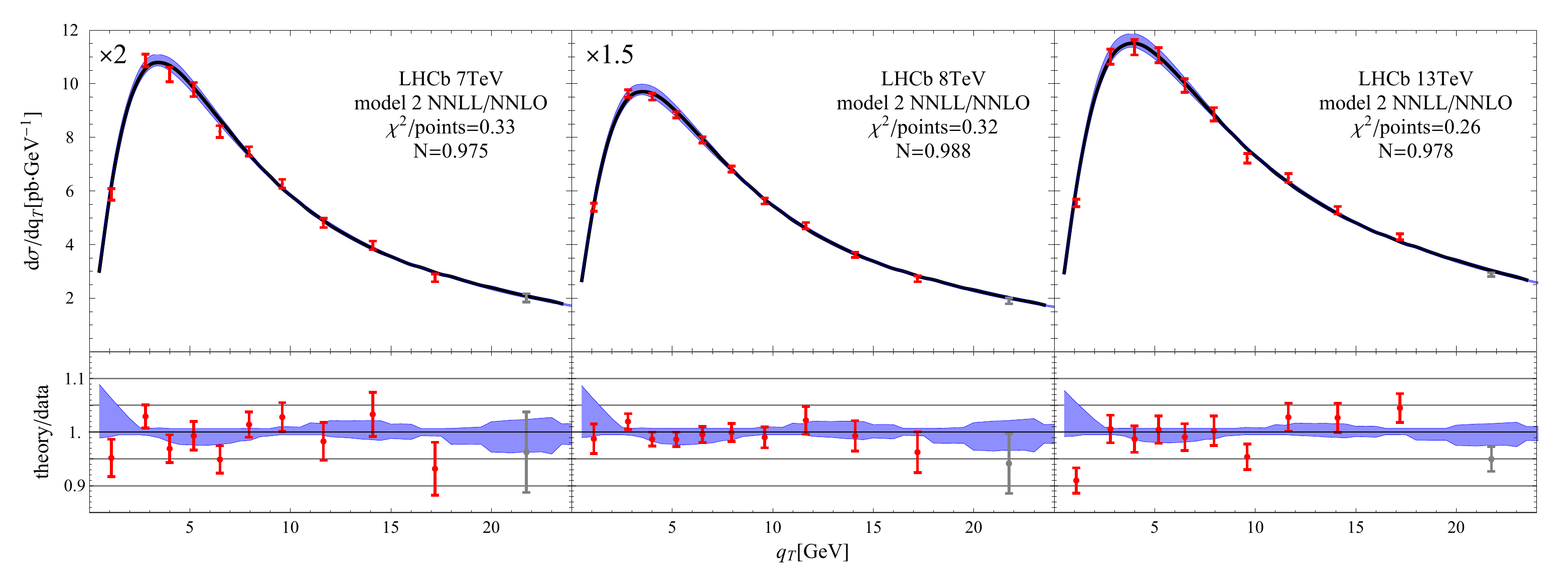}
\caption{\label{fig:LHCb_M2_NNLO}The comparison of the data for Z-boson production collected at LHCb experiment to the fit of \textit{model 2} at NNLL/NNLO. Red data points are those which included in the fit (i.e. with $\delta_T=0.2$). Gray data points are those which are not include in the fit (i.e. $\delta_T>0.2$). The blue band is the theoretical uncertainty obtained from the variation of scales.} 
\end{figure}

\begin{figure}[t]
\centering
\includegraphics[width=0.6\textwidth]{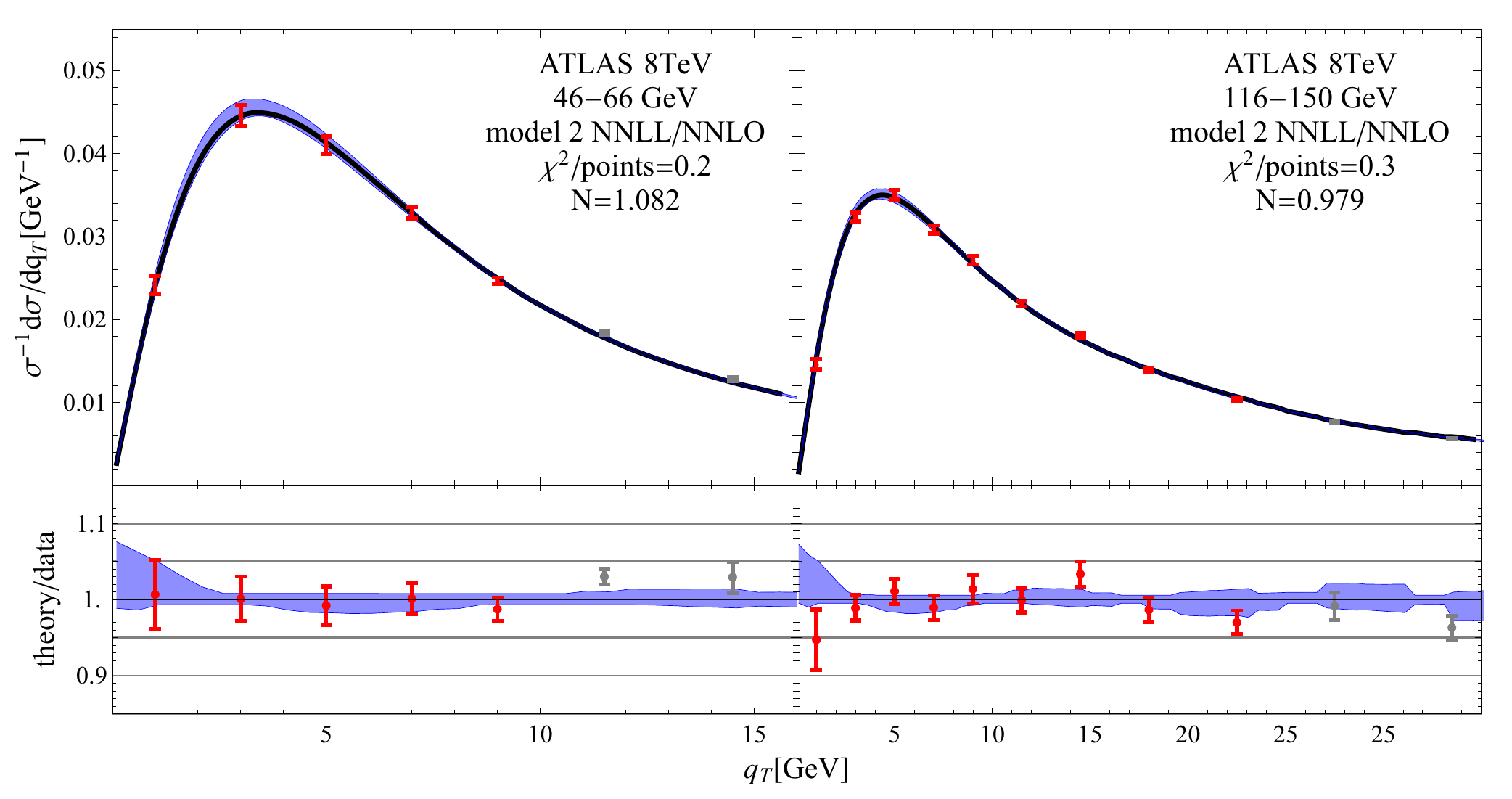}
\caption{\label{fig:ATLAS_DY_M2_NNLO} The comparison of the data for Drell-Yan reaction collected at ATLAS to the fit of \textit{model 2} at NNLL/NNLO. Red data points are those which included in the fit (i.e. with $\delta_T<0.2$). Gray data points are those which are not include in the fit (i.e. $\delta_T>0.25$). The blue band is the theoretical uncertainty obtained from the variation of scales.} 
\end{figure}

\begin{figure}[t]
\centering
\includegraphics[width=0.9\textwidth]{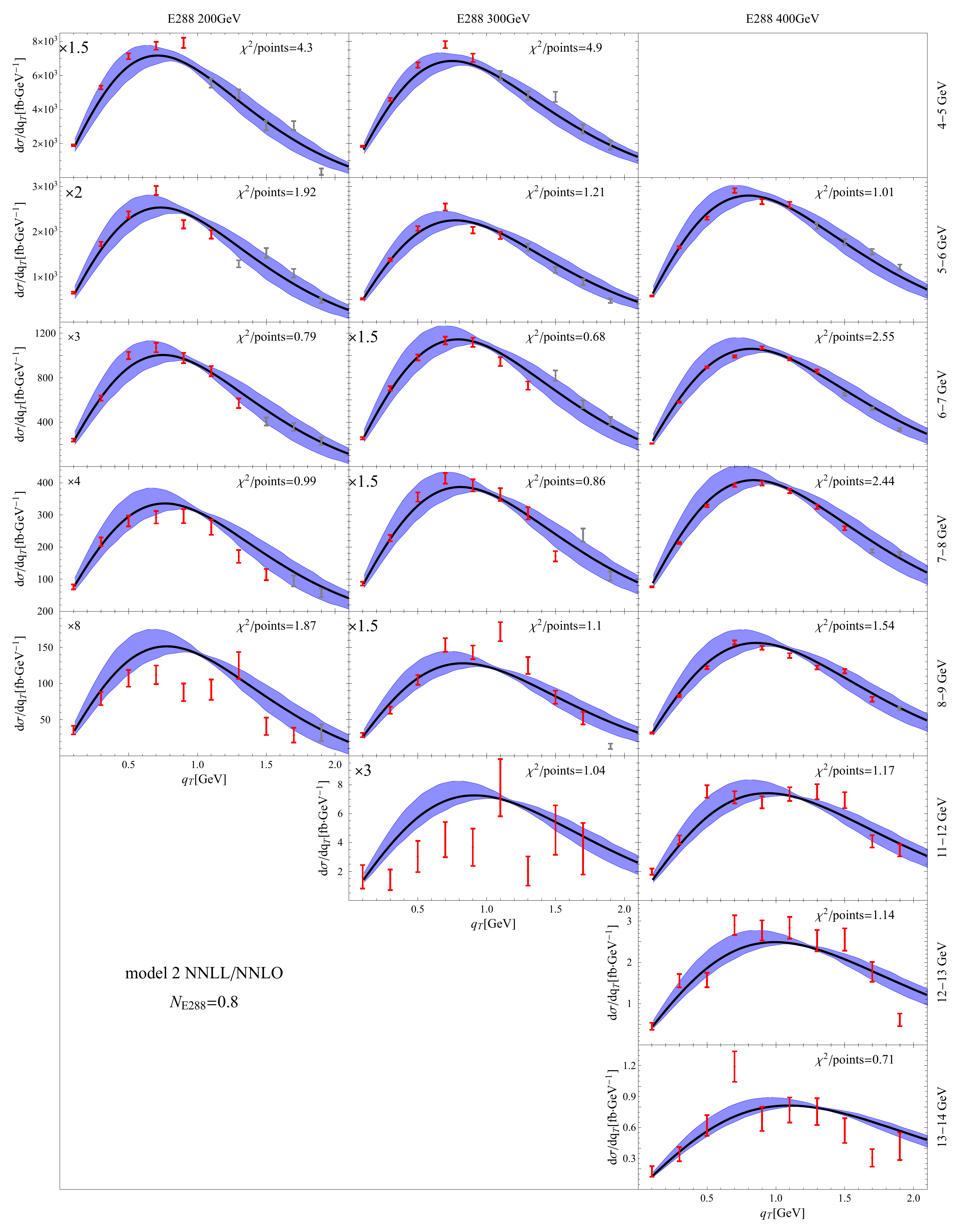}
\caption{\label{fig:E288_M2_NNLO}The comparison of the data for Drell-Yan reaction collected at E288 experiment to the fit of \textit{model 2} at NNLL/NNLO. Red data points are those which included in the fit (i.e. with $\delta_T=0.2$). Gray data points are those which are not include in the fit (i.e. $\delta_T>0.2$). The blue band is the theoretical uncertainty obtained from the variation of scales.} 
\end{figure}

In tab.~\ref{tab:partial_chi} we present the distribution of values for $\chi^2$ among experiments. One can see that the most stringent constraints  come from the Z-boson production data of ATLAS and D0 run2. This is due to the small experimental uncertainty of these data points. At the low-energy, the main tension is presented by the 4-5 GeV bins, while the rest are distributed more-or-less homogeneously. It probably indicates the influence of generic factorization violating terms. The plots of  the theoretical curves (at NNLL/NNLO for model 1) and the data points for individual experiments are shown in figures \ref{fig:Tevatron_M2_NNLO}-\ref{fig:E288_M2_NNLO}. The plots for different models and at other orders can be found in \cite{web}.

\section{Conclusion}
\label{sec:discussion}

The unpolarized Drell-Yan process at small-$q_T$ offers the simplest application of the TMD factorization formalism, and as such it has been studied by many groups. In this work, we have revised the main points of the practical implementation of TMD factorization, and reveal some new aspects  of the TMD phenomenology. Altogether it allows us to critically reanalyze the available Drell-Yan data and to extract consistently the unpolarized TMDPDFs, within some approximation. The primary aim of our analysis is to answer some general questions for the TMD approach such as: Up to which $q_T$ the TMD factorization works? What is the best asymptotical behavior of a TMD distributions? How convergent is TMD formalism at higher orders of perturbative expansion? The answers to these questions are naturally affected by the used prescriptions for the practical implementation of the TMD formalism. Even so, these important issues of TMD phenomenology are undiscussed in the literature or discussed very superficially. Implementing consistently the TMD factorization formalism, we are able to  fit a large set of Drell-Yan data points which ranges from low ($Q=4\;\GeV$) to high ($Q=116$-$150\;\GeV$) dilepton invariant masses on a wide interval of center of mass energies  and using a limited set of parameters (two for the non-perturbative part of TMDPDF and one for the non-perturbative part of the TMD evolution). 
 
In this work, we have formulated and used the $\zeta$-prescription, which is one of the main new theoretical contributions of this article. The $\zeta$-prescription consists of a particular choice of the rapidity evolution scale $\zeta=\zeta_\mu$, which depends on $\mu$, $b$ and the parton flavor (quark or gluon). This choice corresponds to the equi-evolution line in the space of TMD scales, and thus a TMD distribution is $\mu$-independent along this line. As a consequence, all logarithms related to the TMD evolution, which are essentially double logarithms, are eliminated from the small-$b$ OPE. It significantly improves the perturbative convergence and the radius of convergences for the small-$b$ OPE. The value of $\zeta_\mu$ is dictated by the differential equation (\ref{th:zeta_def}), which can be solved order-by-order in the perturbation theory. We stress that the $\zeta$-prescription does not strictly solve the problem of large-$b$ logarithms, which are still present in the matching coefficients. However, these logarithms are not related anymore to the TMD scales. Moreover, these logarithms are accompanied by the $x$-dependent coefficients which preserve the integral over $x$ in accordance with the probability interpretation of PDFs. Note, that $\zeta$-prescription is universal for all TMDs of the leading dynamical twist, due to the universality of TMD ultraviolet and rapidity renormalization factors. There are multiple possibilities to apply $\zeta$-prescription, see some discussion in appendix~\ref{app:zeta}. In this work, we have used the simplest one, which can certainly be improved. A further study of $\zeta$-prescription will be done elsewhere.

Within our implementation of TMD factorization and TMD distributions, which has a generic form, we have three independent perturbative series: one for hard matching, one for rapidity evolution, and one for small-$b$ matching. To defend the approach we provide the estimation of the perturbative uncertainty by variation of associated scales by factor 2, see sec.~\ref{sec:HEE_unser}. We have considered three successive perturbative orders (from NLL/NLO to NNLL/NNLO, see table~\ref{tab:orders}), and demonstrate that the theory uncertainties and the agreement with the data improve with the increase of the perturbative order. The agreement of the theory with the experiment resulting in our fit is a consequence of the $\zeta$-prescription to a large extent. The lowest possible combination of perturbative order, namely NLL/LO, produces very large theoretical error-bands 
 and thus has been excluded from the present study.

Our analysis  shows that data are very selective  about the non-perturbative part of the TMDs and only  well-behaved models can accommodate the fit. The best  models for the non-perturbative part of TMDPDF  that we have found are formulated in sec.~\ref{sec:models}. They have a common non-perturbative structure, namely
\begin{eqnarray}
F(x,\vec b)\simeq \int_x^1 \frac{dz}{z}C(z,\mathbf{L}_\mu)f_{NP}(z,\vec b)f(x/z,\mu),
\end{eqnarray}
where $f$ is the PDF, $C$ is the small-$b$ matching coefficient and $f_{NP}$ is a non-perturbative input. We have found that the best agreement with  data is given when the function $f_{NP}$ behaves as
\begin{eqnarray}\nn
\text{at small $b$:}&\qquad& f_{NP}\simeq e^{-\lambda_2 \vec b^2},
\\
\label{eq:fnpf}
\text{at large $b$:}&\qquad& f_{NP}\simeq e^{-\lambda_1 |\vec b|}.
\end{eqnarray}

We have considered two ansatzes which respect eq.~(\ref{eq:fnpf}), see sec.~\ref{sec:models} and eq.~(\ref{MODEL1})-(\ref{MODEL2}). The models have different behavior in the intermediate $\vec b$ region, in particular, model 2 has $z$ dependence. Nonetheless, the models produce nearly identical values of $\chi^2$ and of parameters $\lambda_{1,2}$. It implies that the parameters $\lambda_{1,2}$  that largely determine the shape of TMDPDF have a precise physical meaning. The values of parameters are reasonable $\lambda_1\sim 1.5\; m_\pi$ and $\lambda_2\sim 0.5$ GeV$^2$. We also study the influence of the parameter $g_K$, that parameterizes the non-perturbative part of TMD evolution. We have found that this parameter is significant at lower order (in our case NLL/NLO) and less significant at higher orders. Moreover, at NNLL/NNLO it is compatible with zero within the error-bars. We supplement our extraction with the estimation of the theoretical and statistical errorbars. The obtained theoretical uncertainty on the extracted parameters is notably larger in comparison to the experimental one (see fig.~\ref{fig:parameters_nogK} and \ref{fig:parameters_gK}). It can indicate the purity of the models (say, the independence on the flavor, or insufficient parameterization), or the general weakness of the theory. 

Another aspect that we point out, is the  practical limitation of TMD factorization. To make the discussion quantitative we introduce the parameter $\delta_T$, which is the highest allowed ratio $q_T/Q$ accounted in the fit. Clearly, at very low $\delta_T$ the TMD formalism should perfectly work, e.g. provide small values of $\chi^2$-distribution. Our expectation is that within the domain of the TMD-factorization the value of $\chi^2/d.o.f.$ is largely constant and starts to grow outside of this domain. Indeed, for the best models, the observed picture agrees with the expectation. In this way, we have shown that TMD factorization as it is, i.e. in the absence of $Y$-term, is capable of describing the data with $q_T\lesssim 0.2\ Q$, i.e. $\delta_T=0.2$. With some risk, one can prolong it to $\delta_T=0.25$. After $\delta_T=0.25$ the TMD factorization loses any agreement with the experiment. This analysis is unique, or at least we do not know any analog in the literature.  

The fit and the plots of the data has been done with the help of \texttt{arTeMiDe}, version 1.1, available at \cite{web}. This is a code package for the numerical evaluation of TMD distributions and related cross-sections. It has a flexible structure and allows to consider an arbitrary combination of perturbative orders up to NNLO for coefficient functions and N$^3$LO for evolution factors.  In the current version, it evaluates only unpolarized TMDPDFs, but we expect to update it for polarized cases and TMD fragmentation functions, as well as, to include the $Y$-term, in the future.  

\acknowledgments We thank Miguel G. Echevarria for the help on the initial stages of this work, and V. Bertone for useful comments on \texttt{arTeMiDe} code.  I.S. also thanks Stefano Melis for  discussions about some technical points. I.S. is supported by the Spanish MECD grant FPA2014-53375-C2-2-P and  FPA2016-75654-C2-2-P and the group UPARCOS.

\clearpage

\appendix

\section{Hard coefficient function}
\label{app:CV}

The hard coefficient function $|C_V(\mu,Q)|^2$ can be obtained from the expression for the quark form factor. At NNLO, it can found in \cite{Kramer:1986sg,Matsuura:1988sm}. Here, we present the combined expression in Drell-Yan kinematic
\begin{eqnarray}
|C_V(\mu,Q)|^2&=&1+2a_sC_F\(-\mathbf{l}_{Q^2}^2-3\mathbf{l}_{Q^2}-8+\frac{7\pi^2}{6}\)+2a_s^2C_F\Bigg\{C_F\Big[
\mathbf{l}_{Q^2}^4+6\mathbf{l}_{Q^2}^3
\\\nn&&+\(25-\frac{7\pi^2}{3}\)\mathbf{l}_{Q^2}^2+\(\frac{93}{2}-5\pi^2-24\zeta_3\)\mathbf{l}_{Q^2}
+\frac{511}{8}-\frac{83\pi^2}{6}-30\zeta_3+\frac{67\pi^4}{60}\Big]
\\\nn&&+C_A\Big[-\frac{11}{9}\mathbf{l}_{Q^2}^3+\mathbf{l}_{Q^2}^2\(-\frac{233}{18}+\frac{\pi^2}{3}\)+\mathbf{l}_{Q^2}\(-\frac{2545}{54}+\frac{22\pi^2}{9}+26\zeta_3\)
\\\nn&&-\frac{51157}{648}+\frac{1061}{108}\pi^2+\frac{313}{9}\zeta_3-\frac{4\pi^4}{45}\Big]\\\nn&&+T_FN_f\Big[\frac{4}{9}\mathbf{l}_{Q^2}^3+\frac{38}{9}\mathbf{l}_{Q^2}^2+\mathbf{l}_{Q^2}\(\frac{418}{27}-\frac{8\pi^2}{9}\)
+\frac{4085}{162}-\frac{91\pi^2}{27}+\frac{4}{9}\zeta_3\Big]\Bigg\}+\mathcal{O}(a_s^3).
\end{eqnarray}

\section{$\zeta$-prescription and expressions for coefficient functions}
\label{app:zeta_main}

In this appendix, we elaborate the details of the $\zeta$-prescription and expression for the coefficient function. Throughout the paper, we use the following notation for logarithms
\begin{eqnarray}
\mathbf{L}_X=\ln\(\frac{\vec b^2 X}{4 e^{-2\gamma_E}}\),\qquad \mathbf{l}_X=\ln\(\frac{\mu^2}{X}\).
\end{eqnarray}
For  convenience we introduce the following notation for the perturbative coefficient of anomalous dimensions
\begin{eqnarray}
\Gamma^f=\sum_{n=0}^\infty a_s^{n+1}\Gamma_n^f,\qquad \gamma_V^f=\sum_{n=1}^\infty a_s^n \gamma_V^{f(n)},
\qquad \mathcal{D}^f(\mu,\vec b)=\sum_{n=1}^\infty  a_s^n \sum_{k=0}^n \mathbf{L}_\mu^k d_f^{(n,k)}.
\end{eqnarray}
The LO terms are
\begin{eqnarray}
\Gamma^q_0=4C_F,\quad \Gamma_0^g=4C_A,\quad \gamma_V^{q(1)}=-6C_F,\quad \gamma_V^{g(1)}=-2\beta_0, \quad d^{(1,1)}_f=\frac{\Gamma^f_0}{2},
\quad d^{(1,0)}_f=0,
\end{eqnarray}
where $\beta_0=\frac{11}{3}C_A-\frac{2}{3}N_f$ is the leading order QCD $\beta$-function. The higher order terms can be found e.g. in ref.~\cite{Echevarria:2016scs}.

\subsection{Derivation of $\zeta$-value}
\label{app:zeta}
The $\zeta$-prescription is defined as a curve in $(\mu,\zeta)$-plane along which a TMD distribution has no evolution.  In other words, at $\zeta=\zeta_\mu$
\begin{eqnarray}
\mu^2 \frac{d F(x,\vec b;\mu,\zeta_\mu)}{d\mu^2}=0.
\end{eqnarray}
This equation can be rewritten as
\begin{eqnarray}
\(\mu^2\frac{\partial}{\partial\mu^2}+\(\frac{\mu^2}{\zeta}\frac{d \zeta}{d\mu^2}\) \zeta\frac{\partial}{\partial\zeta}\)F(x,\vec b;\mu,\zeta)=0.
\end{eqnarray}
Using the explicit expressions for the anomalous dimensions eq.~(\ref{th:evol_mu}), and introducing the intermediate function $\mathbf{l}_{\zeta_\mu}=f(\mathbf{L}_\mu)$ we obtain
\begin{eqnarray}
\mathcal{D}(\mathbf{L}_\mu)f'(\mathbf{L}_\mu)+\frac{\Gamma^f}{2}f(\mathbf{L}_\mu)-\frac{\gamma_V^f}{2}-\mathcal{D}(\mathbf{L}_\mu)=0.
\end{eqnarray}
Solving this equation order-by-order in  perturbation theory, we obtain
\begin{eqnarray}\nn
\mathbf{l}_{\zeta_\mu}&&=\frac{\mathbf{L}_\mu}{2}+\frac{\gamma_V^{f(1)}}{\Gamma_0^f}+\frac{c_1^f}{\mathbf{L}_\mu}
\\\nn &&+
a_s\Big[\frac{\beta_0}{12}\mathbf{L}_\mu^2+\frac{\gamma_V^{f(2)}+d_f^{(2,0)}}{\Gamma_0^f}-\frac{\gamma_V^{f(1)}\Gamma_1^f}{(\Gamma_0^f)^2}
+\frac{c^f_1\beta_0}{2}+\frac{c^f_2}{\mathbf{L}_\mu}-\frac{2c_1^fd^{(2,0)}_f}{\Gamma_0^f \mathbf{L}_\mu^2}\Big]
\\&&\label{app:zeta_complete}+a_s^2\Big[\frac{\beta_0^2}{48}\mathbf{L}_\mu^3+\frac{\beta_1 \Gamma^f_0+\beta_0\Gamma_1^f}{12\Gamma_0^f}\mathbf{L}_\mu^2
+\frac{\beta_0}{6\Gamma_0^f}(c^f_1\beta_0\Gamma_0^f+5d^{(2,0)}_f)\mathbf{L}_\mu+\frac{\gamma_V^{f(1)}(\Gamma_1^f)^2}{(\Gamma_0^f)^3}
\\\nn&&-\frac{\gamma_V^{f(2)}\Gamma_1^f+\gamma_V^{f(1)}\Gamma_2^f+d^{(2,0)}_f\Gamma_1^f}{(\Gamma_0^f)^2}+\frac{\gamma_V^{f(3)}+d^{(3,0)}_f+\beta_0\Gamma_1^f\frac{c_1^f}{2}}{\Gamma_0^f}+\frac{c^f_1\beta_1+c_2^f\beta_0}{2}
\\&&\nn \frac{c_3^f}{\mathbf{L}_\mu}+\frac{2c_1^fd^{(2,0)}_f\beta_0}{\Gamma_0^f}\frac{\ln(\mathbf{L}_\mu)}{\mathbf{L}_\mu}
+\frac{2c_1^fd^{(2,0)}_f\Gamma_1^f}{(\Gamma_0^f)^2\mathbf{L}_\mu^2}-2\frac{c_2^fd^{(2,0)}_f+c_1^fd_f^{(3,0}}{\Gamma_0^f \mathbf{L}_\mu^2}+\frac{4c_1^f(d_f^{(2,0})^2}{(\Gamma_0^f)^2\mathbf{L}_\mu^3}\Big]
\\&&\nn\qquad\qquad\qquad\qquad\qquad\qquad\qquad\qquad\qquad\qquad\qquad\qquad\qquad\qquad+\mathcal{O}(a_s^3),
\end{eqnarray}
where $c_{1,2,3}^f$ are integration constants. To derive this expression we have used that the $\mathbf{L}_\mu$-dependence of rapidity anomalous dimension at NNLO has the form
\begin{eqnarray}
\mathcal{D}^f(\mathbf{L}_\mu)&=&a_s \frac{\Gamma_0^f}{2}\mathbf{L}_\mu+a_s^2\(
\frac{\Gamma_0^f \beta_0}{4} \mathbf{L}_\mu^2+\frac{\Gamma_1^f}{2}\mathbf{L}_\mu+d_f^{(2,0)}
\)
\\\nn&&+a_s^3\(\frac{\Gamma_0^f \beta_0^2}{6}\mathbf{L}_\mu^3+\frac{2\Gamma_1^f\beta_0+\Gamma_0^f\beta_1}{4}\mathbf{L}_\mu^2
+\frac{4\beta_0 d_f^{(2,0)}+\Gamma_2^f}{2}\mathbf{L}_\mu+d_f^{(3,0)}\)+\mathcal{O}(a_s^4),
\end{eqnarray}
where the constants $d^{(2,0)}_f$ and $d^{(3,0)}_f$ can be found in \cite{Vladimirov:2016dll}.

The integration constants $c_{1,2,3}$ that appears in eq.~(\ref{app:zeta_complete}) are to be fixed by additional conditions, which would correspond to a selection of a particular curve among the family of equi-evolution curves in $(\mu,\zeta)$-plane. In our current analysis, we set all constants $c_{i}$ to zero for simplicity. It corresponds to the curve that passes though the point $(\mu,\zeta)=(0,0)$. We leave for the future a dedicated study of boundary condition and its influence on the phenomenology. Thus, the explicit NLO expression for $\mathbf{l}_{\zeta_\mu}$ for quark TMDPDF reads
\begin{eqnarray}
\mathbf{l}_{\zeta_\mu}&=&\frac{\mathbf{L}_\mu}{2}-\frac{3}{2}+a_s\Big[\frac{11 C_A-4 T_F N_f}{36}\mathbf{L}_\mu^2
\\\nn&&
+C_F\(-\frac{3}{4}+\pi^2-12\zeta_3\)+C_A\(\frac{649}{108}-\frac{17\pi^2}{12}+\frac{19}{2}\zeta_3\)+T_FN_f\(-\frac{53}{27}+\frac{\pi^2}{3}\)\Big]+\mathcal{O}(a_s^3).
\end{eqnarray}
The explicit NLO expression for $\mathbf{l}_{\zeta_\mu}$ for gluon TMDPDF reads
\begin{eqnarray}
\mathbf{l}_{\zeta_\mu}&=&\frac{\mathbf{L}_\mu}{2}-\frac{11}{6}+\frac{2}{3}\frac{T_FN_f}{C_A}+a_s\Big[\frac{11 C_A-4 T_F N_f}{36}\mathbf{L}_\mu^2
\\\nn&&
+C_A\(\frac{247}{54}-\frac{11\pi^2}{36}-\frac{5\zeta_3}{2}\)+T_FN_f\(-\frac{16}{3}+\frac{\pi^2}{9}\)+\(2C_F+\frac{40}{27}T_FN_f\)\frac{T_fN_f}{C_A}\Big]+\mathcal{O}(a_s^3).
\end{eqnarray}
The expression for the $\zeta_\mu$ then reads 
\begin{eqnarray}
\zeta_\mu=2\frac{\mu }{b}e^{-v_f-\gamma_E},\qquad v_f=\mathbf{l}_{\zeta_\mu}-\frac{\mathbf{L}_\mu}{2}.
\end{eqnarray}

\subsection{Scale dependence and logarithmic part of coefficient function}
\label{app:zeta_in_coeff}

The small-$b$ coefficient function satisfies the pair of equations
\begin{eqnarray}
\mu^2 \frac{d}{d\mu^2}C_{f\ot f'}(x,\vec b;\mu,\zeta)&=&\sum_{ k} C_{f\ot k}(x,\vec b;\mu,\zeta)\otimes \Big[\frac{\delta_{kf'}}{2}\(\Gamma^f \mathbf{l}_\zeta -\gamma_V^f\)-P_{k\ot f'}(x)\Big],
\\
\zeta\frac{d}{d\zeta}C_{f\ot f'}(x,\vec b;\mu,\zeta)&=&-\mathcal{D}^f(\mu,\vec b)C_{f\ot f'}(x,\vec b;\mu,\zeta),
\end{eqnarray}
where $P(x)$ is the DGLAP kernel of the PDF evolution, and $\otimes$ denotes the Mellin convolution in the variable $x$. Using these equations one finds the logarithmic part of the coefficient function. At NNLO the expression for the coefficient function reads
\begin{eqnarray}\label{app:C_logStructure}
C_{f\ot f'}&=&\delta_{ff'}(\bar x)
\\\nn &&+a_s\Big[-\mathbf{L}_\mu P^{(1)}_{f\ot f'}+C_{f\ot f'}^{(1,0)}+\delta_{ff'}(\bar x)\(
-\frac{\Gamma^f_0}{4}\mathbf{L}_\mu^2+\frac{\Gamma^f_0}{2}\mathbf{L}_\mu \mathbf{l}_\zeta-\frac{\gamma_V^{f(1)}}{2}\mathbf{L}_\mu\)\Bigg]
\\\nn &&+a_s^2\Bigg\{\delta_{ff'}(\bar x)\frac{(\Gamma_1^f)^2}{8}\(\frac{1}{4}\mathbf{L}_\mu^4-\mathbf{L}_\mu^3\mathbf{l}_\zeta+\mathbf{L}_\mu^2\mathbf{l}_\zeta^2\)
\\&&\nn ~~~+\frac{\Gamma_0^f}{4}\Big[\(P^{(1)}_{f\ot f'}+\delta_{ff'}(\bar x)\frac{\gamma_V^{f(1)}}{2}\)\(\mathbf{L}_\mu^3-2\mathbf{L}_\mu^2\mathbf{l}_\zeta\)
+\delta_{ff'}(\bar x)\beta_0\(-\frac{2}{3}\mathbf{L}_\mu^3+\mathbf{L}_\mu^2\mathbf{l}_\zeta\)\Big]
\\&&\nn ~~~+\(\delta_{ff'}(\bar x)\Gamma_1^f+\Gamma_0^fC_{f\ot f'}^{(1,0)}\)\frac{2\mathbf{L}_\mu\mathbf{l}_\zeta-\mathbf{L}_\mu^2}{4}
+\Big[\frac{1}{2}P^{(1)}_{f\ot k}\otimes P^{(1)}_{k\ot f'}+\frac{\gamma_V^{f(1)}-\beta_0}{2}P_{f\ot f'}^{(1)}
\\\nn && ~~~
+\delta_{ff'}(\bar x)\gamma_V^{f(1)}\frac{\gamma_V^{f(1)}-2\beta_0}{8}\Big]\mathbf{L}_\mu^2
+\delta_{ff'}(\bar x)d^{(2,0)}_{f}(\mathbf{l}_\zeta-\mathbf{L}_\mu)
\\\nn &&~~~+\Big[-P^{(2)}_{f\ot f'}-C_{f\ot k}^{(1,0)}\otimes P_{k\ot f'}^{(1)}+C_{f\ot f'}^{(1,0)}\(\beta_0-\frac{\gamma_V^{f(1)}}{2}\)-\delta_{ff'}(\bar x)\frac{\gamma_V^{f(2)}}{2}\Big]\mathbf{L}_\mu+C_{f\ot f}^{(2,0)}\Bigg\}
\\\nn &&+\mathcal{O}(a_s^3),
\end{eqnarray}
where we omit the argument $(x)$ of DGLAP kernel $P(x)=\sum_n a_s^n P^{(n)}(x)$ and the finite part of the coefficient function $C^{(n,0)}(x)$, and $\delta_{ff'}(\bar x)=\delta_{ff'}\delta(\bar x)$.

Substituting the NLO expression for $\mathbf{l}_\zeta$, eq.~ (\ref{app:zeta_complete}) into the coefficient function,  eq.~(\ref{app:C_logStructure}) we obtain at NNLO
\begin{eqnarray}\label{app:C_log2}
C_{f\ot f'}&=&\delta_{ff'}(\bar x)
\\\nn &&+a_s\Big[-\mathbf{L}_\mu P^{(1)}_{f\ot f'}+C_{f\ot f'}^{(1,0)}+\frac{c_1^f\Gamma_0^f}{2}\delta_{ff'}(\bar x)\Bigg]
+a_s^2\Bigg[\(\frac{1}{2}P^{(1)}_{f\ot k}\otimes P^{(1)}_{k\ot f'}-\frac{\beta_0}{2}P_{f\ot f'}^{(1)}\)\mathbf{L}_\mu^2
\\\nn &&+\(-P^{(2)}_{f\ot f'}-\(C_{f\ot k}^{(1,0)}+\frac{c_1^f\Gamma_0^f}{2}\delta_{fk}(\bar x)\)\otimes P_{k\ot f'}^{(1)}+\beta_0\(C^{(1,0)}_{f\ot f'}+\delta_{ff'}(\bar x)\frac{c_1^f\Gamma_0^f}{2}\)\)\mathbf{L}_\mu
\\\nn&& +C_{f\ot f}^{(2,0)}+\frac{c_1^f\Gamma_0^f}{2}C_{f\ot f'}^{(1,0)}(x)+\delta_{ff'}(\bar x)
\(\frac{\gamma_V^{f(1)}d_f^{(2,0)}}{\Gamma_0^f}+\frac{c_1^f\Gamma_1^f}{2}+\frac{c_2^f\Gamma_0^f}{2}+\frac{(c_1^f\Gamma_0^f)^2}{8}\)\Bigg]+\mathcal{O}(a_s^3).
\end{eqnarray}
Note, that despite the fact that the solution for $\zeta$-prescription in eq.~(\ref{app:zeta_complete}) has inverse powers of $\mathbf{L}_\mu$, the coefficient function has not. It is easy to check that this expression convoluted with PDF is renormalization-invariant, 
\begin{eqnarray}
\mu^2\frac{d}{d\mu^2}C_{q\ot k}\otimes f_{k\ot h}(x)=0\, .
\end{eqnarray}

\subsection{Expression for NNLO coefficient function in $\zeta$-prescription}
\label{app:coef}
The NNLO coefficient functions are cumbersome structures, which contain logarithms and polylogarithms of order 2 and 3 and  their straight numerical evaluation  is costly. To speed up the evaluation of convolutions within \texttt{arTeMiDe}, we have used an approximate expression for the coefficient function. A similar method for  higher-order expressions has been suggested in ref.~\cite{vanNeerven:1999ca} and it is  widely used in NNLO+ phenomenology of PDFs. We parameterize  the NNLO coefficient function by 17 terms
\begin{eqnarray}\nn
C(\mathbf{L}_{\mu},x)&=&A_1 \delta(\bar x)+A_2\(\frac{1}{1-x}\)_++A_3\(\frac{\ln \bar x}{1-x}\)_++A_4 \ln \bar x+A_5 \ln^2 \bar x+A_6 \ln^3 \bar x
\\ &&\label{app:C_param}
+B_1\ln x+B_2 \ln^2 x+B_3 \ln^3 x+B_4\frac{1}{x}+B_5\frac{\ln x}{x}
\\\nn && +c_1+c_2 x+c_3 x^2 +c_4 x^3+c_5 \ln \bar x \ln x+c_6 \ln \bar x \ln^2 x.
\end{eqnarray}
Here, the coefficients $A$ represent the singular at $x\to 1$ terms, and are evaluated exactly. The coefficients $B$ represent singular at $x\to 0$ terms, and also evaluated exactly. The coefficients $c$ represent interpolation functions between asymptotics. These coefficients are fit numerically. The relative precision of the approximation is $\sim 10^{-3}$. The convolution integral receives the main numerical contributions at singular points, while the rest are minor corrections. So, we find that the relative accuracy of the convolution is better then $10^{-6}$, which is far beyond any currently needed accuracy. The values of coefficients A, B, and c are given in the tables~\ref{tab:A},~\ref{tab:B} and~\ref{tab:c}.

\begin{sidewaystable}
\begin{tabular}{c|p{6cm}|p{5cm}|p{1cm}|p{2.5cm}|p{2.6cm}|p{0.5cm}}
& $A_1$ & $A_2$ & $A_3$ & $A_4$ & $A_5$ & $A_6$
\\\hline\hline
$\hat C_{q\ot q}$ & \specialcell{
$\mathbf{L}_\mu^2\(-14-\frac{64\pi^2}{27}+\frac{4N_f}{3}\)$ \\$+ \mathbf{L}_\mu 
\!\!\(N_f\(\frac{4}{3}+\frac{20\pi^2}{27}\)\!-\!14\!-\!\frac{70\pi^2}{9}\!+\!\frac{16\zeta_3}{3}\)$ \\ 
$+N_f\!\!\(\frac{352}{243}+\frac{10\pi^2}{27}+\frac{56\zeta_3}{27}\)$ \\ $
-\frac{2416}{81}-\frac{67\pi^2}{9}+\frac{448\zeta_3}{9}+\frac{20\pi^4}{81}$}
& \specialcell{
$\mathbf{L}_\mu^2\(-8+\frac{16N_f}{9}\)$\\ $+\mathbf{L}_\mu\(-\frac{1072}{9}+\frac{176\pi^2}{27}+\frac{160N_f}{27}\)$\\
$+\frac{448N_f}{81}-\frac{3232}{27}+112\zeta_3$
}
&\specialcell{
$\frac{256}{9}\mathbf{L}_\mu^2$
}
&
\specialcell{$ -\frac{256}{9}\mathbf{L}_\mu^2$ \\ $-\frac{256}{9}\mathbf{L}_\mu+\frac{200}{9} $}
&
$ -\frac{64}{9} $
& 
$0 $
\\\hline
$\hat C_{q\ot g}$  & $ 0$& $ 0$& $ 0$& 
$\frac{26}{3}\mathbf{L}_\mu^2-\frac{5}{3} $& $ \frac{10}{3}\mathbf{L}_\mu $& $\frac{5}{9} $
\\\hline
$\hat C_{g\ot q}$ & $ 0$& $0 $& $0 $& 
\specialcell{$\frac{208}{9}\mathbf{L}_\mu^2$\\$+\mathbf{L}_\mu\!\(\!\frac{32N_f}{9}-\frac{208}{9}\!\)$\\$+\frac{32N_f}{27}-\frac{184}{9}$}&
$-\frac{80}{9}\mathbf{L}_\mu+\frac{8N_f}{9}-\frac{44}{9} $& $-\frac{40}{27} $
\\\hline
$\hat C_{g\ot g}$ & 
\specialcell{$-12\pi^2\mathbf{L}_\mu^2$ \\$+ \mathbf{L}_\mu 
\(-96-108\zeta_3+\frac{32N_f}{3}\)$ \\ 
$+N_f\(\frac{548}{27}+5\zeta_2-\frac{28\zeta_3}{3}\)-\frac{56N_f^2}{81}$
\\ $-112-\frac{201\zeta_2}{2}+154\zeta_3+\frac{225\zeta_4}{4}$}
&
\specialcell{
$\mathbf{L}_\mu^2\(66-4N_f\)$\\ $+\mathbf{L}_\mu\(-268+108\zeta_2+\frac{40N_f}{3}\)$\\
$+\frac{112N_f}{9}-\frac{808}{3}+252\zeta_3$
} 
 & $144\mathbf{L}_\mu^2 $ 
 & 
 \specialcell{$-144 \mathbf{L}_\mu^2-144\mathbf{L}_\mu$\\$+ 6-2N_f$}&
  $-36 $& $0 $
\\\hline 
$\hat C_{q\ot q'}$ & $ 0$& $0 $& $0 $& $ 0$& $ 0$& $ 0$
\\\hline 
$\hat C_{q\ot \bar q}$ & $0 $& $0 $& $0 $& $ 0$& $0 $& $0 $
\end{tabular}
\caption{\label{tab:A}Exact values of the coefficients of the unpolarized TMD matching coefficient in the parameterization (\ref{app:C_param}), for the terms singular at $x\to 1$.}
\end{sidewaystable}
\begin{sidewaystable}
\begin{tabular}{c|p{5cm}|p{3cm}|p{3cm}|p{5cm}|p{1cm}}
& $B_1$ & $B_2$ & $B_3$ & $B_4$ & $B_5$
\\\hline\hline
$\hat C_{q\ot q}$ & \specialcell{$-\frac{32}{9}\mathbf{L}_\mu^2
+\mathbf{L}_\mu\(\frac{-152}{9}+\frac{16N_f}{9}\)$ \\
$+\frac{40N_f}{27}-8$
}
&
$-\frac{40\mathbf{L}_\mu}{9}-2+\frac{4N_f}{9} $ &
$-\frac{20}{27}$
& 0 & 0
\\
\hline 
$\hat C_{q\ot g}$ & $\frac{14}{3}\mathbf{L}_\mu^2-\frac{22}{3}\mathbf{L}_\mu+\frac{34}{3}$
&
$\frac{14}{3}\mathbf{L}_\mu-\frac{7}{6}$
& 
$\frac{7}{9}$
&
$4\mathbf{L}_\mu^2-\frac{51}{3}\mathbf{L}_\mu+\frac{172}{9}-\frac{4\pi^2}{3}$
 & $0$
\\
\hline 
$\hat C_{g\ot q}$ & $-\frac{224}{9}\mathbf{L}_\mu^2+\frac{448}{9}\mathbf{L}_\mu-\frac{200}{9}$
&
$-\frac{224}{9}\mathbf{L}_\mu+\frac{112}{9}$
& 
$-\frac{112}{27}$
&
\specialcell{$\(\frac{104}{3}-\frac{64N_f}{9}\)\mathbf{L}_\mu^2$\\$+\(-16+16\zeta_2+\frac{320}{27}N_f\)\mathbf{L}_\mu$
\\$+\frac{896N_f}{81}-\frac{12640}{27}+\frac{352\zeta_2}{3}+192\zeta_3$}
 & $-32\mathbf{L}_\mu^2$
\\
\hline 
$\hat C_{g\ot g}$ & \specialcell{$\mathbf{L}_\mu^2\(-72+\frac{16N_f}{3}\)$
\\ $+\mathbf{L}_\mu\(12+24N_f\)$ \\
$+\frac{74N_f}{3}-293$
}
&
\specialcell{$\mathbf{L}_\mu \(-72+\frac{16N_f}{3}\)$\\ $+3+6N_f$} &
$-12+\frac{8N_f}{9}$
& 
\specialcell{$\mathbf{L}_\mu^2\(-198-\frac{4N_f}{9}\)$\\$+\mathbf{L}_\mu\(36\zeta_2+\frac{244N_f}{9}\)$\\$+\frac{226N_f}{9}-\frac{3160}{3}+264\zeta_2+432\zeta_3$}
 & $-72\mathbf{L}_\mu^2$
\\\hline
$\hat C_{q\ot q'}$ & $\frac{8}{3}\mathbf{L}_\mu^2-\frac{8}{3}\mathbf{L}_\mu+\frac{8}{3} $& $\frac{8}{3}\mathbf{L}_\mu-\frac{2}{3} $& $\frac{4}{9} $
& $\frac{16}{9}\mathbf{L}_\mu^2-\frac{208}{27}\mathbf{L}_\mu+\frac{688}{81}-\frac{32\zeta_2}{9} $& $0$
\\\hline
$\hat C_{q\ot \bar q}$ & $\frac{16}{9}\mathbf{L}_\mu-\frac{4}{3} $& $\frac{8}{9}\mathbf{L}_\mu $& $\frac{4}{27} $
& $0 $& $0$
\end{tabular}
\caption{\label{tab:B}Exact values of the coefficients of the unpolarized TMD matching coefficient in the parameterization (\ref{app:C_param}), for the terms singular at $x\to 1$.}

\begin{tabular}{c|p{3.3cm}|p{3.5cm}|p{3.3cm}|p{3.1cm}|p{3.1cm}|p{2.9cm}}
& $c_1$ & $c_2$ & $c_3$ & $c_4$ & $c_5$ & $c_6$
\\\hline\hline
$\hat C_{q\ot q}$ & \small\specialcell{
$\mathbf{L}_\mu^2(-3.27-0.89N_f)$
\\
$+\mathbf{L}_\mu(-11.16-1.17N_f) $
\\
$-15.59-3.64N_f $
}
&\small
\specialcell{
$\mathbf{L}_\mu^2(27.44-0.89N_f)$
\\
$+\mathbf{L}_\mu(95.11-8.10N_f) $
\\
$-26.19-4.04N_f $
}
&\small
\specialcell{
$-8.25\mathbf{L}_\mu^2$
\\
$+\mathbf{L}_\mu(-2.75-0.12N_f) $
\\
$-61.36-0.62N_f $
}
&\small
\specialcell{
$6.26\mathbf{L}_\mu^2$
\\
$+\mathbf{L}_\mu(10.83-0.08N_f) $
\\
$-18.60-0.20N_f $
}
&\small
\specialcell{
$-13.56\mathbf{L}_\mu^2 $
\\
$+\mathbf{L}_\mu(1.23-0.11N_f) $
\\
$-32.70-0.07N_f $
}
&\small
\specialcell{
$-3.08\mathbf{L}_\mu^2 $
\\
$+\mathbf{L}_\mu(-1.53-0.26N_f)$
\\
$+12.63-0.17N_f $
}
\\\hline
$\hat C_{q\ot g}$ & \small
\specialcell{$2.90\mathbf{L}_\mu^2+21.88\mathbf{L}_\mu$\\$+ 5.33$}
&\small
\specialcell{$-33.59\mathbf{L}_\mu^2-32.82\mathbf{L}_\mu$\\$-37.80$}
&\small
\specialcell{$69.07\mathbf{L}_\mu^2+115.50\mathbf{L}_\mu$\\$+ 55.94$}
&\small
\specialcell{$-40.23\mathbf{L}_\mu^2-99.18\mathbf{L}_\mu$\\$+ 44.59$}
&\small
\specialcell{$14.97\mathbf{L}_\mu^2-18.01\mathbf{L}_\mu$\\$-10.96$}
&\small
\specialcell{$8.91\mathbf{L}_\mu^2-12.95\mathbf{L}_\mu$\\$-6.52$}
\\\hline
$\hat C_{g\ot q}$ & \small
\specialcell{
$\mathbf{L}_\mu^2(-93.12+7.11N_f)$
\\
$+\mathbf{L}_\mu(-15.38-19.00N_f)$
\\
$-43.40-16.98N_f $
}
&\small
\specialcell{
$\mathbf{L}_\mu^2(174.51-3.56N_f)$
\\
$+\mathbf{L}_\mu(-176.55+21.89N_f)$
\\
$+173.18+9.60N_f $
}
&\small
\specialcell{
$-74.33\mathbf{L}_\mu^2$
\\
$+\mathbf{L}_\mu(46.66-3.76N_f) $
\\
$-88.43-2.20N_f$
}
&\small
\specialcell{
$21.88\mathbf{L}_\mu^2$
\\
$+\mathbf{L}_\mu(-16.82+0.25N_f)$
\\
$-8.72-1.15N_f $
}
&\small
\specialcell{
$-9.03\mathbf{L}_\mu^2$
\\
$+\mathbf{L}_\mu(-24.55-2.33N_f)$
\\
$-40.28-0.55N_f $
}
&\small
\specialcell{
$-9.84\mathbf{L}_\mu^2$
\\
$+\mathbf{L}_\mu(28.84-0.50N_f)$
\\
$-35.67-0.06N_f$
}
\\\hline
$\hat C_{g\ot g}$ 
&\small
 \specialcell{
$\mathbf{L}_\mu^2(-66.99+10.73N_f)$
\\
$+\mathbf{L}_\mu(-281.6-1.04N_f)$
\\
$+480.95+5.64N_f $
}
&\small
\specialcell{
$\mathbf{L}_\mu^2(688.48.26-12.73N_f)$
\\
$+\mathbf{L}_\mu(619.7-33.12N_f)$
\\
$+705.78-41.19N_f $
}
&\small
\specialcell{
$\mathbf{L}_\mu^2(-838.04+9.53N_f)$
\\
$+\mathbf{L}_\mu(-744.81+8.86N_f) $
\\
$-233.01-2.04N_f$
}
&\small
\specialcell{
$\mathbf{L}_\mu^2(418.09-3.04N_f)$
\\
$+\mathbf{L}_\mu(438.44-15.06N_f) $
\\
$+154.84-3.95N_f $
}
&\small
\specialcell{
$\mathbf{L}_\mu^2(-229.01-0.91 N_f)$
\\
$+\mathbf{L}_\mu(-124.39-3.21N_f)$
\\
$+72.41-3.04N_f $
}
&\small
\specialcell{
$\mathbf{L}_\mu^2(-109.33+0.99 N_f)$
\\
$+\mathbf{L}_\mu(101.43-0.94N_f)$
\\
$-67.23+3.27N_f$
}
\\\hline
$\hat C_{q\ot q'}$ &\small 
\specialcell{$1.37\mathbf{L}_\mu^2+10.59\mathbf{L}_\mu$\\$-3.53$}
&\small
\specialcell{$-4.37\mathbf{L}_\mu^2-5.18\mathbf{L}_\mu$\\$+4.87$}
&\small
\specialcell{$2.76\mathbf{L}_\mu^2+3.38\mathbf{L}_\mu$\\$- 11.53$}
&\small
\specialcell{$-1.52\mathbf{L}_\mu^2-1.12\mathbf{L}_\mu$\\$+ 7.47$}
&\small
\specialcell{$-0.46\mathbf{L}_\mu^2+1.11\mathbf{L}_\mu$\\$+2.38$}
&\small
\specialcell{$0.49\mathbf{L}_\mu^2-4.01\mathbf{L}_\mu$\\$+0.85$}
\\\hline
$\hat C_{q\ot \bar q}$ 
& \small
$0.56\mathbf{L}_\mu-3.33$
& \small
$2.11\mathbf{L}_\mu+5.85$
&\small
$-6.95\mathbf{L}_\mu-4.78$
&\small
$4.23\mathbf{L}_\mu+2.24$
&\small
$1.14\mathbf{L}_\mu+0.50$
&\small
$0.41\mathbf{L}_\mu+0.08$
\end{tabular}
\caption{\label{tab:c}Approximate values of the coefficients of the unpolarized TMD matching coefficient in the parameterization (\ref{app:C_param}), for the regular in $0<x<1$ terms.}

\end{sidewaystable}
\clearpage

\bibliographystyle{JHEP}
\bibliography{TMD_ref}

\providecommand{\href}[2]{#2}\begingroup\raggedright\begin{thebibliography}{10}

\bibitem{Collins:2011zzd}
J.~Collins, \emph{{Foundations of perturbative QCD}}.
\newblock Cambridge University Press, 2013.

\bibitem{GarciaEchevarria:2011rb}
M.~G. Echevarria, A.~Idilbi and I.~Scimemi, \emph{{Factorization Theorem For
  Drell-Yan At Low $q_T$ And Transverse Momentum Distributions
  On-The-Light-Cone}},
  \href{http://dx.doi.org/10.1007/JHEP07(2012)002}{\emph{JHEP} {\bfseries 07}
  (2012) 002}, [\href{https://arxiv.org/abs/1111.4996}{{\ttfamily 1111.4996}}].

\bibitem{Echevarria:2012js}
M.~G. Echevarria, A.~Idilbi and I.~Scimemi, \emph{{Soft and Collinear
  Factorization and Transverse Momentum Dependent Parton Distribution
  Functions}},
  \href{http://dx.doi.org/10.1016/j.physletb.2013.09.003}{\emph{Phys. Lett.}
  {\bfseries B726} (2013) 795--801},
  [\href{https://arxiv.org/abs/1211.1947}{{\ttfamily 1211.1947}}].

\bibitem{Echevarria:2014rua}
M.~G. Echevarria, A.~Idilbi and I.~Scimemi, \emph{{Unified treatment of the QCD
  evolution of all (un-)polarized transverse momentum dependent functions:
  Collins function as a study case}},
  \href{http://dx.doi.org/10.1103/PhysRevD.90.014003}{\emph{Phys. Rev.}
  {\bfseries D90} (2014) 014003},
  [\href{https://arxiv.org/abs/1402.0869}{{\ttfamily 1402.0869}}].

\bibitem{Gaunt:2014ska}
J.~R. Gaunt, \emph{{Glauber Gluons and Multiple Parton Interactions}},
  \href{http://dx.doi.org/10.1007/JHEP07(2014)110}{\emph{JHEP} {\bfseries 07}
  (2014) 110}, [\href{https://arxiv.org/abs/1405.2080}{{\ttfamily 1405.2080}}].

\bibitem{Becher:2010tm}
T.~Becher and M.~Neubert, \emph{{{Drell-Yan} Production at Small $q_T$,
  Transverse Parton Distributions and the Collinear Anomaly}},
  \href{http://dx.doi.org/10.1140/epjc/s10052-011-1665-7}{\emph{Eur. Phys. J.}
  {\bfseries C71} (2011) 1665},
  [\href{https://arxiv.org/abs/1007.4005}{{\ttfamily 1007.4005}}].

\bibitem{Chiu:2012ir}
J.-Y. Chiu, A.~Jain, D.~Neill and I.~Z. Rothstein, \emph{{A Formalism for the
  Systematic Treatment of Rapidity Logarithms in Quantum Field Theory}},
  \href{http://dx.doi.org/10.1007/JHEP05(2012)084}{\emph{JHEP} {\bfseries 05}
  (2012) 084}, [\href{https://arxiv.org/abs/1202.0814}{{\ttfamily 1202.0814}}].

\bibitem{Mantry:2010bi}
S.~Mantry and F.~Petriello, \emph{{Transverse Momentum Distributions in the
  Non-Perturbative Region}},
  \href{http://dx.doi.org/10.1103/PhysRevD.84.014030}{\emph{Phys. Rev.}
  {\bfseries D84} (2011) 014030},
  [\href{https://arxiv.org/abs/1011.0757}{{\ttfamily 1011.0757}}].

\bibitem{Echevarria:2015usa}
M.~G. Echevarria, I.~Scimemi and A.~Vladimirov, \emph{{Transverse momentum
  dependent fragmentation function at next-to–next-to–leading order}},
  \href{http://dx.doi.org/10.1103/PhysRevD.93.011502,
  10.1103/PhysRevD.94.099904}{\emph{Phys. Rev.} {\bfseries D93} (2016) 011502},
  [\href{https://arxiv.org/abs/1509.06392}{{\ttfamily 1509.06392}}].

\bibitem{Echevarria:2015byo}
M.~G. Echevarria, I.~Scimemi and A.~Vladimirov, \emph{{Universal transverse
  momentum dependent soft function at NNLO}},
  \href{http://dx.doi.org/10.1103/PhysRevD.93.054004}{\emph{Phys. Rev.}
  {\bfseries D93} (2016) 054004},
  [\href{https://arxiv.org/abs/1511.05590}{{\ttfamily 1511.05590}}].

\bibitem{Echevarria:2016scs}
M.~G. Echevarria, I.~Scimemi and A.~Vladimirov, \emph{{Unpolarized Transverse
  Momentum Dependent Parton Distribution and Fragmentation Functions at
  next-to-next-to-leading order}},
  \href{http://dx.doi.org/10.1007/JHEP09(2016)004}{\emph{JHEP} {\bfseries 09}
  (2016) 004}, [\href{https://arxiv.org/abs/1604.07869}{{\ttfamily
  1604.07869}}].

\bibitem{Scimemi:2016ffw}
I.~Scimemi and A.~Vladimirov, \emph{{Power corrections and renormalons in
  Transverse Momentum Distributions}},
  \href{http://dx.doi.org/10.1007/JHEP03(2017)002}{\emph{JHEP} {\bfseries 03}
  (2017) 002}, [\href{https://arxiv.org/abs/1609.06047}{{\ttfamily
  1609.06047}}].

\bibitem{Landry:1999an}
F.~Landry, R.~Brock, G.~Ladinsky and C.~P. Yuan, \emph{{New fits for the
  nonperturbative parameters in the CSS resummation formalism}},
  \href{http://dx.doi.org/10.1103/PhysRevD.63.013004}{\emph{Phys. Rev.}
  {\bfseries D63} (2001) 013004},
  [\href{https://arxiv.org/abs/hep-ph/9905391}{{\ttfamily hep-ph/9905391}}].

\bibitem{Qiu:2000hf}
J.-w. Qiu and X.-f. Zhang, \emph{{Role of the nonperturbative input in QCD
  resummed Drell-Yan $Q_{T}$ distributions}},
  \href{http://dx.doi.org/10.1103/PhysRevD.63.114011}{\emph{Phys. Rev.}
  {\bfseries D63} (2001) 114011},
  [\href{https://arxiv.org/abs/hep-ph/0012348}{{\ttfamily hep-ph/0012348}}].

\bibitem{Landry:2002ix}
F.~Landry, R.~Brock, P.~M. Nadolsky and C.~P. Yuan, \emph{{Tevatron Run-1 $Z$
  boson data and Collins-Soper-Sterman resummation formalism}},
  \href{http://dx.doi.org/10.1103/PhysRevD.67.073016}{\emph{Phys. Rev.}
  {\bfseries D67} (2003) 073016},
  [\href{https://arxiv.org/abs/hep-ph/0212159}{{\ttfamily hep-ph/0212159}}].

\bibitem{Watt:2003vf}
G.~Watt, A.~D. Martin and M.~G. Ryskin, \emph{{Unintegrated parton
  distributions and electroweak boson production at hadron colliders}},
  \href{http://dx.doi.org/10.1103/PhysRevD.70.014012,
  10.1103/PhysRevD.70.079902}{\emph{Phys. Rev.} {\bfseries D70} (2004) 014012},
  [\href{https://arxiv.org/abs/hep-ph/0309096}{{\ttfamily hep-ph/0309096}}].

\bibitem{Catani:2007vq}
S.~Catani and M.~Grazzini, \emph{{An NNLO subtraction formalism in hadron
  collisions and its application to Higgs boson production at the LHC}},
  \href{http://dx.doi.org/10.1103/PhysRevLett.98.222002}{\emph{Phys. Rev.
  Lett.} {\bfseries 98} (2007) 222002},
  [\href{https://arxiv.org/abs/hep-ph/0703012}{{\ttfamily hep-ph/0703012}}].

\bibitem{Bozzi:2010xn}
G.~Bozzi, S.~Catani, G.~Ferrera, D.~de~Florian and M.~Grazzini,
  \emph{{Production of Drell-Yan lepton pairs in hadron collisions:
  Transverse-momentum resummation at next-to-next-to-leading logarithmic
  accuracy}},
  \href{http://dx.doi.org/10.1016/j.physletb.2010.12.024}{\emph{Phys. Lett.}
  {\bfseries B696} (2011) 207--213},
  [\href{https://arxiv.org/abs/1007.2351}{{\ttfamily 1007.2351}}].

\bibitem{Su:2014wpa}
P.~Sun, J.~Isaacson, C.~P. Yuan and F.~Yuan, \emph{{Universal Non-perturbative
  Functions for SIDIS and Drell-Yan Processes}},
  \href{https://arxiv.org/abs/1406.3073}{{\ttfamily 1406.3073}}.

\bibitem{DAlesio:2014mrz}
U.~D'Alesio, M.~G. Echevarria, S.~Melis and I.~Scimemi, \emph{{Non-perturbative
  QCD effects in $q_{T}$ spectra of Drell-Yan and Z-boson production}},
  \href{http://dx.doi.org/10.1007/JHEP11(2014)098}{\emph{JHEP} {\bfseries 11}
  (2014) 098}, [\href{https://arxiv.org/abs/1407.3311}{{\ttfamily 1407.3311}}].

\bibitem{Catani:2015vma}
S.~Catani, D.~de~Florian, G.~Ferrera and M.~Grazzini, \emph{{Vector boson
  production at hadron colliders: transverse-momentum resummation and leptonic
  decay}}, \href{http://dx.doi.org/10.1007/JHEP12(2015)047}{\emph{JHEP}
  {\bfseries 12} (2015) 047},
  [\href{https://arxiv.org/abs/1507.06937}{{\ttfamily 1507.06937}}].

\bibitem{Bacchetta:2017gcc}
A.~Bacchetta, F.~Delcarro, C.~Pisano, M.~Radici and A.~Signori,
  \emph{{Extraction of partonic transverse momentum distributions from
  semi-inclusive deep-inelastic scattering, Drell-Yan and Z-boson production}},
   \href{https://arxiv.org/abs/1703.10157}{{\ttfamily 1703.10157}}.

\bibitem{Guzey:2012jp}
V.~Guzey, M.~Guzzi, P.~M. Nadolsky, M.~Strikman and B.~Wang, \emph{{Massive
  neutral gauge boson production as a probe of nuclear modifications of parton
  distributions at the LHC}},
  \href{http://dx.doi.org/10.1140/epja/i2013-13035-6}{\emph{Eur. Phys. J.}
  {\bfseries A49} (2013) 35},
  [\href{https://arxiv.org/abs/1212.5344}{{\ttfamily 1212.5344}}].

\bibitem{Guzzi:2013aja}
M.~Guzzi, P.~M. Nadolsky and B.~Wang, \emph{{Nonperturbative contributions to a
  resummed leptonic angular distribution in inclusive neutral vector boson
  production}}, \href{http://dx.doi.org/10.1103/PhysRevD.90.014030}{\emph{Phys.
  Rev.} {\bfseries D90} (2014) 014030},
  [\href{https://arxiv.org/abs/1309.1393}{{\ttfamily 1309.1393}}].

\bibitem{Schweitzer:2010tt}
P.~Schweitzer, T.~Teckentrup and A.~Metz, \emph{{Intrinsic transverse parton
  momenta in deeply inelastic reactions}},
  \href{http://dx.doi.org/10.1103/PhysRevD.81.094019}{\emph{Phys. Rev.}
  {\bfseries D81} (2010) 094019},
  [\href{https://arxiv.org/abs/1003.2190}{{\ttfamily 1003.2190}}].

\bibitem{Schweitzer:2012hh}
P.~Schweitzer, M.~Strikman and C.~Weiss, \emph{{Intrinsic transverse momentum
  and parton correlations from dynamical chiral symmetry breaking}},
  \href{http://dx.doi.org/10.1007/JHEP01(2013)163}{\emph{JHEP} {\bfseries 01}
  (2013) 163}, [\href{https://arxiv.org/abs/1210.1267}{{\ttfamily 1210.1267}}].

\bibitem{Collins:1981va}
J.~C. Collins and D.~E. Soper, \emph{{Back-To-Back Jets: Fourier Transform from
  B to K-Transverse}},
  \href{http://dx.doi.org/10.1016/0550-3213(82)90453-9}{\emph{Nucl. Phys.}
  {\bfseries B197} (1982) 446--476}.

\bibitem{Collins:2014jpa}
J.~Collins and T.~Rogers, \emph{{Understanding the large-distance behavior of
  transverse-momentum-dependent parton densities and the Collins-Soper
  evolution kernel}},
  \href{http://dx.doi.org/10.1103/PhysRevD.91.074020}{\emph{Phys. Rev.}
  {\bfseries D91} (2015) 074020},
  [\href{https://arxiv.org/abs/1412.3820}{{\ttfamily 1412.3820}}].

\bibitem{Ito:1980ev}
A.~S. Ito et~al., \emph{{Measurement of the Continuum of Dimuons Produced in
  High-Energy Proton - Nucleus Collisions}},
  \href{http://dx.doi.org/10.1103/PhysRevD.23.604}{\emph{Phys. Rev.} {\bfseries
  D23} (1981) 604--633}.

\bibitem{Aad:2015auj}
{\scshape ATLAS} collaboration, G.~Aad et~al., \emph{{Measurement of the
  transverse momentum and $\phi ^*_{\eta }$ distributions of Drell–Yan lepton
  pairs in proton–proton collisions at $\sqrt{s}=8$ TeV with the ATLAS
  detector}},
  \href{http://dx.doi.org/10.1140/epjc/s10052-016-4070-4}{\emph{Eur. Phys. J.}
  {\bfseries C76} (2016) 291},
  [\href{https://arxiv.org/abs/1512.02192}{{\ttfamily 1512.02192}}].

\bibitem{web}
``\texttt{arTeMiDe} web-page, https://teorica.fis.ucm.es/artemide/.''

\bibitem{Kang:2012am}
Z.-B. Kang and J.-W. Qiu, \emph{{Nuclear modification of vector boson
  production in proton-lead collisions at the LHC}},
  \href{http://dx.doi.org/10.1016/j.physletb.2013.03.030}{\emph{Phys. Lett.}
  {\bfseries B721} (2013) 277--283},
  [\href{https://arxiv.org/abs/1212.6541}{{\ttfamily 1212.6541}}].

\bibitem{Drell:1970wh}
S.~D. Drell and T.-M. Yan, \emph{{Massive Lepton Pair Production in
  Hadron-Hadron Collisions at High-Energies}},
  \href{http://dx.doi.org/10.1103/PhysRevLett.25.316}{\emph{Phys. Rev. Lett.}
  {\bfseries 25} (1970) 316--320}.

\bibitem{Altarelli:1978id}
G.~Altarelli, R.~K. Ellis and G.~Martinelli, \emph{{Leptoproduction and
  Drell-Yan Processes Beyond the Leading Approximation in Chromodynamics}},
  \href{http://dx.doi.org/10.1016/0550-3213(78)90085-8,
  10.1016/0550-3213(78)90067-6}{\emph{Nucl. Phys.} {\bfseries B143} (1978)
  521}.

\bibitem{Tangerman:1994eh}
R.~D. Tangerman and P.~J. Mulders, \emph{{Intrinsic transverse momentum and the
  polarized Drell-Yan process}},
  \href{http://dx.doi.org/10.1103/PhysRevD.51.3357}{\emph{Phys. Rev.}
  {\bfseries D51} (1995) 3357--3372},
  [\href{https://arxiv.org/abs/hep-ph/9403227}{{\ttfamily hep-ph/9403227}}].

\bibitem{Kramer:1986sg}
G.~Kramer and B.~Lampe, \emph{{Two Jet Cross-Section in e+ e- Annihilation}},
  \href{http://dx.doi.org/10.1007/BF01679868}{\emph{Z. Phys.} {\bfseries C34}
  (1987) 497}.

\bibitem{Matsuura:1988sm}
T.~Matsuura, S.~C. van~der Marck and W.~L. van Neerven, \emph{{The Calculation
  of the Second Order Soft and Virtual Contributions to the Drell-Yan
  Cross-Section}},
  \href{http://dx.doi.org/10.1016/0550-3213(89)90620-2}{\emph{Nucl. Phys.}
  {\bfseries B319} (1989) 570--622}.

\bibitem{Idilbi:2006dg}
A.~Idilbi, X.-d. Ji and F.~Yuan, \emph{{Resummation of threshold logarithms in
  effective field theory for DIS, Drell-Yan and Higgs production}},
  \href{http://dx.doi.org/10.1016/j.nuclphysb.2006.07.002}{\emph{Nucl. Phys.}
  {\bfseries B753} (2006) 42--68},
  [\href{https://arxiv.org/abs/hep-ph/0605068}{{\ttfamily hep-ph/0605068}}].

\bibitem{Collins:1984kg}
J.~C. Collins, D.~E. Soper and G.~F. Sterman, \emph{{Transverse Momentum
  Distribution in Drell-Yan Pair and W and Z Boson Production}},
  \href{http://dx.doi.org/10.1016/0550-3213(85)90479-1}{\emph{Nucl. Phys.}
  {\bfseries B250} (1985) 199--224}.

\bibitem{Davies:1984sp}
C.~T.~H. Davies, B.~R. Webber and W.~J. Stirling, \emph{{Drell-Yan
  Cross-Sections at Small Transverse Momentum}},
  \href{http://dx.doi.org/10.1016/0550-3213(85)90402-X}{\emph{Nucl. Phys.}
  {\bfseries B256} (1985) 413}.

\bibitem{Ellis:1997ii}
R.~K. Ellis and S.~Veseli, \emph{{$W$ and $Z$ transverse momentum
  distributions: Resummation in $q_{T}$ space}},
  \href{http://dx.doi.org/10.1016/S0550-3213(97)00655-X}{\emph{Nucl. Phys.}
  {\bfseries B511} (1998) 649--669},
  [\href{https://arxiv.org/abs/hep-ph/9706526}{{\ttfamily hep-ph/9706526}}].

\bibitem{Olive:2016xmw}
{\scshape Particle Data Group} collaboration, C.~Patrignani et~al.,
  \emph{{Review of Particle Physics}},
  \href{http://dx.doi.org/10.1088/1674-1137/40/10/100001}{\emph{Chin. Phys.}
  {\bfseries C40} (2016) 100001}.

\bibitem{Bozzi:2008bb}
G.~Bozzi, S.~Catani, G.~Ferrera, D.~de~Florian and M.~Grazzini,
  \emph{{Transverse-momentum resummation: A Perturbative study of Z production
  at the Tevatron}},
  \href{http://dx.doi.org/10.1016/j.nuclphysb.2009.02.014}{\emph{Nucl. Phys.}
  {\bfseries B815} (2009) 174--197},
  [\href{https://arxiv.org/abs/0812.2862}{{\ttfamily 0812.2862}}].

\bibitem{Vladimirov:2017ksc}
A.~Vladimirov, \emph{{Structure of rapidity divergences in soft factors}},
  \href{https://arxiv.org/abs/1707.07606}{{\ttfamily 1707.07606}}.

\bibitem{Moch:2005id}
S.~Moch, J.~A.~M. Vermaseren and A.~Vogt, \emph{{The Quark form-factor at
  higher orders}},
  \href{http://dx.doi.org/10.1088/1126-6708/2005/08/049}{\emph{JHEP} {\bfseries
  08} (2005) 049}, [\href{https://arxiv.org/abs/hep-ph/0507039}{{\ttfamily
  hep-ph/0507039}}].

\bibitem{Vladimirov:2016dll}
A.~A. Vladimirov, \emph{{Soft-/rapidity- anomalous dimensions correspondence}},
  \href{http://dx.doi.org/10.1103/PhysRevLett.118.062001}{\emph{Phys. Rev.
  Lett.} {\bfseries 118} (2017) 062001},
  [\href{https://arxiv.org/abs/1610.05791}{{\ttfamily 1610.05791}}].

\bibitem{Li:2016ctv}
Y.~Li and H.~X. Zhu, \emph{{Bootstrapping Rapidity Anomalous Dimensions for
  Transverse-Momentum Resummation}},
  \href{http://dx.doi.org/10.1103/PhysRevLett.118.022004}{\emph{Phys. Rev.
  Lett.} {\bfseries 118} (2017) 022004},
  [\href{https://arxiv.org/abs/1604.01404}{{\ttfamily 1604.01404}}].

\bibitem{Becher:2013iya}
T.~Becher and G.~Bell, \emph{{Enhanced nonperturbative effects through the
  collinear anomaly}},
  \href{http://dx.doi.org/10.1103/PhysRevLett.112.182002}{\emph{Phys. Rev.
  Lett.} {\bfseries 112} (2014) 182002},
  [\href{https://arxiv.org/abs/1312.5327}{{\ttfamily 1312.5327}}].

\bibitem{Gehrmann:2014yya}
T.~Gehrmann, T.~Luebbert and L.~L. Yang, \emph{{Calculation of the transverse
  parton distribution functions at next-to-next-to-leading order}},
  \href{http://dx.doi.org/10.1007/JHEP06(2014)155}{\emph{JHEP} {\bfseries 06}
  (2014) 155}, [\href{https://arxiv.org/abs/1403.6451}{{\ttfamily 1403.6451}}].

\bibitem{Laenen:2000de}
E.~Laenen, G.~F. Sterman and W.~Vogelsang, \emph{{Higher order QCD corrections
  in prompt photon production}},
  \href{http://dx.doi.org/10.1103/PhysRevLett.84.4296}{\emph{Phys. Rev. Lett.}
  {\bfseries 84} (2000) 4296--4299},
  [\href{https://arxiv.org/abs/hep-ph/0002078}{{\ttfamily hep-ph/0002078}}].

\bibitem{Kulesza:2002rh}
A.~Kulesza, G.~F. Sterman and W.~Vogelsang, \emph{{Joint resummation in
  electroweak boson production}},
  \href{http://dx.doi.org/10.1103/PhysRevD.66.014011}{\emph{Phys. Rev.}
  {\bfseries D66} (2002) 014011},
  [\href{https://arxiv.org/abs/hep-ph/0202251}{{\ttfamily hep-ph/0202251}}].

\bibitem{Aybat:2011zv}
S.~M. Aybat and T.~C. Rogers, \emph{{TMD Parton Distribution and Fragmentation
  Functions with QCD Evolution}},
  \href{http://dx.doi.org/10.1103/PhysRevD.83.114042}{\emph{Phys. Rev.}
  {\bfseries D83} (2011) 114042},
  [\href{https://arxiv.org/abs/1101.5057}{{\ttfamily 1101.5057}}].

\bibitem{Catani:2009sm}
S.~Catani, L.~Cieri, G.~Ferrera, D.~de~Florian and M.~Grazzini, \emph{{Vector
  boson production at hadron colliders: a fully exclusive QCD calculation at
  NNLO}}, \href{http://dx.doi.org/10.1103/PhysRevLett.103.082001}{\emph{Phys.
  Rev. Lett.} {\bfseries 103} (2009) 082001},
  [\href{https://arxiv.org/abs/0903.2120}{{\ttfamily 0903.2120}}].

\bibitem{Affolder:1999jh}
{\scshape CDF} collaboration, T.~Affolder et~al., \emph{{The transverse
  momentum and total cross section of $e^+e^-$ pairs in the $Z$ boson region
  from $p\bar{p}$ collisions at $\sqrt{s} = 1.8$ TeV}},
  \href{http://dx.doi.org/10.1103/PhysRevLett.84.845}{\emph{Phys. Rev. Lett.}
  {\bfseries 84} (2000) 845--850},
  [\href{https://arxiv.org/abs/hep-ex/0001021}{{\ttfamily hep-ex/0001021}}].

\bibitem{Abbott:1999wk}
{\scshape D0} collaboration, B.~Abbott et~al., \emph{{Measurement of the
  inclusive differential cross section for $Z$ bosons as a function of
  transverse momentum in $\bar{p}p$ collisions at $\sqrt{s} = 1.8$ TeV}},
  \href{http://dx.doi.org/10.1103/PhysRevD.61.032004}{\emph{Phys. Rev.}
  {\bfseries D61} (2000) 032004},
  [\href{https://arxiv.org/abs/hep-ex/9907009}{{\ttfamily hep-ex/9907009}}].

\bibitem{Abbott:1999yd}
{\scshape D0} collaboration, B.~Abbott et~al., \emph{{Differential production
  cross section of $Z$ bosons as a function of transverse momentum at $\sqrt{s}
  = 1.8$ TeV}},
  \href{http://dx.doi.org/10.1103/PhysRevLett.84.2792}{\emph{Phys. Rev. Lett.}
  {\bfseries 84} (2000) 2792--2797},
  [\href{https://arxiv.org/abs/hep-ex/9909020}{{\ttfamily hep-ex/9909020}}].

\bibitem{Aaltonen:2012fi}
{\scshape CDF} collaboration, T.~Aaltonen et~al., \emph{{Transverse momentum
  cross section of $e^+e^-$ pairs in the $Z$-boson region from $p\bar{p}$
  collisions at $\sqrt{s}=1.96$ TeV}},
  \href{http://dx.doi.org/10.1103/PhysRevD.86.052010}{\emph{Phys. Rev.}
  {\bfseries D86} (2012) 052010},
  [\href{https://arxiv.org/abs/1207.7138}{{\ttfamily 1207.7138}}].

\bibitem{Abazov:2007ac}
{\scshape D0} collaboration, V.~M. Abazov et~al., \emph{{Measurement of the
  shape of the boson transverse momentum distribution in $p \bar{p} \to Z /
  \gamma^{*} \to e^+ e^- + X$ events produced at $\sqrt{s}$=1.96-TeV}},
  \href{http://dx.doi.org/10.1103/PhysRevLett.100.102002}{\emph{Phys. Rev.
  Lett.} {\bfseries 100} (2008) 102002},
  [\href{https://arxiv.org/abs/0712.0803}{{\ttfamily 0712.0803}}].

\bibitem{Harland-Lang:2014zoa}
L.~A. Harland-Lang, A.~D. Martin, P.~Motylinski and R.~S. Thorne, \emph{{Parton
  distributions in the LHC era: MMHT 2014 PDFs}},
  \href{http://dx.doi.org/10.1140/epjc/s10052-015-3397-6}{\emph{Eur. Phys. J.}
  {\bfseries C75} (2015) 204},
  [\href{https://arxiv.org/abs/1412.3989}{{\ttfamily 1412.3989}}].

\bibitem{Ridder:2016nkl}
A.~Gehrmann-De~Ridder, T.~Gehrmann, E.~W.~N. Glover, A.~Huss and T.~A. Morgan,
  \emph{{The NNLO QCD corrections to Z boson production at large transverse
  momentum}}, \href{http://dx.doi.org/10.1007/JHEP07(2016)133}{\emph{JHEP}
  {\bfseries 07} (2016) 133},
  [\href{https://arxiv.org/abs/1605.04295}{{\ttfamily 1605.04295}}].

\bibitem{Aad:2014xaa}
{\scshape ATLAS} collaboration, G.~Aad et~al., \emph{{Measurement of the
  $Z/\gamma^*$ boson transverse momentum distribution in $pp$ collisions at
  $\sqrt{s}$ = 7 TeV with the ATLAS detector}},
  \href{http://dx.doi.org/10.1007/JHEP09(2014)145}{\emph{JHEP} {\bfseries 09}
  (2014) 145}, [\href{https://arxiv.org/abs/1406.3660}{{\ttfamily 1406.3660}}].

\bibitem{Chatrchyan:2011wt}
{\scshape CMS} collaboration, S.~Chatrchyan et~al., \emph{{Measurement of the
  Rapidity and Transverse Momentum Distributions of $Z$ Bosons in $pp$
  Collisions at $\sqrt{s}=7$ TeV}},
  \href{http://dx.doi.org/10.1103/PhysRevD.85.032002}{\emph{Phys. Rev.}
  {\bfseries D85} (2012) 032002},
  [\href{https://arxiv.org/abs/1110.4973}{{\ttfamily 1110.4973}}].

\bibitem{Khachatryan:2016nbe}
{\scshape CMS} collaboration, V.~Khachatryan et~al., \emph{{Measurement of the
  transverse momentum spectra of weak vector bosons produced in proton-proton
  collisions at $\sqrt{s}$ = 8 TeV}}, {\emph{Submitted to: JHEP} (2016) },
  [\href{https://arxiv.org/abs/1606.05864}{{\ttfamily 1606.05864}}].

\bibitem{Aaij:2015gna}
{\scshape LHCb} collaboration, R.~Aaij et~al., \emph{{Measurement of the
  forward $Z$ boson production cross-section in $pp$ collisions at $\sqrt{s}=7$
  TeV}}, \href{http://dx.doi.org/10.1007/JHEP08(2015)039}{\emph{JHEP}
  {\bfseries 08} (2015) 039},
  [\href{https://arxiv.org/abs/1505.07024}{{\ttfamily 1505.07024}}].

\bibitem{Aaij:2015zlq}
{\scshape LHCb} collaboration, R.~Aaij et~al., \emph{{Measurement of forward W
  and Z boson production in $pp$ collisions at $ \sqrt{s}=8 $ TeV}},
  \href{http://dx.doi.org/10.1007/JHEP01(2016)155}{\emph{JHEP} {\bfseries 01}
  (2016) 155}, [\href{https://arxiv.org/abs/1511.08039}{{\ttfamily
  1511.08039}}].

\bibitem{Aaij:2016mgv}
{\scshape LHCb} collaboration, R.~Aaij et~al., \emph{{Measurement of the
  forward Z boson production cross-section in pp collisions at $\sqrt{s} = 13$
  TeV}}, \href{http://dx.doi.org/10.1007/JHEP09(2016)136}{\emph{JHEP}
  {\bfseries 09} (2016) 136},
  [\href{https://arxiv.org/abs/1607.06495}{{\ttfamily 1607.06495}}].

\bibitem{vanNeerven:1999ca}
W.~L. van Neerven and A.~Vogt, \emph{{NNLO evolution of deep inelastic
  structure functions: The Nonsinglet case}},
  \href{http://dx.doi.org/10.1016/S0550-3213(99)00668-9}{\emph{Nucl. Phys.}
  {\bfseries B568} (2000) 263--286},
  [\href{https://arxiv.org/abs/hep-ph/9907472}{{\ttfamily hep-ph/9907472}}].

\bibitem{Ogata_quadrature}
H.Ogata, \emph{{A numerical integration formula based on the Bessl functions}},
  \href{http://dx.doi.org/10.2977/prims/1145474602}{\emph{publ. RIMS. Kyoto
  Univ.} {\bfseries 41} (2005) 949--970}.

\bibitem{James:1975dr}
F.~James and M.~Roos, \emph{{Minuit: A System for Function Minimization and
  Analysis of the Parameter Errors and Correlations}},
  \href{http://dx.doi.org/10.1016/0010-4655(75)90039-9}{\emph{Comput. Phys.
  Commun.} {\bfseries 10} (1975) 343--367}.

\bibitem{James:1994vla}
F.~James, \emph{{MINUIT Function Minimization and Error Analysis: Reference
  Manual Version 94.1}}, .

\bibitem{Becher:2011xn}
T.~Becher, M.~Neubert and D.~Wilhelm, \emph{{Electroweak Gauge-Boson Production
  at Small $q_T$: Infrared Safety from the Collinear Anomaly}},
  \href{http://dx.doi.org/10.1007/JHEP02(2012)124}{\emph{JHEP} {\bfseries 02}
  (2012) 124}, [\href{https://arxiv.org/abs/1109.6027}{{\ttfamily 1109.6027}}].

\end{thebibliography}\endgroup
\end{document}